\documentclass[useAMS,usenatbib]{mn2e}
\usepackage{graphicx,amssymb, txfonts,pstricks}

\def\p1{Paper~I}

\def\kms {$\rm km\,s^{-1}$}

\title[Stellar kinematics of nearby AGNs]{Gemini NIFS survey of feeding and feedback processes in nearby Active Galaxies: I - Stellar kinematics}

\author[Riffel, R. A. et al.]{Rogemar A. Riffel$^{1}$\thanks{E-mail: rogemar@ufsm.br}, Thaisa Storchi-Bergmann$^{2}$, Rogerio Riffel$^{2}$,  Luis G. Dahmer-Hahn$^2$, 
\newauthor Marlon R. Diniz$^1$, Astor J. Sch\"onell$^2$,  Natacha Z. Dametto$^2$ \\
$^{1}$ Departamento de F\'\i sica, Centro de Ci\^encias Naturais e Exatas, Universidade Federal de Santa Maria, 97105-900, Santa Maria, RS, Brazil \\ 
$^{2}$ Departamento de Astronomia, Instituto de F\'\i sica, Universidade Federal do Rio Grande do Sul, CP 15051, 91501-970, Porto Alegre, RS, Brazil \\
}

\begin{document}

\date{Accepted 1988 December 15. Received 1988 December 14; in original form 1988 October 11}

\pagerange{\pageref{firstpage}--\pageref{lastpage}} \pubyear{2014}

\maketitle

\label{firstpage}

\begin{abstract}

We use the Gemini Near-Infrared Integral Field Spectrograph (NIFS) to map the stellar kinematics
of the inner few hundred parsecs of a sample of 16 nearby Seyfert galaxies, at
a spatial resolution of  tens of parsecs and spectral resolution of 40~\kms. We 
find that the line-of-sight (LOS) velocity fields for most galaxies are well reproduced 
by rotating disk models. The kinematic position angle (PA) derived for the LOS
velocity field is consistent with the large scale photometric PA.
 The residual velocities are correlated with the hard X-ray luminosity,  suggesting that more luminous AGN have a larger impact in the surrounding stellar dynamics. The central velocity dispersion values are usually higher than the rotation velocity amplitude, what we attribute to the strong contribution of bulge kinematics
in these inner regions. For 50\% of the galaxies, we find an inverse correlation between the velocities and the $h_3$ Gauss-Hermitte moment, 
implying red wings in the blueshifted side and blue wings in the redshifted side of the velocity field, attributed to the movement of the bulge stars lagging the rotation.
Two of the 16 galaxies (NGC 5899 and Mrk 1066) show an S-shape zero velocity line, attributed to the gravitational potential of a nuclear bar. Velocity dispersion ($\sigma$) maps show rings of low-$\sigma$  values ($\sim50-80$\,\kms) for 4 objects and ``patches" of low-$\sigma$  for 6 galaxies at 150--250 pc from the nucleus, attributed
to young/ intermediate age stellar populations.

\end{abstract}

\begin{keywords}
galaxies: kinematics and dynamics -- galaxies: active -- galaxies: ISM -- infrared: galaxies
\end{keywords}

\begin{table*}
\centering
\small
\caption{Sample: (1) Galaxy's name; (2) redshift; (3) nuclear activity; (4) Hubble type as quoted in NED; (5) Spectral Resolution in $\AA$; (6) Angular Resolution; (7) Spectral Band; (8) Gemini project code; (9) total exposure time; (10) hard X-ray luminosity (14-195 keV) from the Swift-BAT 60-month catalogue \citep{ajello12} ;(11) reference for the stellar kinematics. }
\vspace{0.3cm}
\begin{tabular}{l l l l l l l l l l l}
\hline
1 	         & 2           & 3  &4     & 5  & 6 & 7 &8 & 9 & 10 & 11   \\
Galaxy 	         & $z$     & Act.   & Hub. Type & Sp. R. & An. R. & B & Project ID & Exp. T. &  log $L_X$ [erg s$^{-1}$] & Ref.  \\
\hline	
         NGC788  &  0.014 & Sy2   & SA0/a?(s)	    &3.5 &0\farcs13 & K  &GN-2015B-Q-29  &4400 & 43.2 & a \\
NGC1052$^\dagger$&  0.005 & Sy2   & E4  	    &3.4 &0\farcs14 & Kl &GN-2010B-Q-25  &2400 & 41.9 & a \\
        NGC2110  &  0.008 & Sy2   & SAB0$^-$	    &3.4 &0\farcs15 & Kl &GN-2010B-Q-25  &3600 & 43.3 & b\\
        NGC3227  &  0.004 & Sy1.5 & SAB(s)a pec     &3.4 &0\farcs13 & Kl &GN-2016A-Q-6   &2400 & 42.3 & a\\
        NGC3516  &  0.009 & Sy1.5 & (R)SB0$^0$?(s)  &3.6 &0\farcs15 & K  &GN-2015A-Q-3   &4500 & 42.7 & a\\
NGC4051$\dagger$ & 0.002  & Sy1   & SAB(rs)bc	    &3.2 &0\farcs17 & K  &GN-2006A-SV-123&4500 & 41.5 & c\\
        NGC4235  &  0.008 & Sy1   & SA(s)a edge-on  &3.4 &0\farcs13 & Kl &GN-2016A-Q-6   &4000 & 42.3 & a\\
        NGC4388  &  0.008 & Sy2   & SA(s)b? edge-on &3.5 &0\farcs14 & K  &GN-2015A-Q-3   &800  & 43.3 & a\\
        NGC5506  &  0.006 & Sy1.9 & Sa pec edge-on  &3.5 &0\farcs15 & K  &GN-2015A-Q-3   &4000 & 43.1 & a\\
NGC5548$\dagger$ & 0.017  & Sy1   & (R')SA0/a(s)    &3.5 &0\farcs20 & Kl &GN-2012A-Q-57  &2700 & 43.4 & a\\
NGC5899$\dagger$ &  0.009 & Sy2   & SAB(rs)c	    &3.5 & 0\farcs12& Kl &GN-2013A-Q-48  &4600 & 42.1 & a\\
NGC5929$\dagger$ & 0.008  & Sy2   & Sab? pec	    &3.2 &0\farcs12 & Kl &GN-2011A-Q-43  &6000 & --   & d\\
Mrk607$\dagger$  & 0.009  & Sy2   & Sa? edge-on     & 3.5& 0\farcs12& Kl &GN-2012B-Q-45  &6000 & --   & a \\
         Mrk766  &  0.013 & Sy1.5 & (R')SB(s)a?     &3.5 &0\farcs19 & Kl &GN-2010A-Q-42  &3300 & 42.8 & a\\
Mrk1066$\dagger$ & 0.012  & Sy2   & (R)SB0$^+$(s)   & 3.3&0\farcs15 & Kl &GN-2008B-Q-30  &4800 & --   & e\\
Mrk1157$\dagger$ & 0.015  & Sy2   &(R')SB0/a	    &3.5 &0\farcs12 &Kl  &GN-2009B-Q-27  &3300 & --   & f\\
\hline														       
\multicolumn{11}{l}{$\dagger$These galaxies do not follow all selection criteria and are part of the complementary sample.} \\														       
\multicolumn{11}{l}{a -- This work;  b -- \citet{n2110}; c -- \citet{n4051}; d -- \citet{n5929};  e -- \citet{m1066_kin};}   \\
\multicolumn{11}{l}{ f -- \citet{m1157_kin}}  \\
\end{tabular}
 \label{tab_obs}
\end{table*}

\section{Introduction}

Active Galactic Nuclei (AGN) characterize a critical phase in galaxy evolution in which its nuclear super-massive black hole (SMBH) is being fed due to gas accretion to the nuclear region. Once the accretion disk surrounding the SMBH is formed, feedback processes begin to occur, comprising jets of relativistic particles emitted from the inner rim of the accretion disk,  winds emanating from outer regions of the disk and radiation emitted by the hot gas in the disk or by its corona \citep{frank02,elvis00,ciotti10}. The AGN feeding and feedback processes couple the growth of the SMBHs and their host galaxies, and are claimed to explain the correlation between the mass of the SMBH and the mass of the galaxy bulge \citep{ferrarese05,somerville08,kormendy13}. The feeding processes are a necessary condition to trigger the nuclear activity, while the feedback processes are now a fundamental ingredient in galaxy evolution models: without AGN feedback, the models cannot reproduce the properties of the massive galaxies -- these galaxies end up forming too many stars \citep{springel05,fabian12,terrazas16}.  

The study of feeding and feedback processes in AGNs requires a detailed mapping of the gas and stellar kinematics in the vicinity of the central engine. Usually these studies are based on high spatial resolution Integral Field Spectroscopy \citep[IFS, e.g.][]{fathi06,barbosa09,n1068-kin,sanchez09,hicks09,hicks13,sb07,sb09,sb10,mazzalay14,m79,m1066_kin,n5929_let,n2110}. To isolate and quantify  gas streaming motions towards the center of galaxies or gas outflows from the nucleus through the gas kinematics, it is necessary to properly map the motions of the gas due to the gravitational potential of the galaxy. A way to map the gravitational potential of the galaxies is by two-dimensional mapping of the stellar kinematics. So far, studies available for nearby galaxies show that the stellar kinematics usually present regular rotation within the inner kiloparsec \citep[e.g.][]{barbosa06,ganda06} and thus can be used to isolate possible non-circular motions from the gas kinematics \citep[e.g.][]{n4051,m1066_kin,n2110}.  

Adaptive optics IFS at 10-m class telescopes provides an unprecedented possibility to map the stellar kinematics of nearby galaxies at spatial resolutions of a few tens of parsecs. So  far, adaptive optics systems are available mainly in near-infrared (near-IR) wavelengths, where the dust extinction at the central regions of galaxies is less important than at optical bands. In addition, strong absorption CO bands are present in the spectra of galaxies, originated in giant and super-giant stars that dominate the continuum emission in the central regions \citep[e.g.][]{maraston05,rogerio15}. Thus, near-IR IFS at 10-m telescopes provides a unique tool to map the stellar kinematics at central region of active galaxies, at unprecedented spatial resolution,  by fitting the CO absorption band-heads \citep[e.g.][]{rogemar15}.  

In this paper, we map the stellar kinematics of the inner 3$^{\prime\prime}\times$3$^{\prime\prime}$ of a sample of 16 nearby Seyfert galaxies. This work is part of a larger project in which our group AGNIFS ({\it AGN Integral Field Spectroscopy}) aims to study the feeding and feedback processes of a sample of 30 nearby Seyfert galaxies, selected by their X-ray luminosity. The kinematic maps presented here will be used to isolate gas non-circular motions in future works by our group and to quantify gas inflows and outflows. 

This paper is organized as follows: section~\ref{obs} presents the sample and a description of the observations and data reduction procedure, while the spectral fitting procedure is presented in sec.~\ref{fit}. The resulting stellar kinematics maps are presented and discussed in section~\ref{stelkin} and the kinematic derived parameters are compared with those characterizing the large scale disks of the host galaxies in section~\ref{comp}. Finally, section~\ref{conc} presents the main conclusions of this work.

\begin{figure}
\begin{tabular}{c c}
\includegraphics[scale=0.22]{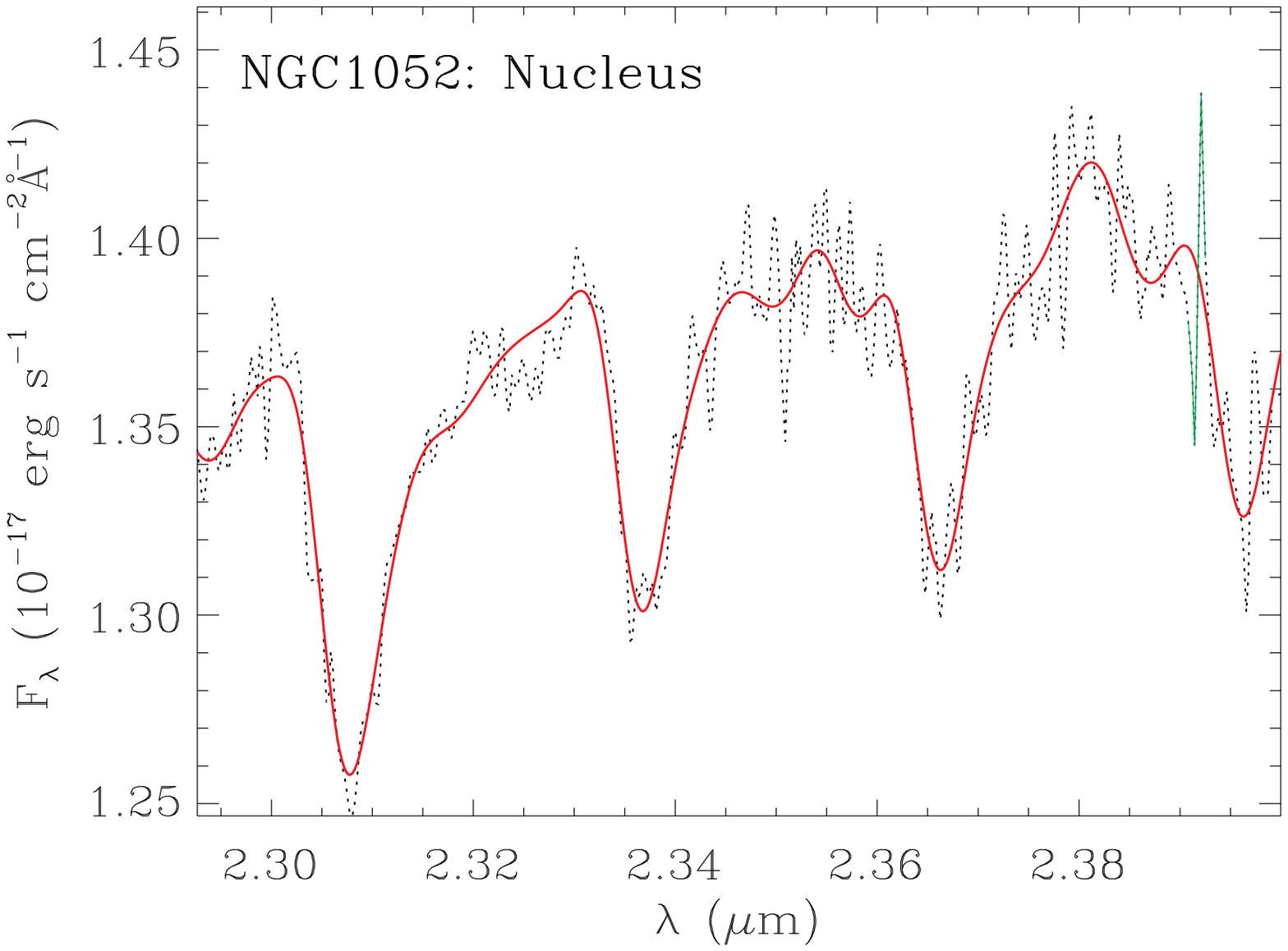} & 
\includegraphics[scale=0.22]{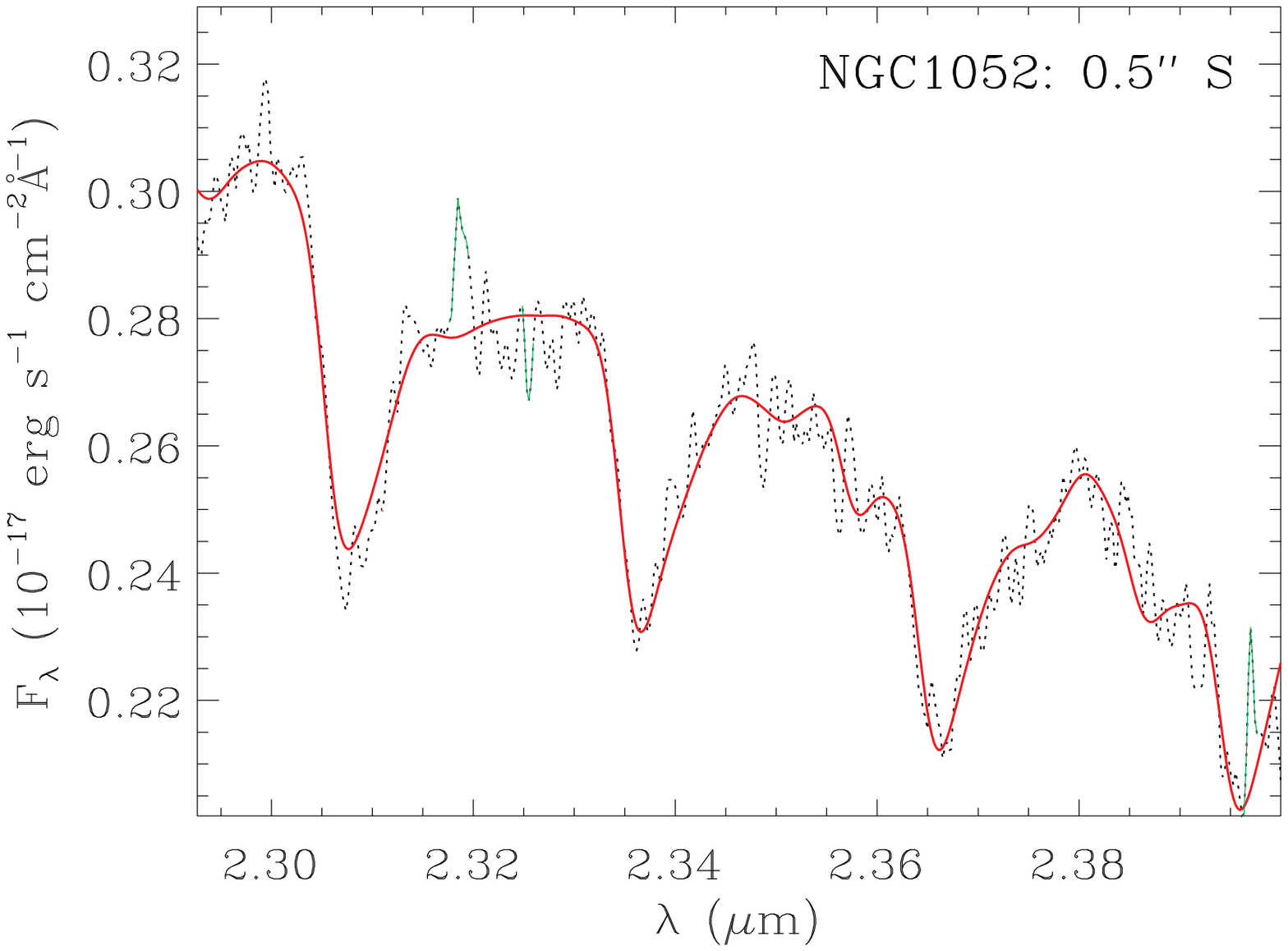}\\

\includegraphics[scale=0.22]{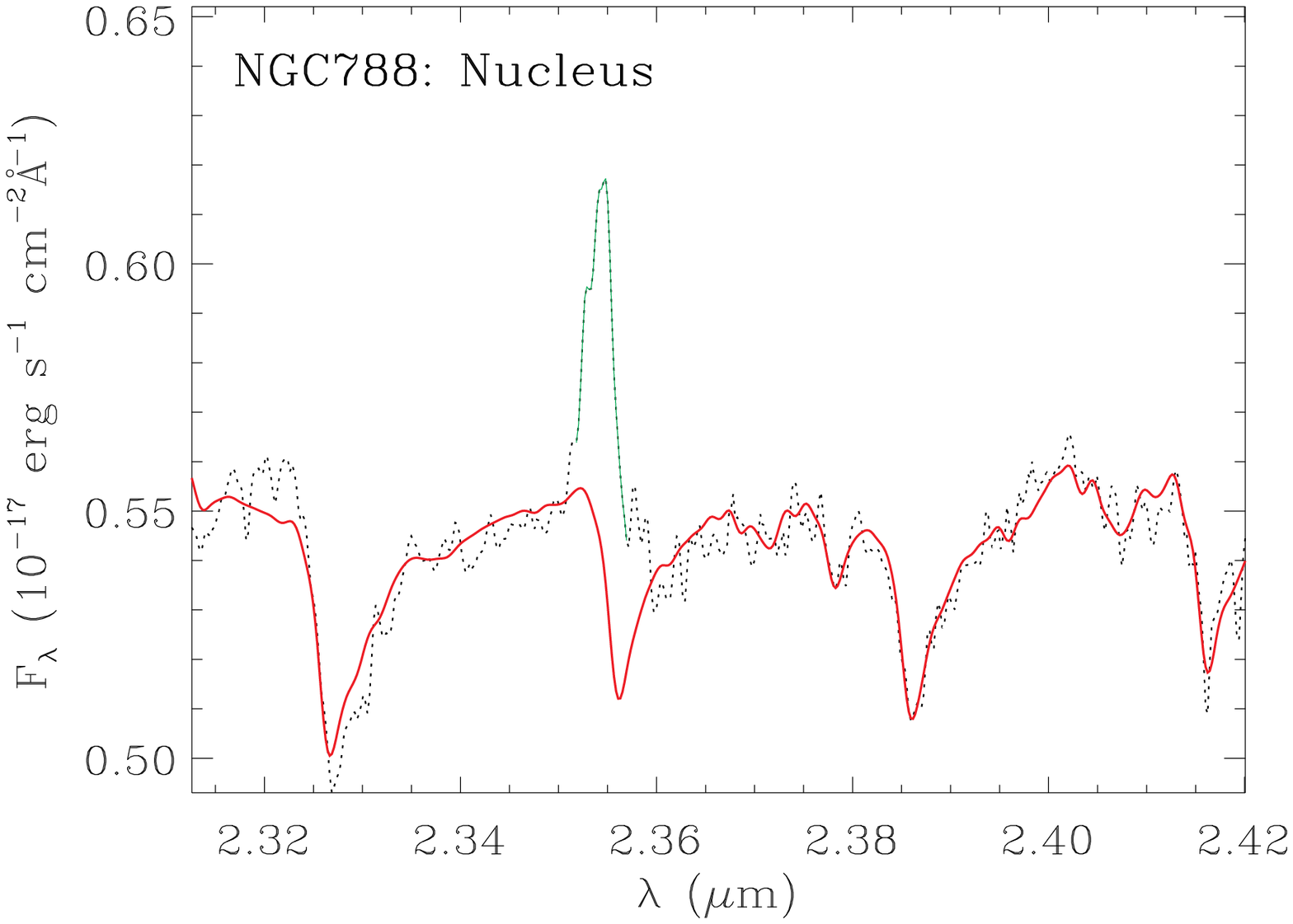} & 
\includegraphics[scale=0.225]{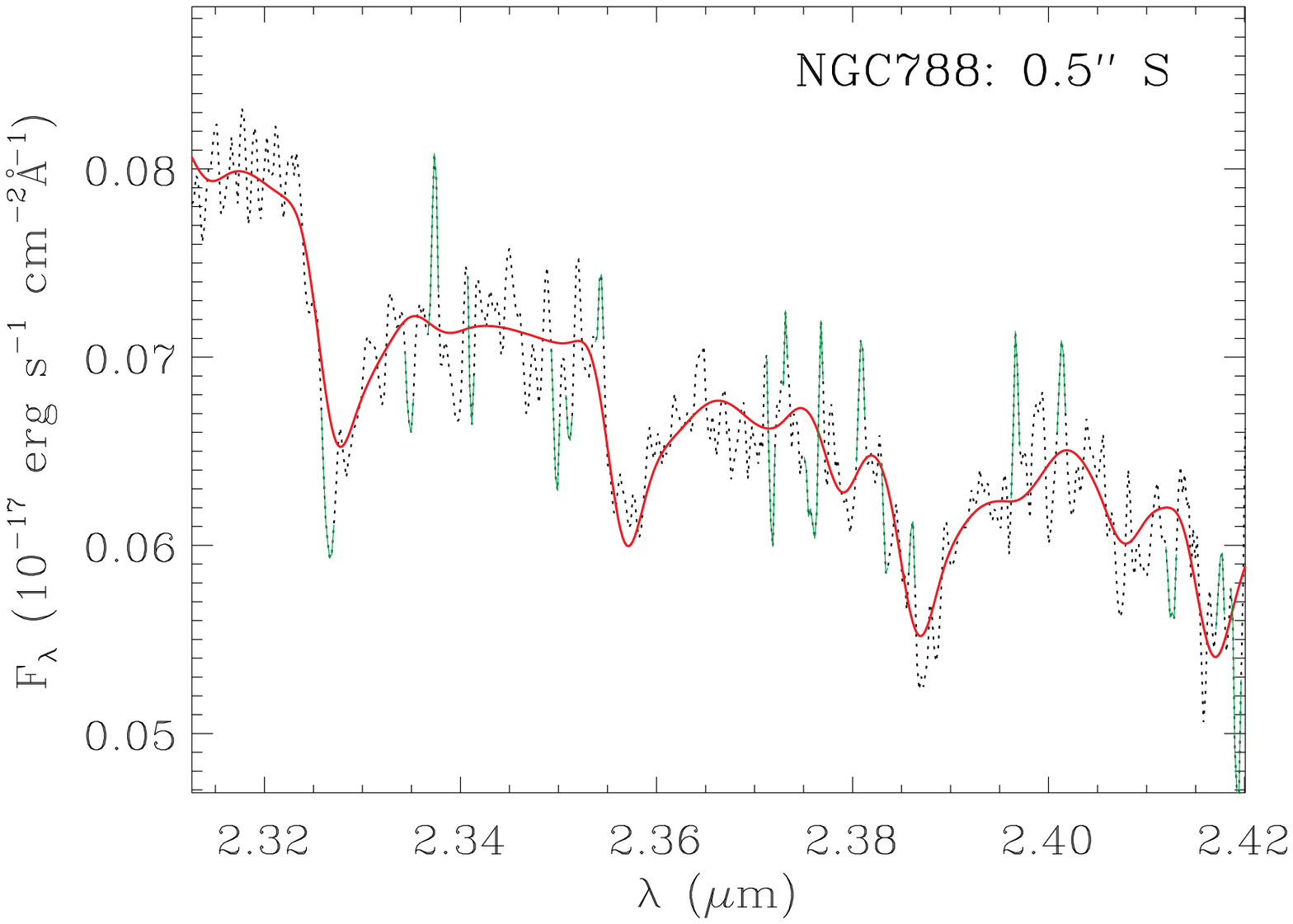}\\

\includegraphics[scale=0.22]{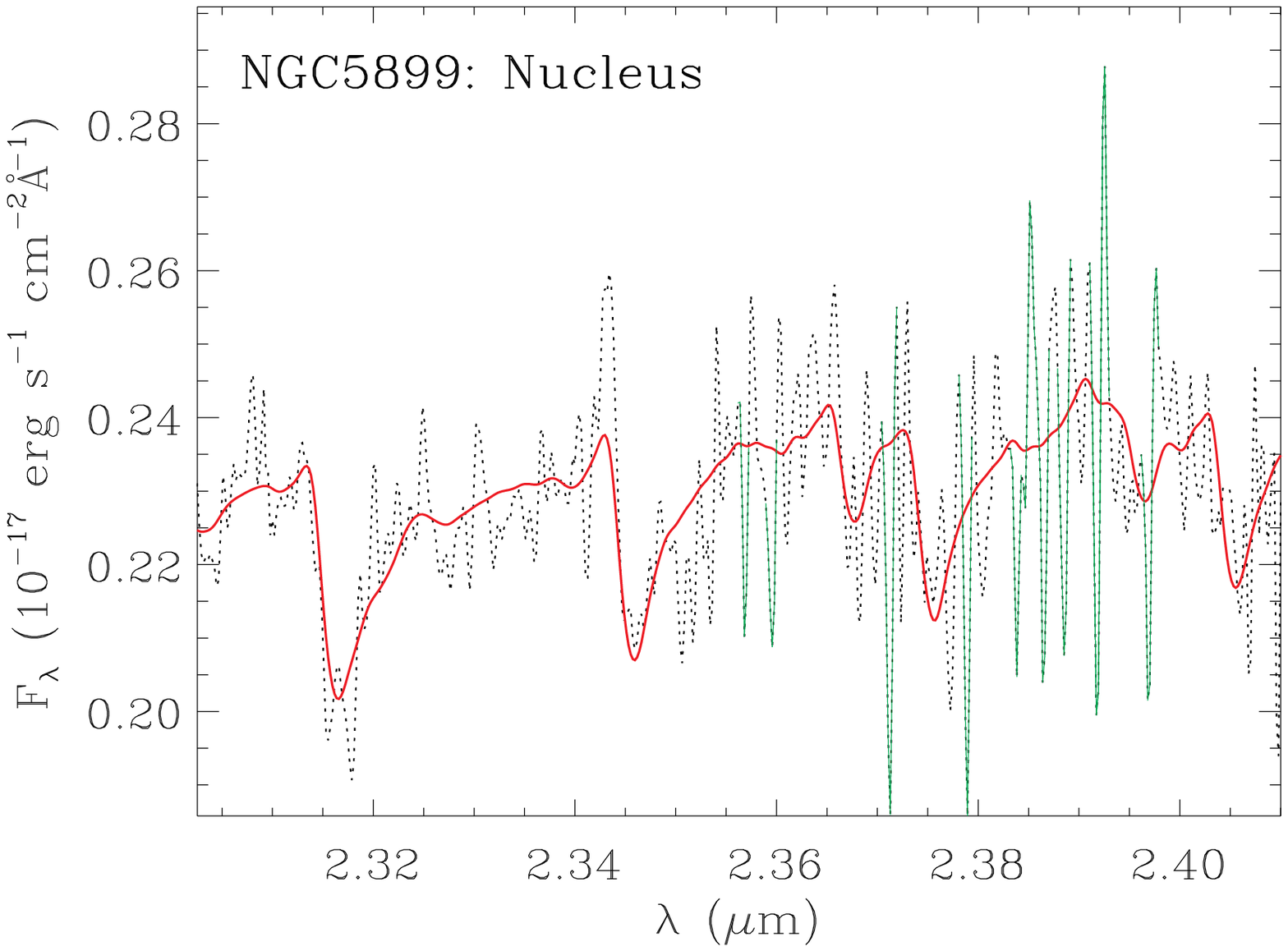} & 
\includegraphics[scale=0.225]{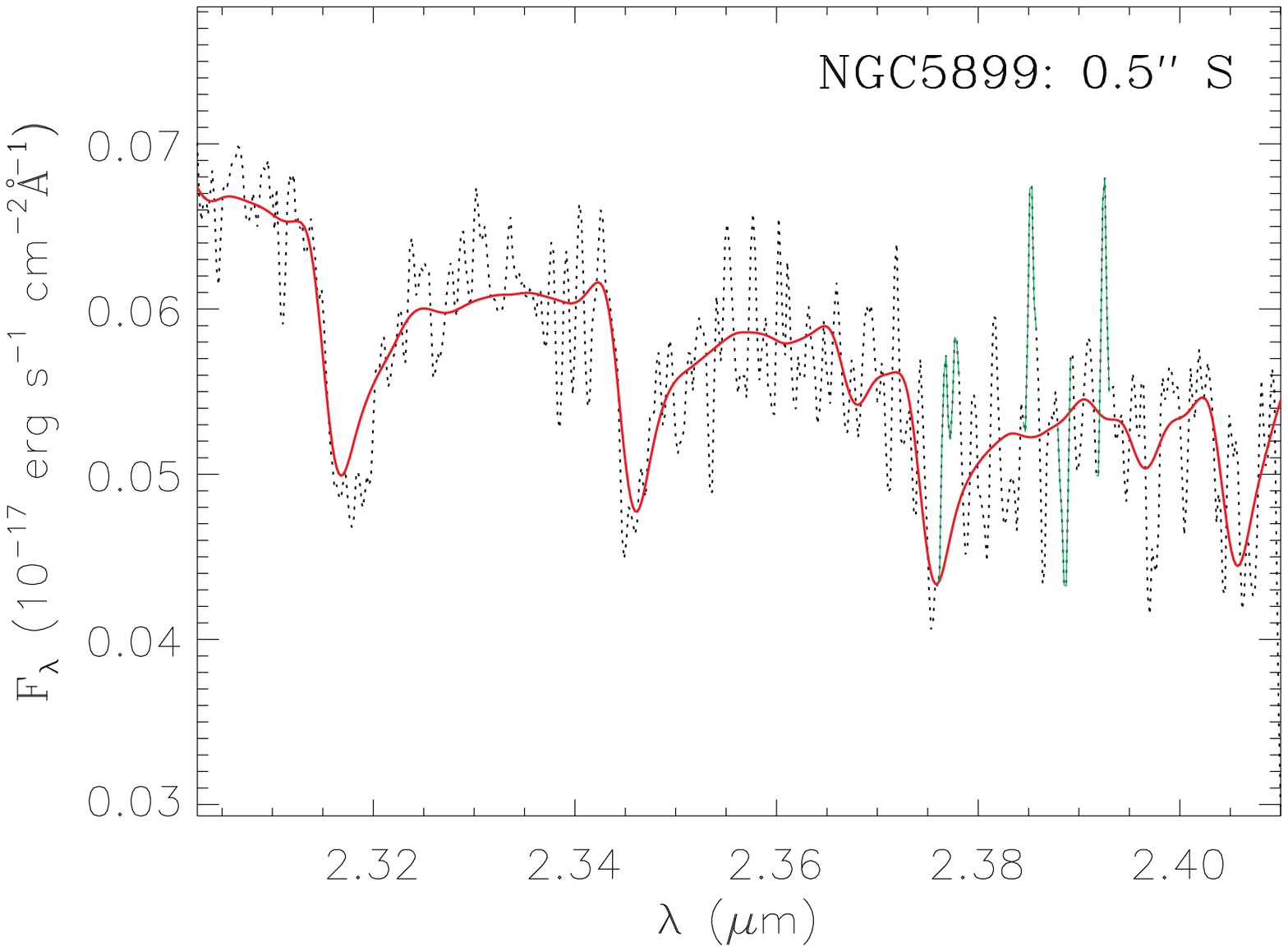}

\end{tabular}
\caption{Examples of fits for the nuclear spectrum (left) and typical extra nuclear spectrum (right) for NGC\,1052 (top panels), NGC\,788 (center panels) and NGC\,5899 (bottom panels). The observed spectra are shown as black dotted lines, the best fitted models as red continuous lines and the masked regions (following the criteria explained in the text) during the fits are shown in green.}
\label{sample_fits}
\end{figure}

\begin{figure*}
\includegraphics[scale=0.7]{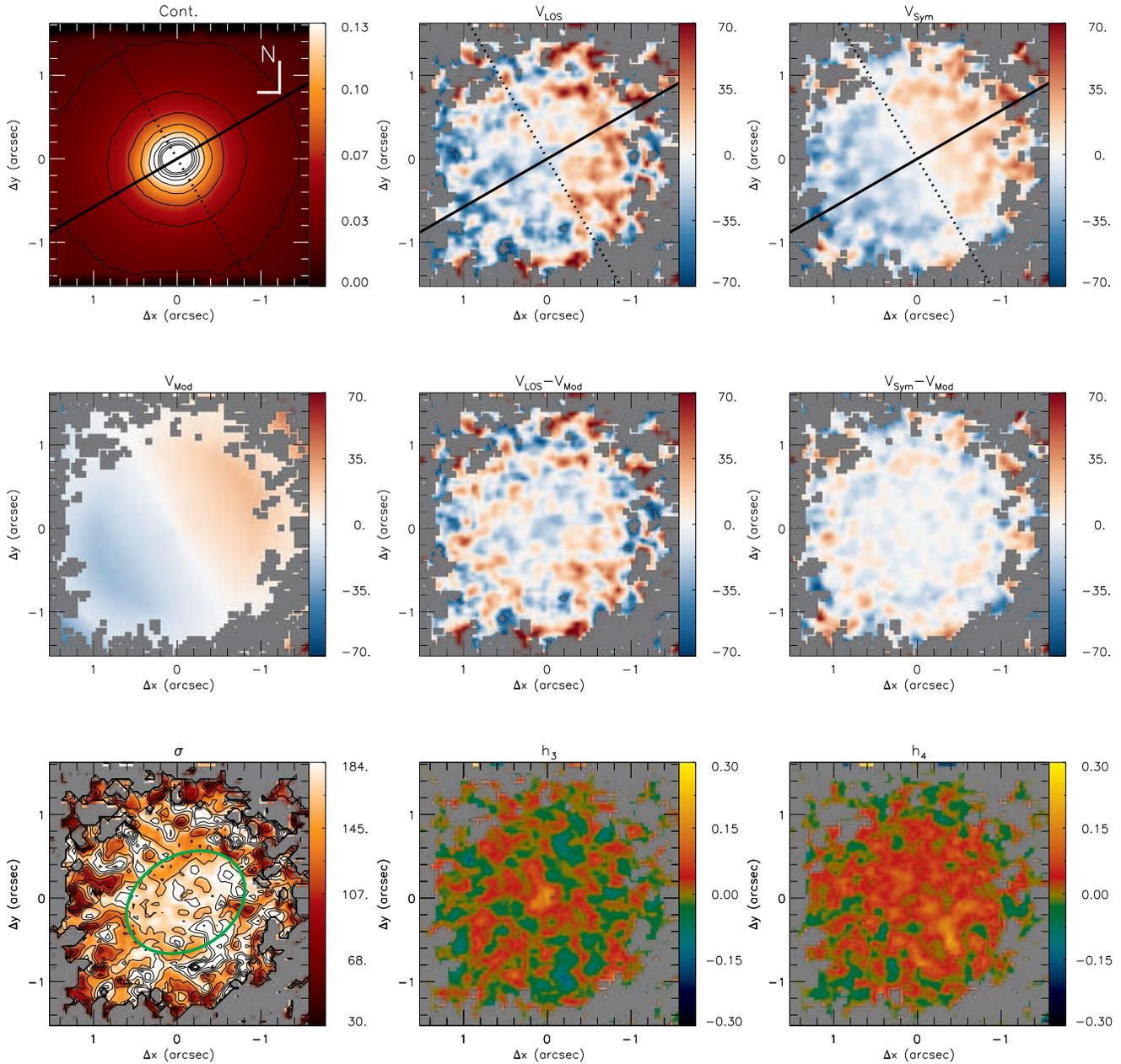}
\caption{NGC\,788: Top-left: K-band continuum image obtained by averaging the spectra, with the color bar shown in units of 10$^{-17}$ erg\,s$^{-1}$\,cm$^{-2}$\,$\AA^{-1}$; top-middle: stellar velocity field; top-right: symmetrized velocity field; middle-left: rotating disk model; middle-middle: residual map for the symmetrized velocity field; middle-right: residual map for the observed velocity field; bottom-left: stellar velocity dispersion; bottom-middle: h$_{3}$ Gauss-Hermite moment and bottom-right: h$_{4}$ Gauss-Hermite moment. Gray regions are masked locations and correspond to regions where the signal-to-noise of the spectra was not high enough to get reliable fits. The continuous line identifies the orientation of the line of nodes and the dotted line marks the orientation of the minor axis of the galaxy. North is up and East left in all maps. The color bar for velocity, model, residual maps and $\sigma$ show the velocities in units of \kms\ and the systemic velocity of the galaxy was subtracted from the observed velocities.}
\label{n788}
\end{figure*}

\begin{figure*}
\includegraphics[scale=0.7]{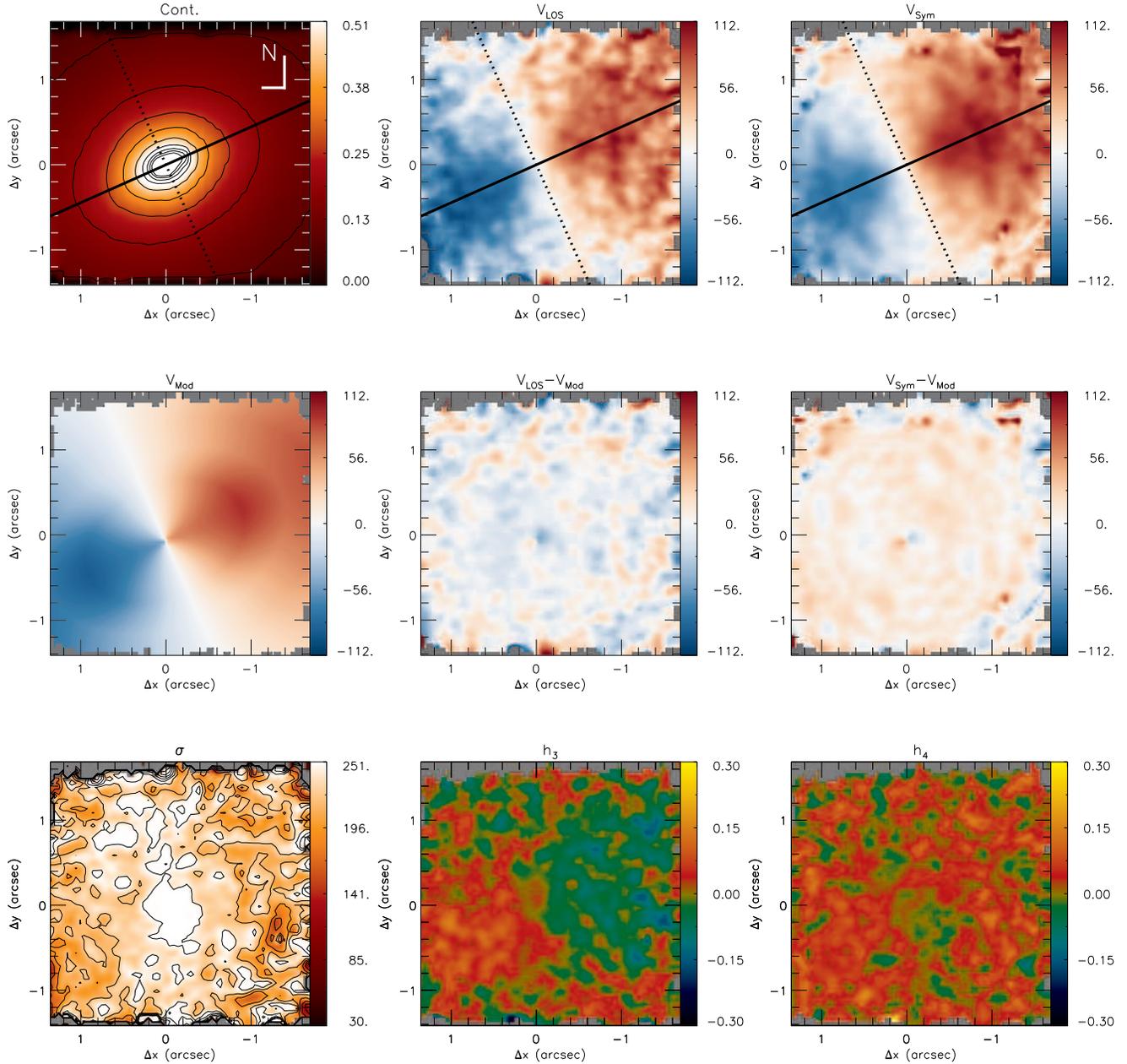}
\caption{Same as Fig.~\ref{n788} for NGC\,1052.}
\label{n1052}
\end{figure*}

\begin{figure*}
\includegraphics[scale=0.7]{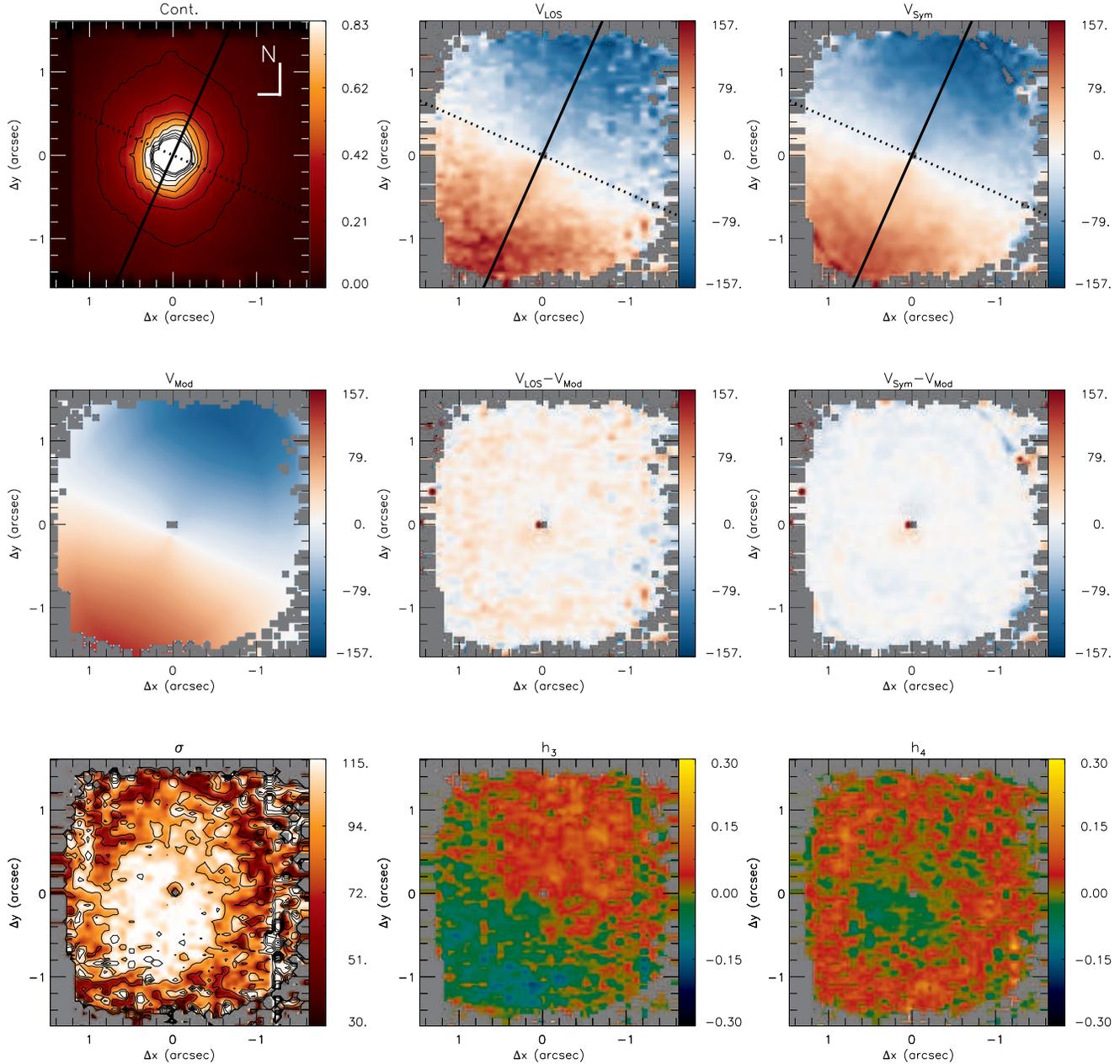}
\caption{Same as Fig.~\ref{n788} for NGC\,3227.}
\label{n3227}
\end{figure*}

\begin{figure*}
\includegraphics[scale=0.7]{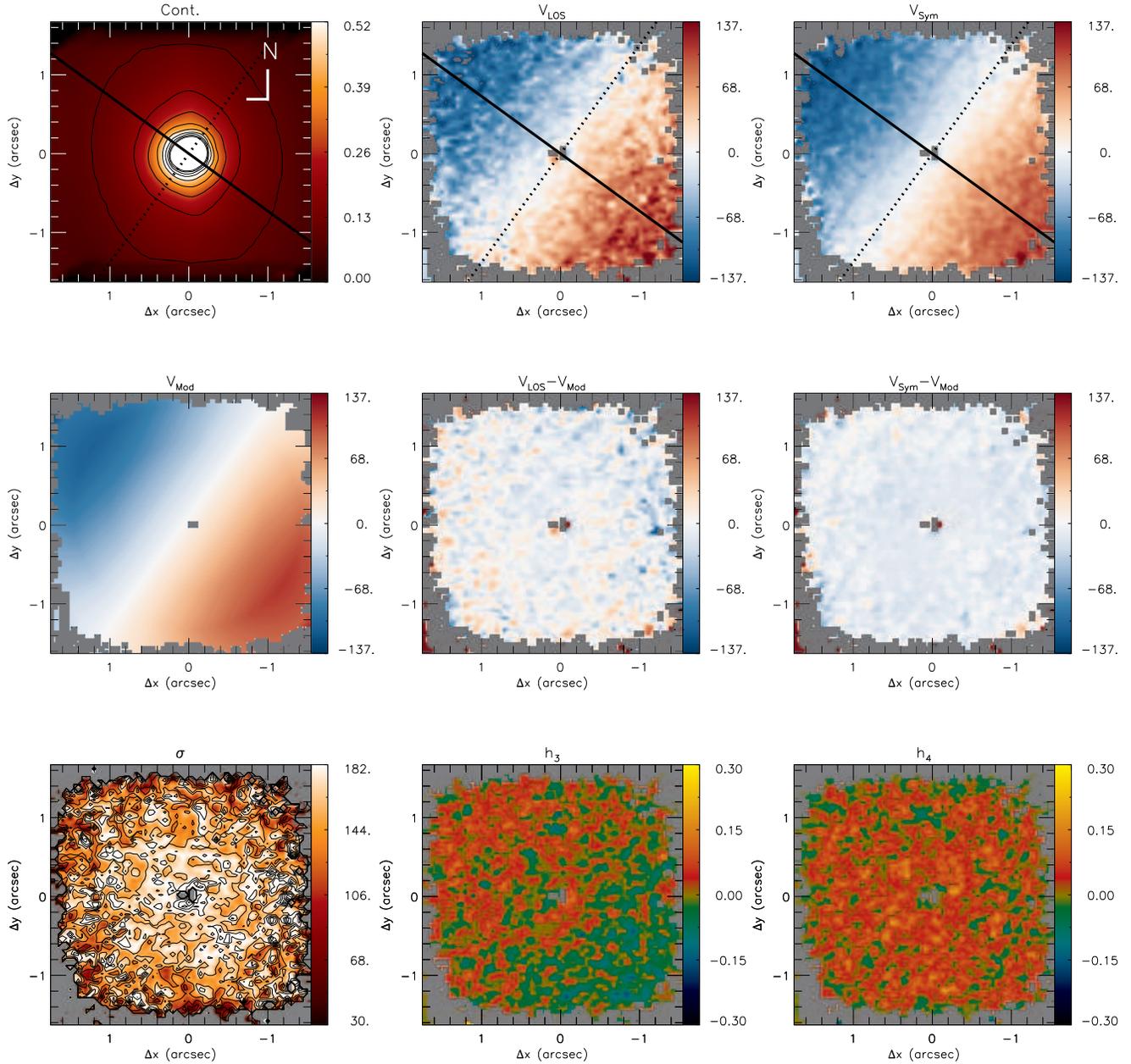}
\caption{Same as Fig.~\ref{n788} for NGC\,3516.}
\label{n3516}
\end{figure*}

\begin{figure*}
\includegraphics[scale=0.7]{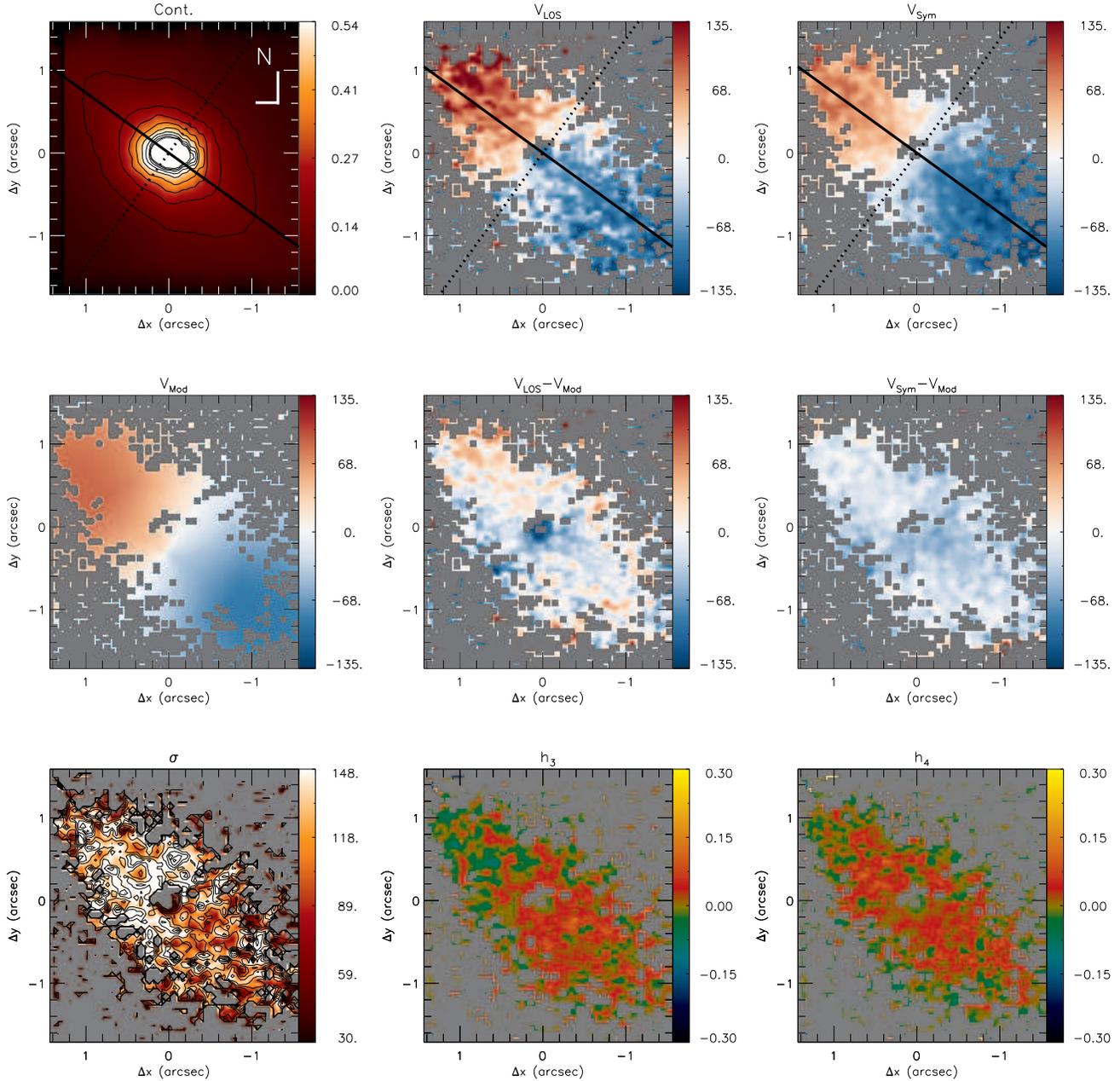}
\caption{Same as Fig.~\ref{n788} for NGC\,4235.}
\label{n4235}
\end{figure*}

\begin{figure*}
\includegraphics[scale=0.7]{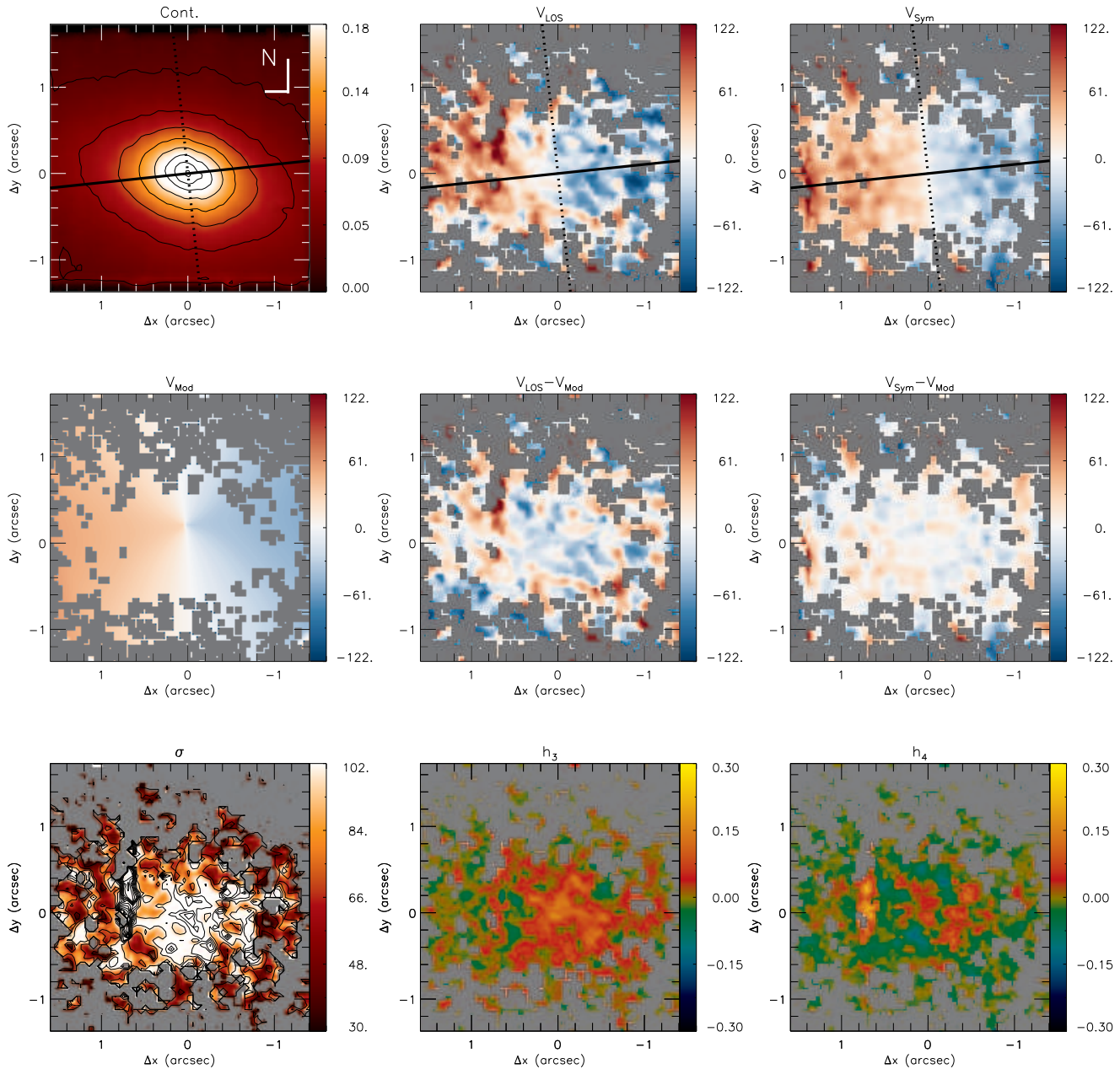}
\caption{Same as Fig.~\ref{n788} for NGC\,4388.}
\label{n4388}
\end{figure*}

\begin{figure*}
\includegraphics[scale=0.7]{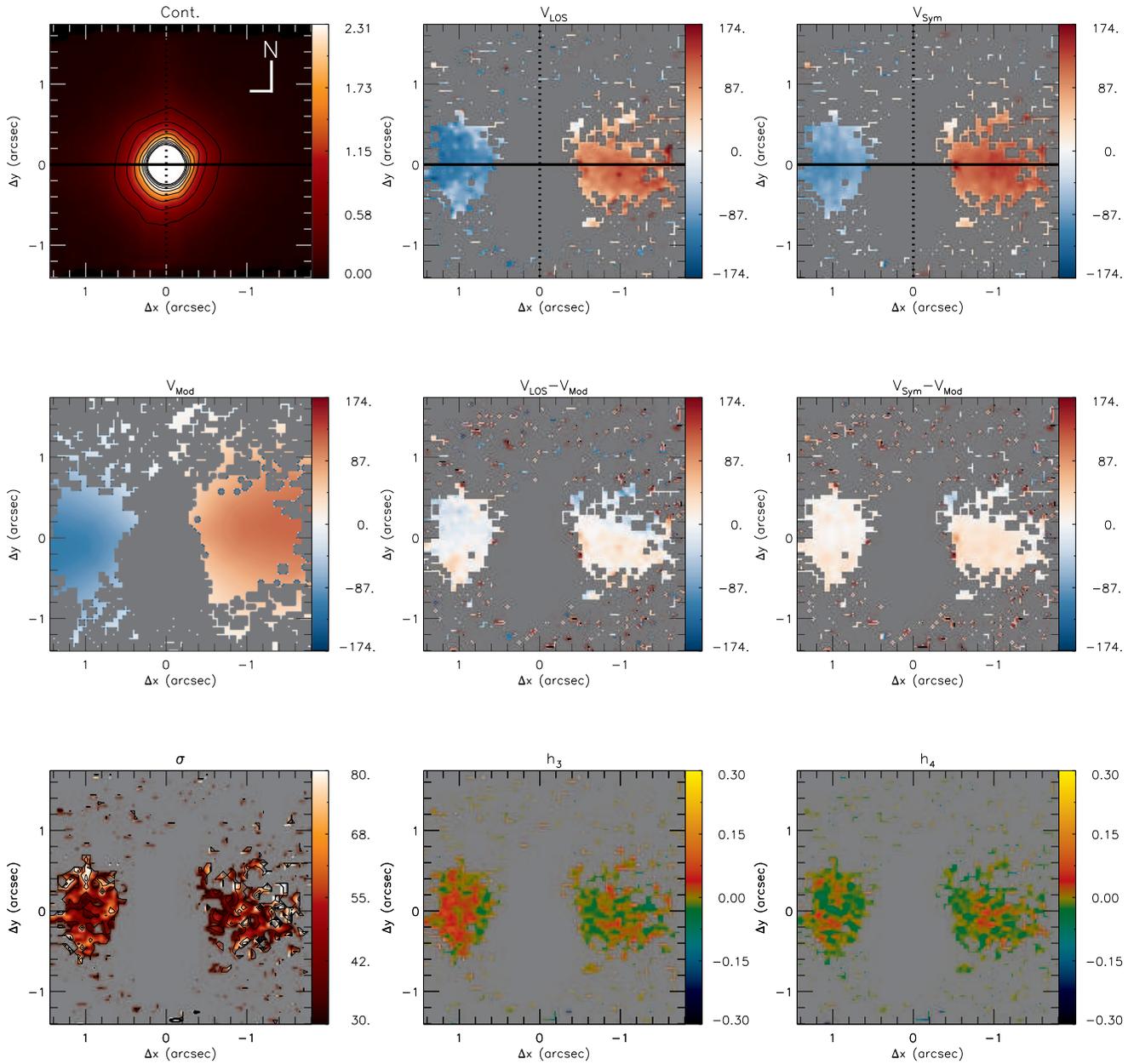}
\caption{Same as Fig.~\ref{n788} for NGC\,5506.}
\label{n5506}
\end{figure*}

\begin{figure*}
\includegraphics[scale=0.7]{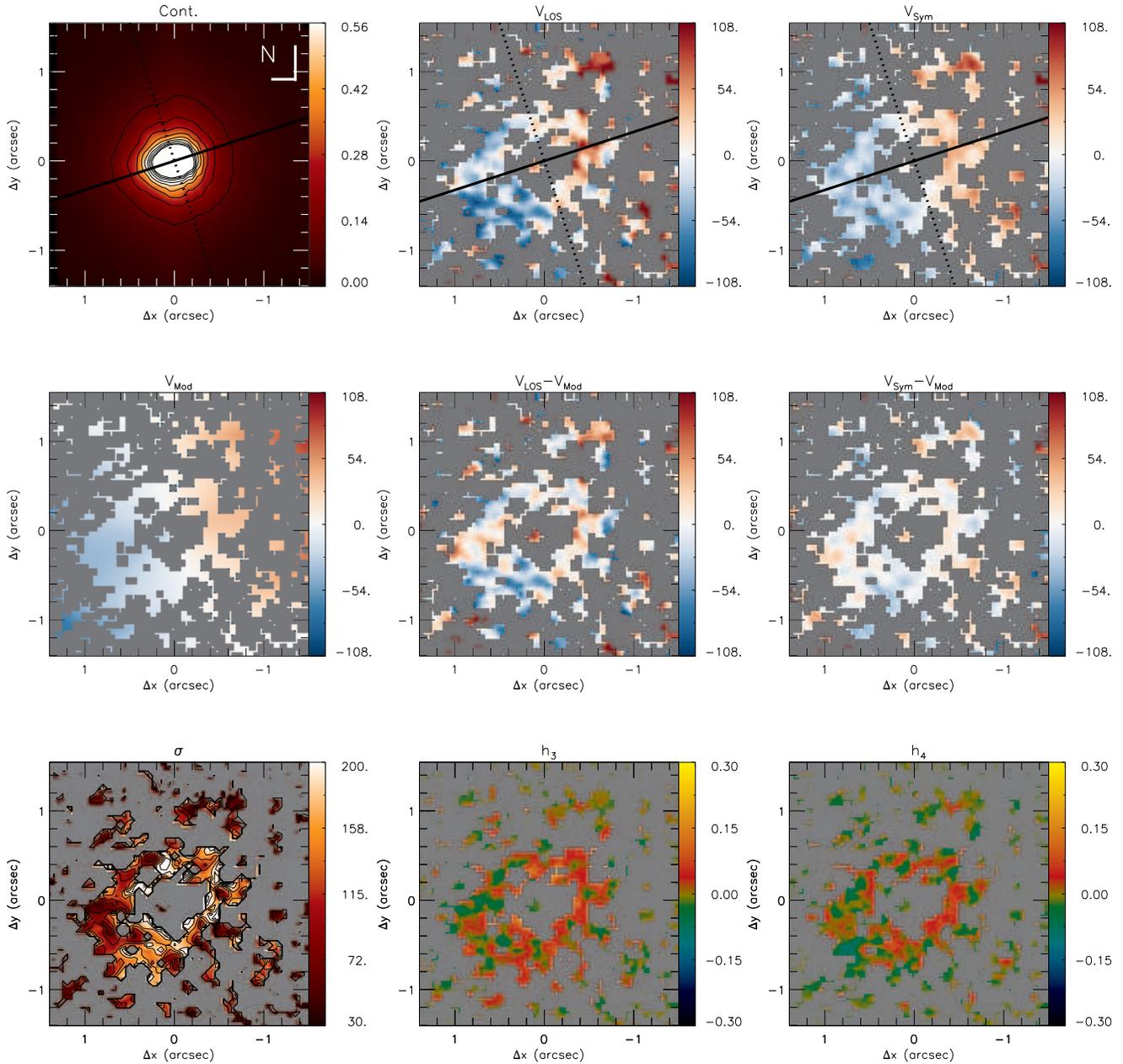}
\caption{Same as Fig.~\ref{n788} for NGC\,5548.}
\label{n5548}
\end{figure*}

\begin{figure*}
\includegraphics[scale=0.7]{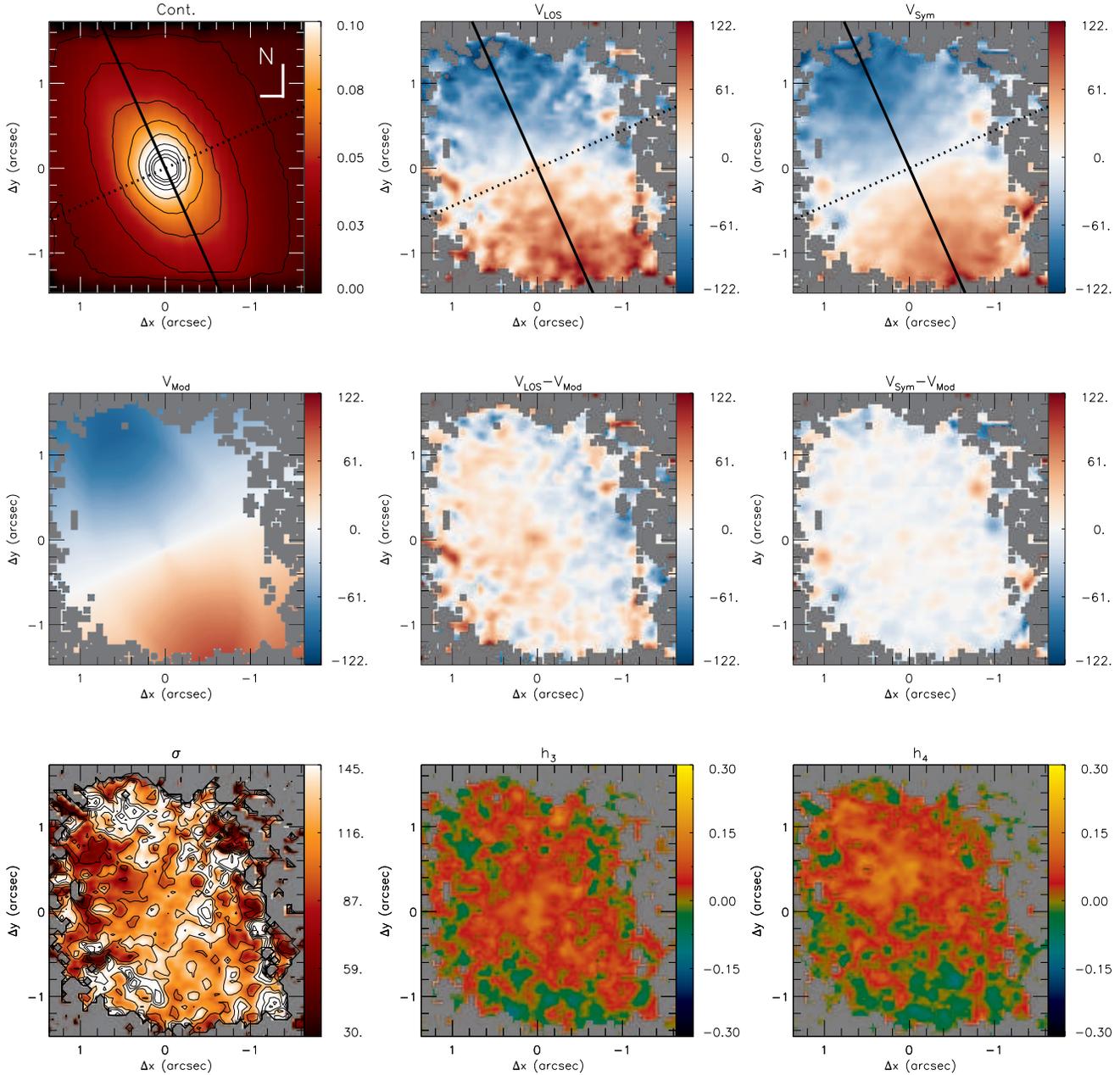}
\caption{Same as Fig.~\ref{n788} for NGC\,5899.}
\label{n5899}
\end{figure*}

\begin{figure*}
\includegraphics[scale=0.7]{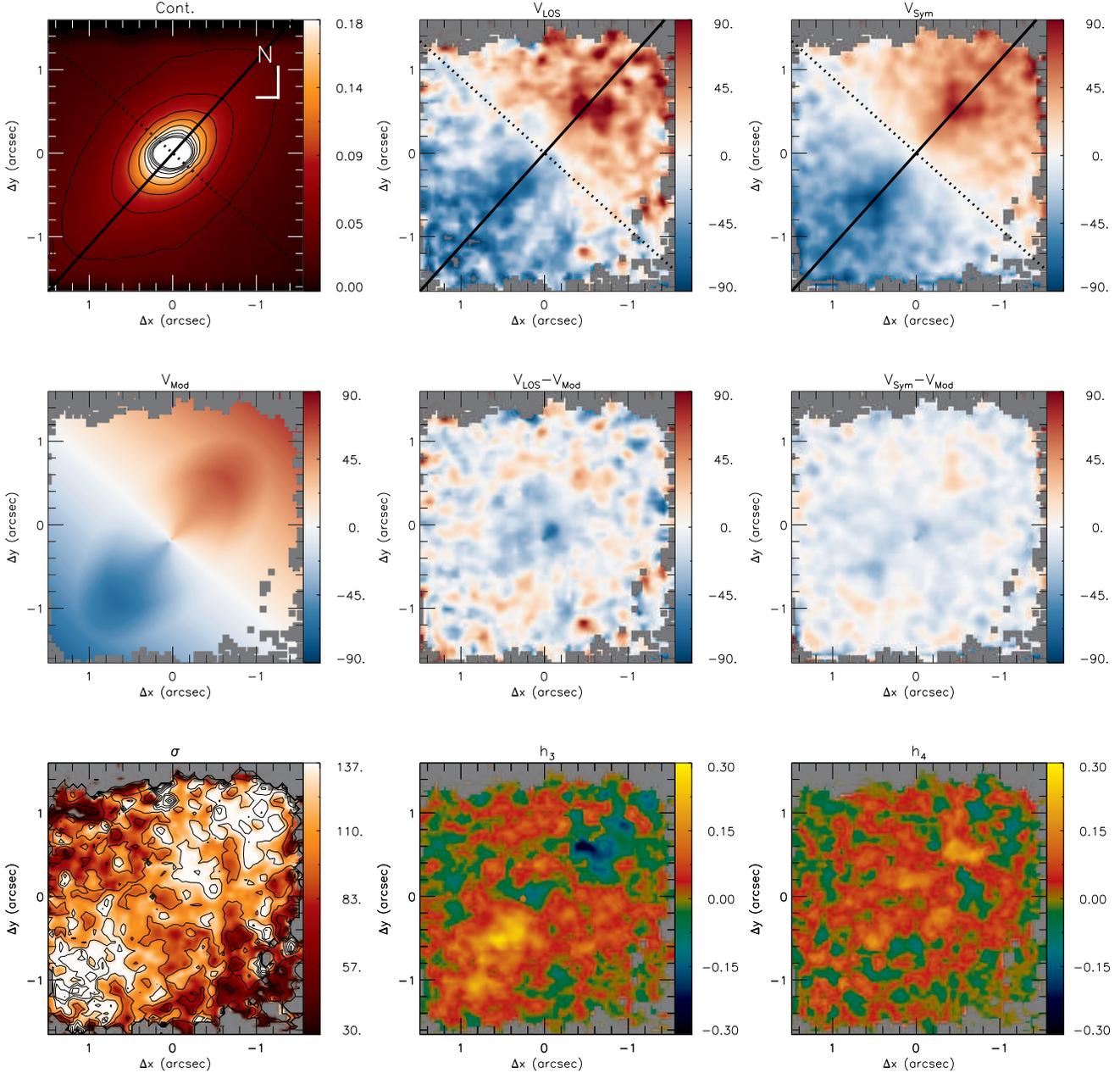}
\caption{Same as Fig.~\ref{n788} for Mrk\,607.}
\label{m607}
\end{figure*}

\begin{figure*}
\includegraphics[scale=0.7]{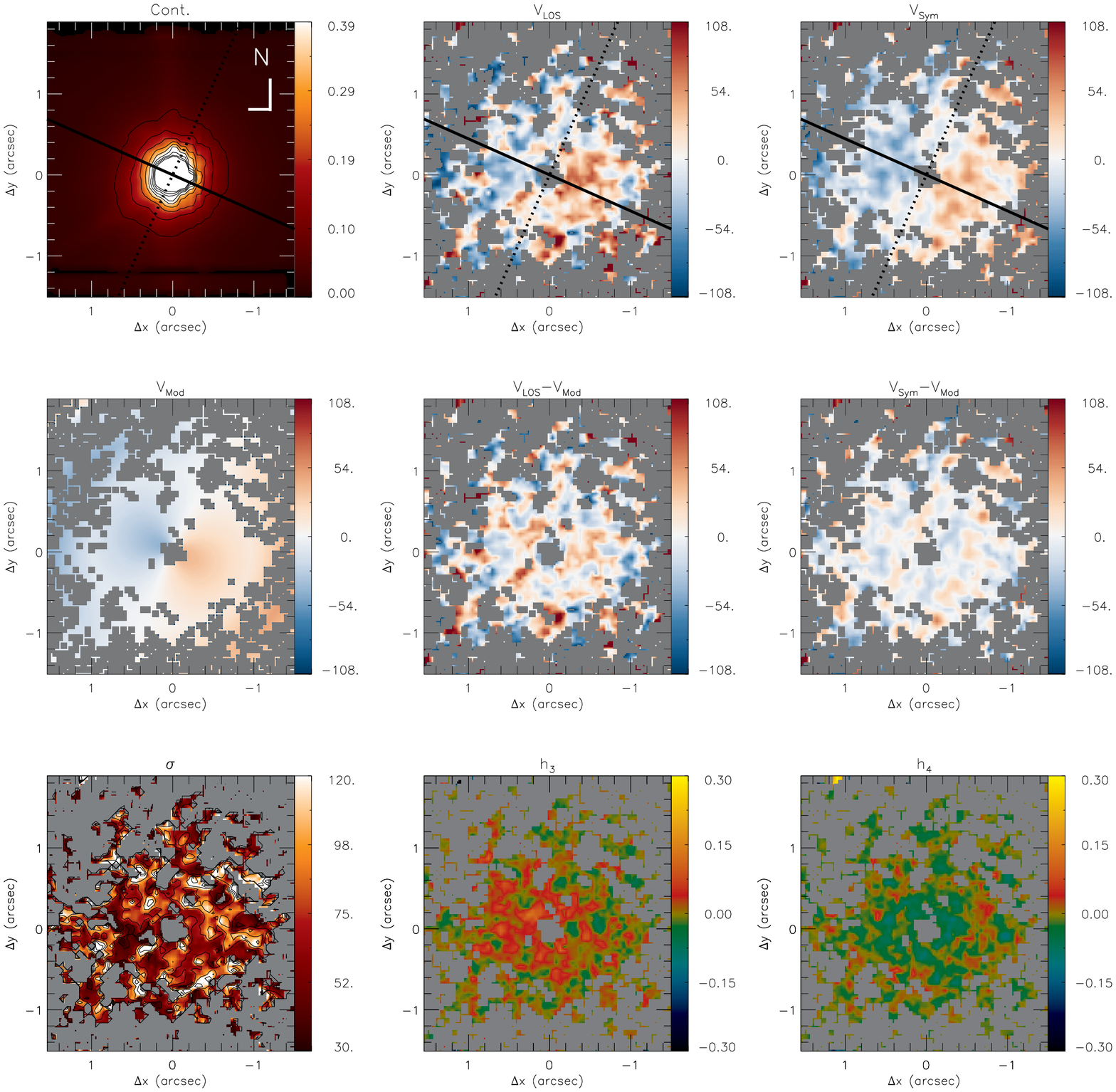}
\caption{Same as Fig.~\ref{n788} for Mrk\,766.}
\label{m766}
\end{figure*}

\section{Sample and observations}\label{obs}

\subsection{The sample}

Our sample comprises 16 AGN host galaxies: 8 from a large Gemini proposal
(PI Storchi-Bergmann) to obtain NIFS (Near-Infrared Integral Field Spectrograph)
observations of 20 AGN selected from the Swift-BAT 60-month
catalogue \citep{ajello12} to have 14--195 keV luminosities  $L_X \ge 10^{41.5}$ erg\,s$^{-1}$, and 8 from previous similar 
NIFS observations by our group of nearby AGN hosts. Four of these 8 galaxies have similar $L_X$ luminosities 
and 4 are not in the Swift-BAT catalog. One additional source (Mrk\,79) was observed in previous run with similar NIFS configuration, 
but the signal-to-noise ratio of the CO absorption banheads was not high enough to allow stellar kinematics measurements. 
By the end of 2018, we should have additional 
observations of another 13 galaxies of the large proposal and thus will have in the end a total of
29 galaxies comprising 20 Swift-BAT selected galaxies plus the 8 galaxies included in the present study (plus Mrk\,79) that  
comprise what we call a ``complementary sample", and identified by the symbol $\dagger$ in Table \ref{tab_obs}.

Additional criteria for the sample is that the redshift is z$\le$0.015, and to be accessible
for NIFS ($-30^\circ< \delta < 73^\circ$). The  $L_X$ criterium above defining
the sample of the large proposal was adopted because the the Swift-BAT 14--195 keV band
measures direct emission from the AGN rather than scattered or re-processed emission, and is much less
sensitive to obscuration in the line-of-sight than soft X-ray or optical
wavelengths, allowing a selection based only on the AGN properties.
In order to assure that we would be able to probe the feeding and feedback 
processes via the gas kinematics we further selected the galaxies having
previously observed extended gas emission \citep[e.g.][]{schmitt00} and [O\,{\sc iii}]$\lambda$5007 luminosities. We have excluded from
the sample a few galaxies that had guiding problems in the observations. A complete characterization of the sample will be presented in a forthcoming paper (Riffel et al., in
preparation).

So far, we have already observed 20 galaxies with NIFS, but only for 16 of them we were able to measure the stellar kinematics by fitting the K-band CO absorption bands at 2.3\,$\mu$m.  We were not able to obtain reliable fits of the remaining four objects (Mrk\,3, Mrk\,79, NGC\,4151 and NGC\,1068) due to the dilution of the K-band CO band-heads due to strong continuum emission. For the latter two, the stellar kinematics has already been studied using the H-band spectra \citep{onken14,sb12}. The 16 galaxies selected for the present study are listed in Table\,\ref{tab_obs}, together with relevant information.

\subsection{Observations}

We used the Gemini Near-Infrared Integral Field Spectrograph \citep[NIFS,][]{mcgregor03} to observe the galaxies listed in Table~\ref{tab_obs}. All observations were obtained using the Gemini North Adaptive Optics system ALTAIR from 2008 to 2016.  The observations followed the standard  Object-Sky-Sky-Object dither sequence, with off-source sky positions since all  targets are extended, and individual exposure times that varied according to the target. The ``HK$_{-}$G0603" filter and the ``Kl$_{-}$G5607" and ``K$_{-}$G5605" gratings were used during the observations.  Most of the observations were performed in the K$_l$ band, with the spectra centred at 2.3~$\mu$m, while for five galaxies the spectra were obtained at the K band and centred at 2.2~$\mu$m. The spectral range for the K$_l$ data is $\sim2.080-2.500\,\mu$m and for the K band is $\sim1.980-2.400\,\mu$m. Both ranges include the $^{12}$CO and $^{13}$CO absorption band-heads at $\sim$2.3~$\mu$m for all galaxies, used to measure the stellar kinematics. 

The spectral resolution ranges between 3.2 and 3.6$\,\AA$, as obtained from the measurement of the full width at half maximum (FWHM) of ArXe lamp lines, used to wavelength calibrate the spectra, close to the central wavelength. This translates to resolutions in the range 42-47 \kms\ in the velocity space. 
 The angular resolution is in the range 0\farcs12--0\farcs20, derived from the FWHM of flux distribution of the standard stars, corresponding to few tens of parsecs at the distance of most galaxies. More details about the observations are shown in Table~\ref{tab_obs}.

\subsection{Data reduction}

The data reduction followed the standard procedure and was accomplished by using tasks specifically developed for NIFS data reduction, contained in the {\sc nifs} package, which is part of {\sc gemini iraf} package, as well as generic {\sc iraf} tasks and {\sc idl} scripts. The data reduction procedure included the trimming of the images, flat-fielding, sky subtraction, wavelength and s-distortion calibrations. The telluric absorptions have been removed using observations of telluric standard stars with A spectral type. These stars were used to flux calibrate the spectra of the galaxies by interpolating a black body function to their spectra. Finally, datacubes were created for each individual exposure at an angular sampling of 0\farcs05$\times$0\farcs05 and combined in a final datacube for each galaxy. All datacubes cover the inner $\sim3^{\prime\prime}\times3^{\prime\prime}$, with exception of the datacube for NGC\,4051 that covers the inner $\sim3^{\prime\prime}\times4^{\prime\prime}$,  via spatial dithering during the observations \citep{n4051}.

 The median value of the signal-to-noise ratio (S/N) in the continuum of our sample ranges from 10 to 30, with the maximum value of up to S/N$\sim$100 observed for the nuclear spectrum of NGC\,4051. The median value of the S/N for the  $^{12}$CO (2-0)  bandhead at 2.29\,$\mu$m is larger than 3 for all galaxies of the sample. 

Detailed descriptions of the observations and data reduction procedures for the galaxies already analyzed are presented in \citet{m1066_exc} for Mrk\,1066, \citet{m1157_kin} for Mrk\,1157, \citet{n2110} for NGC\,2110 and \citet{n4051} for NGC\,4051. For the remaining objects a more detailed discussion will be presented in a forthcoming paper (Riffel et al., in preparation).

\section{Spectral fitting}\label{fit}

The stellar line-of-sight velocity distribution (LOSVD) of each galaxy was obtained by fitting the spectra within the spectral range $\sim$2.26--2.40\,$\mu$m (rest wavelengths), which includes the CO absorption band-heads from $\sim$2.29 to $\sim$2.40\,$\mu$m \citep{winge09}, usually among of the strongest absorption lines in the central region of galaxies \citep[e.g.][]{rogerio06,rogerio15}. The used spectral range also includes weaker absorption lines from Mg\,{\sc i} and Ca\,{\sc i} at 2.26--2.28\,$\mu$m. The spectra were fitted by using the 
penalized Pixel-Fitting {\sc ppxf} method \citep{cappellari04}, that finds the best fit to a galaxy spectrum by convolving  
stellar spectra templates with a given LOSVD, under the assumption that it is reproduced by Gauss-Hermite series. 

The {\sc ppxf} code requires the use of spectral templates that are used to reproduce the galaxy's spectrum. We used the spectra of the Gemini library of late spectral type stars observed with the Gemini Near-Infrared Spectrograph (GNIRS) IFU and  NIFS \citep{winge09} as stellar templates. The spectral resolution of the stellar library is very similar to that of the spectra of the galaxies of our sample and the library includes stars with spectral types from F7 to M5,  minimizing the template mismatch problem. 

The spectral range used in the fit includes the [Ca\,{\sc viii}]$\lambda2.321\,\mu$m emission line, which ``contaminates" the (3-1)$^{12}$CO bandhead and affects the stellar kinematics measurements \citep{davies06,n4051}. The [Ca\,{\sc viii}] is seen at the nucleus and close vicinity, being unresolved for most galaxies of our sample and we have excluded its spectral region from the fit when the line is present. In addition, we used the {\it clean} parameter of {\sc ppxf} to reject all spectral pixels deviating more than 3$\sigma$ from the best fit, in order to exclude possible remaining sky lines and spurious features.

In Figure~\ref{sample_fits} we present examples of typical fits for the galaxies NGC\,1052, NGC\,788 and NGC\,5899. The left panels show the fits for the nuclear spectrum, while extra-nuclear spectra are shown at the right panels, extracted at 0\farcs5 south of the nucleus -- chosen to represent typical extra-nuclear spectra. The observed spectra are shown as black dotted lines, the fits are shown in red and the regions masked during the fits, following the criteria explained above, are shown in green.  As can be seen for NGC\,788, the cleaning procedure properly excluded the region affected by the [Ca\,{\sc viii}]$\lambda2.321\,\mu$m emission line, as well as spurious features. For all galaxies, the standard deviations of the residuals (observed -- model) are similar to that of the galaxy spectra derived within a 200\,\AA\ spectral window, blue-ward to the first CO absorption bandhead, meaning that the spectra are well reproduced by the models.

The {\sc ppxf} code returns as output, measurements for the radial velocity (V$_{\rm LOS}$), stellar velocity dispersion ($\sigma$), and higher order Gauss-Hermite moments ($h_{3}$ and  $h_{4}$) for each spatial position, as well as their associated uncertainties. Using the results of the fit, we have constructed two-dimensional maps for each fitted parameter ($V_{LOS}$, $\sigma$, $h_{3}$ and  $h_{4}$), which are presented in the next section.


\section{Stellar kinematics}\label{stelkin}

Figures~\ref{n788}--\ref{m766} show the resulting maps for the stellar kinematics of our sample. 
 The stellar kinematics for  NGC\,4051, NGC\,2110, NGC\,5929, Mrk\,1066 and Mrk\,1157 was already discussed in previous works by our group (see references in Table~\ref{tab_obs}). Thus, we present the corresponding maps for these galaxies in Figures \ref{n2110_fig}--\ref{mrk1157_fig} of the Appendix\,\ref{appendix_a}, to be published on-line only.

In order to characterize the LOS velocity fields, we have symmetrized the stellar velocity field, using the {\it fit$_-$kinematic$_-$pa}\footnote{This routine was developed by M. Cappellari and is available at \\ http://www-astro.physics.ox.ac.uk/$~$mxc/software} routine. This routine measures the global kinematic position-angle and systemic velocity of the galaxy from integral field observations of the galaxy's kinematics. The method is described in \citet{krajnovic06}. \citet{cappellari07} and \citet{krajnovic11} show examples of application of the method to study the large-scale stellar kinematics of large sample of galaxies of the SAURON and ATLAS$^{\rm 3D}$ surveys. 

In addition, we fitted the LOS velocity fields by a rotating disk model. 
The model was obtained using the DiskFit routine \citep{sellwood15,sellwood10,spekkens07,reese07,barnes03,kuzio12} to fit the symmetrized velocity fields. This code fits non-parametric kinematic models to a given velocity field allowing the inclusion of circular and non-circular motions in a thin disk. Examples of application of this code for the gas and stellar kinematics of the inner region of active galaxies are shown in \citet{fischer15} and \citet{n5929}, respectively. We fitted the symmetrized velocity field, instead of the observed one, as the rotating disk model is expected to be symmetric and small fluctuations in velocity due to higher uncertainties at some locations would result in a worse model for the stellar kinematics. During the fit, we have fixed the kinematical center to the position of the peak of the continuum emission and the systemic velocity and orientation of the line of nodes of the galaxy were fixed as the values obtained from the symmetrization of the velocity field, in order to reduce the number of parameters to be fitted. The  ellipticity and disk inclination were allowed to vary during the fit and the resulting fitted values for each galaxy are shown in Table~\ref{tab_mod}.

Figures~\ref{n788}--\ref{m766} are organized as follows:
\begin{itemize}
\item Top-left panel: K-band image, obtained as the average flux between 2.20 and 2.30 $\mu$m. The continuous line shows the orientation of the kinematic major axis and the dotted line shows the orientation of the minor axis of the galaxy, obtained from the symmetrization of the stellar velocity field. The color bar shows the flux scale in units of  10$^{-17}$ erg\,s$^{-1}$\,cm$^{-2}$\,$\AA^{-1}$.

\item Top-center panel: Measured LOS stellar velocity field obtained from the fit of the spectra using the {\sc ppxf} routine \citep{cappellari04}, following the procedure described in Section~\ref{fit}. 

\item Top-right panel:  Symmetrized velocity field. The color bar is shown in units of \kms\ and the systemic velocity of the galaxy was subtracted.

\item Middle-left panel: Rotating disk model, obtained by fitting the symmetrized velocity field.

\item Middle-center panel: Residual map obtained by subtracting the rotating disk model from the observed velocity field.

\item Middle-right panel: Residual map obtained by subtracting the rotating disk model from the symmetrized velocity field, constructed in order to verify if the velocity field is well reproduced by the model. 

\item Bottom-left panel: Stellar velocity dispersion ($\sigma$) map. The color bar shows the $\sigma$ in \kms\ units.   

\item Bottom-center panel: map for the $h_{3}$ Gauss-Hermite moment. The $h_{3}$ moment measures asymmetric deviations of the LOSVD from a Gaussian velocity distribution \citep{vandermarel93,gerhard93,profit}.  High (positive) $h_3$ values correspond to the presence of red wings in the LOSVD while low (negative) $h_3$ values correspond to the presence of blue wings.

\item Bottom-right panel: map for the $h_{4}$ Gauss-Hermite moment, that measures symmetric deviations of the LOSVD relative to a Gaussian velocity distribution. High (positive) $h_4$ values indicate LOSVD is more ``pointy" than a Gaussian, while low (negative) $h_4$ values indicate profiles more ``boxy" than a Gaussian.

\end{itemize}

In all figures, North is up and East to the left and the gray regions represent masked locations. In these regions it was not possible to obtain good fits due to low signal-to-noise ratio of the spectra or due to the non detection of absorption lines (e.g. due to the dilution of the absorption lines by strong AGN continuum emission). The masked regions correspond to locations where the uncertainty in velocity or velocity dispersion is higher than 30\,\kms, while for most other locations the uncertainties are smaller than 15\,\kms.

A rotating disk pattern is recognized in the LOS velocity field for all galaxies, with a straight zero velocity line for most of them.
 For two galaxies -- Mrk\,1066 \citep{m1066_kin} and NGC\,5899 (Fig.\,\ref{n5899}) -- an  {\it S} shape zero velocity line is observed, which is a signature of the presence of a nuclear bar or  spiral arms in these galaxies \citep[e.g.][]{combes95,emsellem06}. For most galaxies, the maximum amplitude of the rotation curve is expected to be observed outside of the NIFS FoV. The rotation disk signature is clearly seen in the one one-dimensional plots shown in Fig.~\ref{vcuts}, obtained by averaging the velocity and $\sigma$ values within a pseudo-slit with width of 0\farcs25 oriented along the major axis of the galaxy. The deprojected velocity amplitude within the NIFS FoV ranges from $\sim50$ to $\sim$300~\kms.  

For 5 galaxies, the maximum $\sigma$ value is smaller than 100\,\kms, 7 have maximum $\sigma$ in the range 100--150\,\kms\ and 4 show the maximum $\sigma>150$\,\kms. In addition, distinct structures of low-$\sigma$ values ($\sim$50-80\,\kms) are seen in the the maps: low$-\sigma$ rings or partial rings are seen for Mrk\,1066, Mrk\,1157, NGC\,5929 and NGC\,788.
 These rings have been identified by visual inspection and are marked in the corresponding $\sigma$ maps as green ellipses. In all cases the size of the structures is larger than the spatial resolution and the $\sigma$ decrease is larger than the velocity resolution of the data. These structures have been attributed to intermediate-age stellar populations \citep[ages in the range 100-700~Myr; e.g.][]{mrk1066_pop,mrk1157_pop}, with origin in kinematically colder regions that still preserve the kinematics of the gas from which they have been formed. Patches of low-$\sigma$ are seen for Mrk\,607, NGC\,2110, NGC\,3516, NGC\,4051, NGC\,4235 and NGC\,5899, while three objects (NGC\,1052, NGC\,3227 and NGC\,4388) show a centrally peaked $\sigma$ distribution. These rings and partial rings are located at distances from the nucleus ranging between 150 and 250 pc.
Nuclear $\sigma$ values for each galaxy are presented in the last column of Table~\ref{tab_mod}, obtained by fitting the nuclear spectrum integrated within a circular aperture with radius of 75~pc.

Most galaxies show low $h_{3}$ values ($-$0.10$<h_{3}<$0.10), suggesting that their LOSVDs are well represented by a Gaussian velocity distribution. Exceptions are Mrk\,607, Mrk1066 and NGC\,4051 that show $h_{3}$ values larger than 0.15. For 8 galaxies (NGC\,1052, NGC\,2110, NGC\,3227, NGC\,3516, NGC\,4051, NGC\,5506, Mrk\,607 and Mrk\,1066), the $h_{3}$ map is anti-correlated with the velocity field, with positive $h_{3}$ values seen at locations where rotation velocities are observed in blueshifts and negative $h_{3}$ values associated to regions where the rotation velocities are observed in redshifts.  A similar trend (but not so clear) is observed for other four galaxies (NGC\,4235, NGC\,4388, NGC\,5929 and Mrk\,1157). In Fig.~\ref{h3vel} of the Appendix~\ref{appendix} we present plots of the LOS velocity vs. $h_3$ for all galaxies, which show this anti-correlation clearly. %
The anti-correlation seen between $h_{3}$ and the LOS velocity can be interpreted  as due do the contribution of stars rotating slower than those in the galaxy disk, probably due to motion in the galaxy bulge \citep[e.g.][]{emsellem06,ricci16}. 

The $h_{4}$ moment maps show small values at most locations for all galaxies, with $-0.05<h_{4}<0.05$. For some galaxies, that present strong low-$\sigma$ structures (e.g., Mrk\,1066, Mrk\,607 and NGC\,4051), higher positive $h_{4}$ values are observed co-spatial with the low-$\sigma$ regions. This is also seen in the plots presented in Fig.~\ref{h3vel}, that show a trend of higher values of $\sigma$ being observed in regions with negative $h_4$ values, while small $\sigma$ values being associated to positive $h_4$ values.  
We interpret this correlation between low-$\sigma$ and high $h_{4}$ values as an additional support to the presence of young/intermediate age stars at these locations, as more peaked velocity distributions are expected for young stars (as they are located in a thin disk structure).



The rotating disk models reproduce well the observed velocity fields for all galaxies, as seen in the residual velocity maps that show values smaller than 20~\kms\ at most locations. Table~\ref{tab_mod} shows the model parameters for each galaxy. The systemic velocities from the table are relative to the heliocentric frame. The deprojected velocity amplitude of the galaxies of our sample ranges from $\sim$60~\kms\ (for NGC\,4051 - a Sc galaxy) to $\sim$340\,\kms (for NGC\,3516 - a S0 galaxy).

\begin{table*}
\centering
\caption{Kinematic parameters obtained by symmetrize the stellar velocity.  Col. 1: Galaxy name; cols 2-3: systemic velocity ($V_s$) corrected for the heliocentric frame and orientation of the line of nodes $(\Psi_0$) derived from the symmetrization of the velocity fields; cols 4-5: disk ellipticity ($e$) and inclination ($i$) derived by modeling the velocity field using the DiskFit routine; cols: 6-8: orientation of the major axis, ellipticity and inclination of the large scale disk, as available at NED; col 9: the nuclear stellar velocity dispersion measured within a circular aperture with 150~pc diameter.}
\vspace{0.3cm}
\begin{tabular}{l c c c c c c c c}
\hline
        & \multicolumn{2}{c}{FitKinematicPA} & \multicolumn{2}{c}{DiskFit}& \multicolumn{3}{c}{Large Scale Disk} \\ \hline 
Galaxy  & $V_s$ (\kms)& $\Psi_0$($^\circ$) & $e$         & $i$($^\circ$)  &  $\Psi_{0NED}$($^\circ$) &$e_{NED}$ & $i_{NED}$($^\circ$) &$\sigma$(\kms)  \\ \hline           
 NGC788 &  4034  &   120$\pm$3  &   0.07$\pm$0.01 & 20.8$\pm$0.2 &  100 & 0.67 & 42.3    &   187$\pm$4\\ 
NGC1052 &  1442  &   114$\pm$3  &   0.32$\pm$0.01 & 47.5$\pm$0.2 &  120 & 0.71 & 45.6    &  245$\pm$4 \\ 
NGC2110 &  2335  &   156$\pm$3  &   0.26$\pm$0.01 & 42.5$\pm$0.3 &  160 & 0.63 & 38.7    &  238$\pm$5 \\ 
NGC3227 &  1174  &   156$\pm$3  &   0.30$\pm$0.01 & 45.4$\pm$0.2 &  152 & 0.88 & 62.0    &  130$\pm$7 \\ 
NGC3516 &  2631  &	54$\pm$3   &   0.05$\pm$0.01 & 18.2$\pm$0.8 &  7   & 0.69 & 43.9 &  186$\pm$3  \\ 
NGC4051 &  717   &	130$\pm$9  &   0.20$\pm$0.01 & 37.3$\pm$0.9 &  142 & 0.84 & 56.6 &   72$\pm$3   \\ 
NGC4235 &  2276  &	54$\pm$3   &   0.25$\pm$0.01 & 41.2$\pm$0.5 &  50  & 0.94 & 70.1 &  183$\pm$12 \\  
NGC4388 &  2537  &	96$\pm$3   &   0.11$\pm$0.01 & 27.7$\pm$0.3 &  89  & 0.95 & 71.3 &  106$\pm$6   \\ 
NGC5506 &  1878  &	96$\pm$3   &   0.48$\pm$0.01 & 58.7$\pm$0.3 &  90  & 0.97 & 76.1 &   -          \\ 
NGC5548 &  5128  &   108$\pm$3  &   0.51$\pm$0.04 & 60.9$\pm$6.8 &  60  & 0.34 & 19.9    &  276$\pm$22  \\  
NGC5899 &  2616  &	24$\pm$3   &   0.54$\pm$0.01 & 62.7$\pm$0.3 &  25  & 0.92 & 67.7 &  147$\pm$9   \\ 
NGC5929 &  2491  &	30$\pm$9   &   0.51$\pm$0.01 & 60.7$\pm$0.5 &  38  & 0.69 & 43.9 &  134$\pm$5   \\ 
 MRK607 &  2781  &   138$\pm$3  &   0.47$\pm$0.01 & 58.2$\pm$0.1 &  135 & 0.94 & 70.1    &  132$\pm$4  \\ 
 MRK766 &  3855  &	66$\pm$3   &   0.05$\pm$0.02 & 18.2$\pm$3.7 &  110 & 0.77 & 50.2 &   78$\pm$6   \\  
MRK1066 &  3583  &   120$\pm$3  &   0.36$\pm$0.01 & 50.2$\pm$0.3 &  140 & 0.92 & 66.4    &  103$\pm$4 \\ 
MRK1157 &  4483  &   114$\pm$3  &   0.29$\pm$0.01 & 45.1$\pm$0.9 &  95  & 0.88 & 61.3    &   92$\pm$4  \\ 

\hline
\end{tabular}
\label{tab_mod}
\end{table*}

\section{Discussions}

\subsection{Comparison between kinematic and large scale disk parameters}\label{comp}

Table~\ref{tab_mod} shows the parameters derived from the symmetrization (using the FitKinematicPA routine) and from the fit of the rotation disk model (using the DiskFit code) of the velocity fields. These parameters can be compared to those obtained for the large scale disks. The position angle (PA) of the major axis from NED\footnote{NASA/IPAC  Extragalactic  Database available at $http://ned.ipac.caltech.edu/$ } ($\Psi_{0NED}$), shown in Table~\ref{tab_mod} is derived from the K$_S$ band photometry obtained from the  Two Micron All Sky Survey \citep[2MASS,][]{jarrett03}. The ellipticity ($e_{NED}$) and inclination ($i_{NED}$) of the disk are also obtained from the apparent major ($a$) and minor ($b$) axis available on NED database from K$_S$ images, as $e_{NED}=1-\frac{b}{a}$ and $i_{NED}={\rm acos(b/a)}$, respectively.

\begin{figure*}
\includegraphics[scale=0.3]{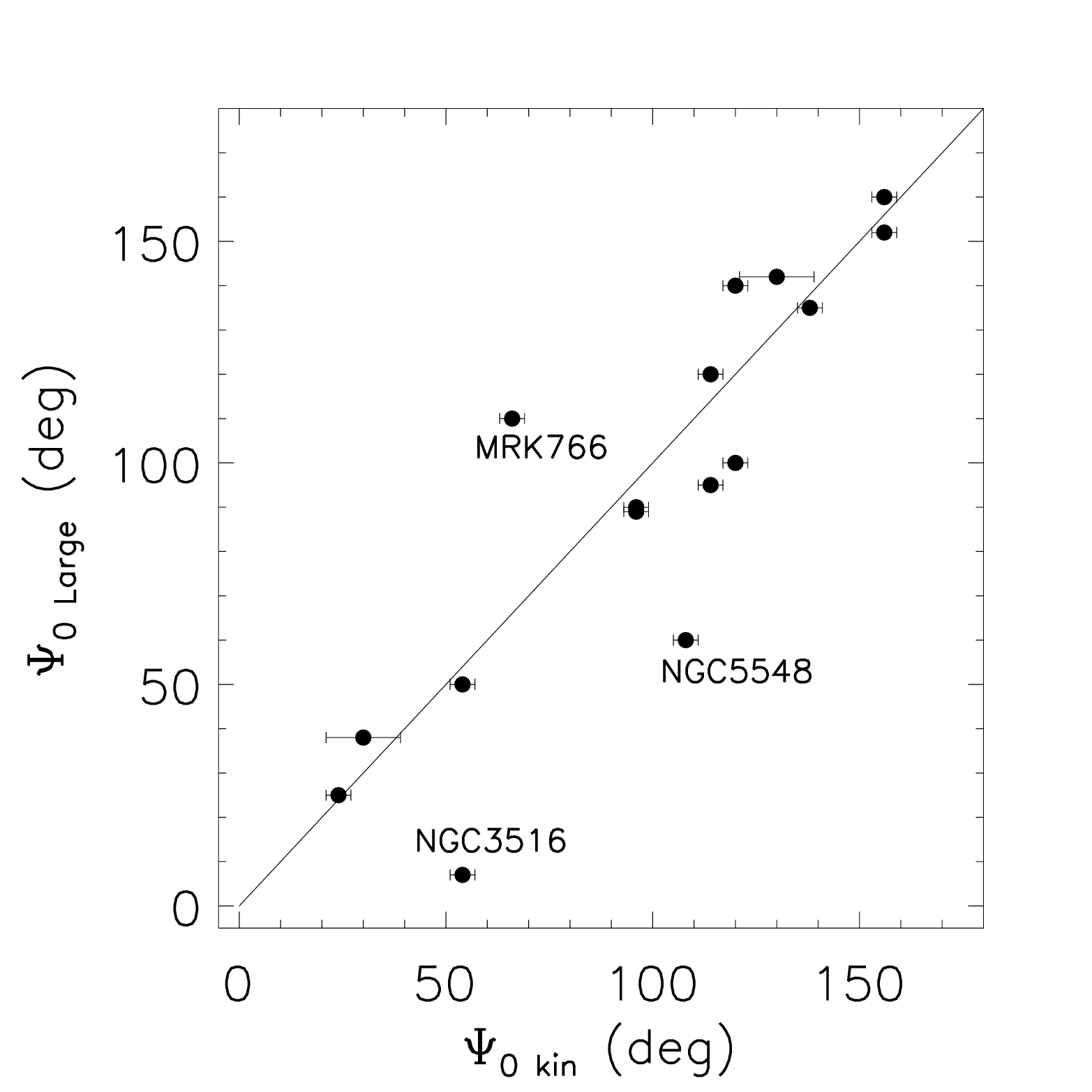}
\includegraphics[scale=0.3]{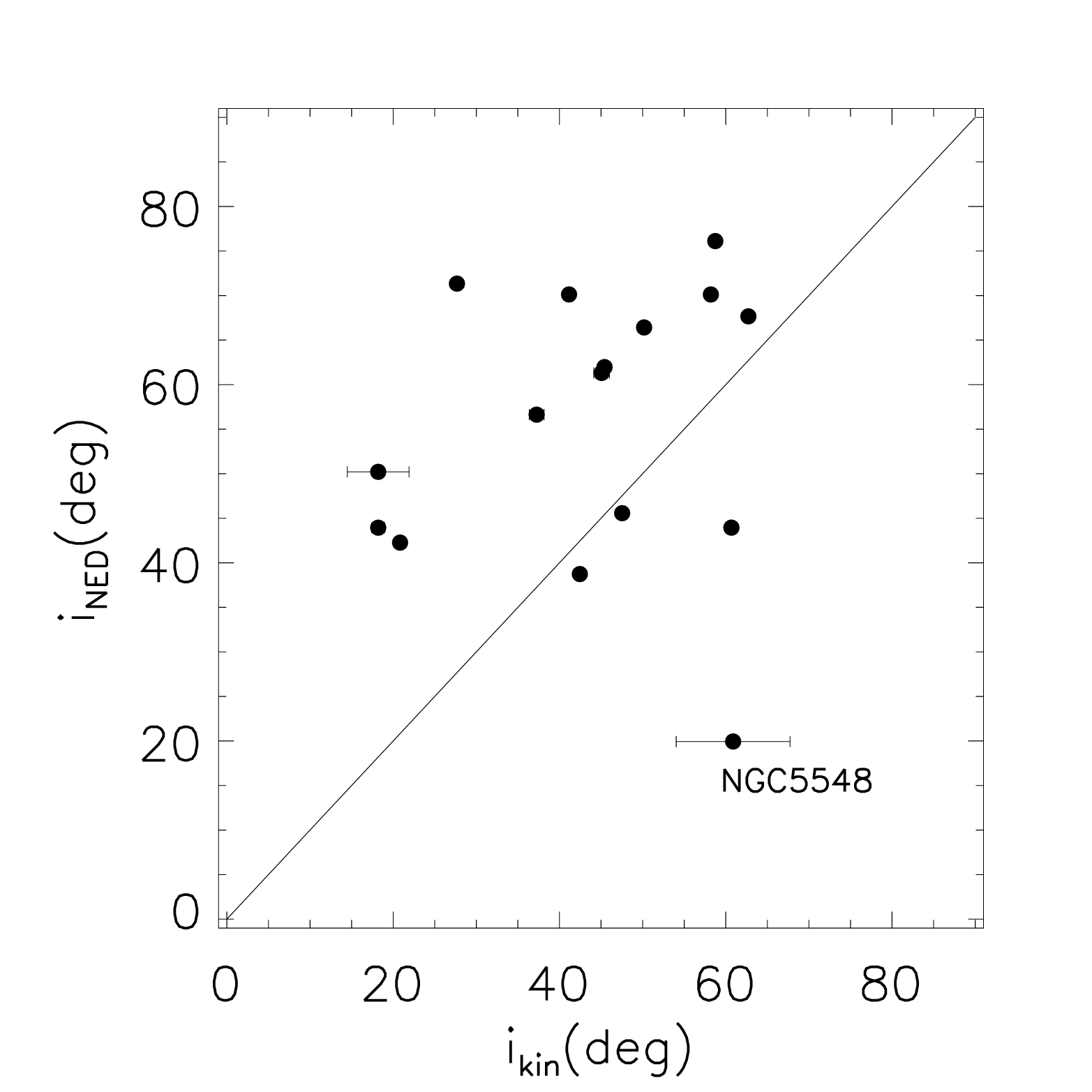}
\includegraphics[scale=0.3]{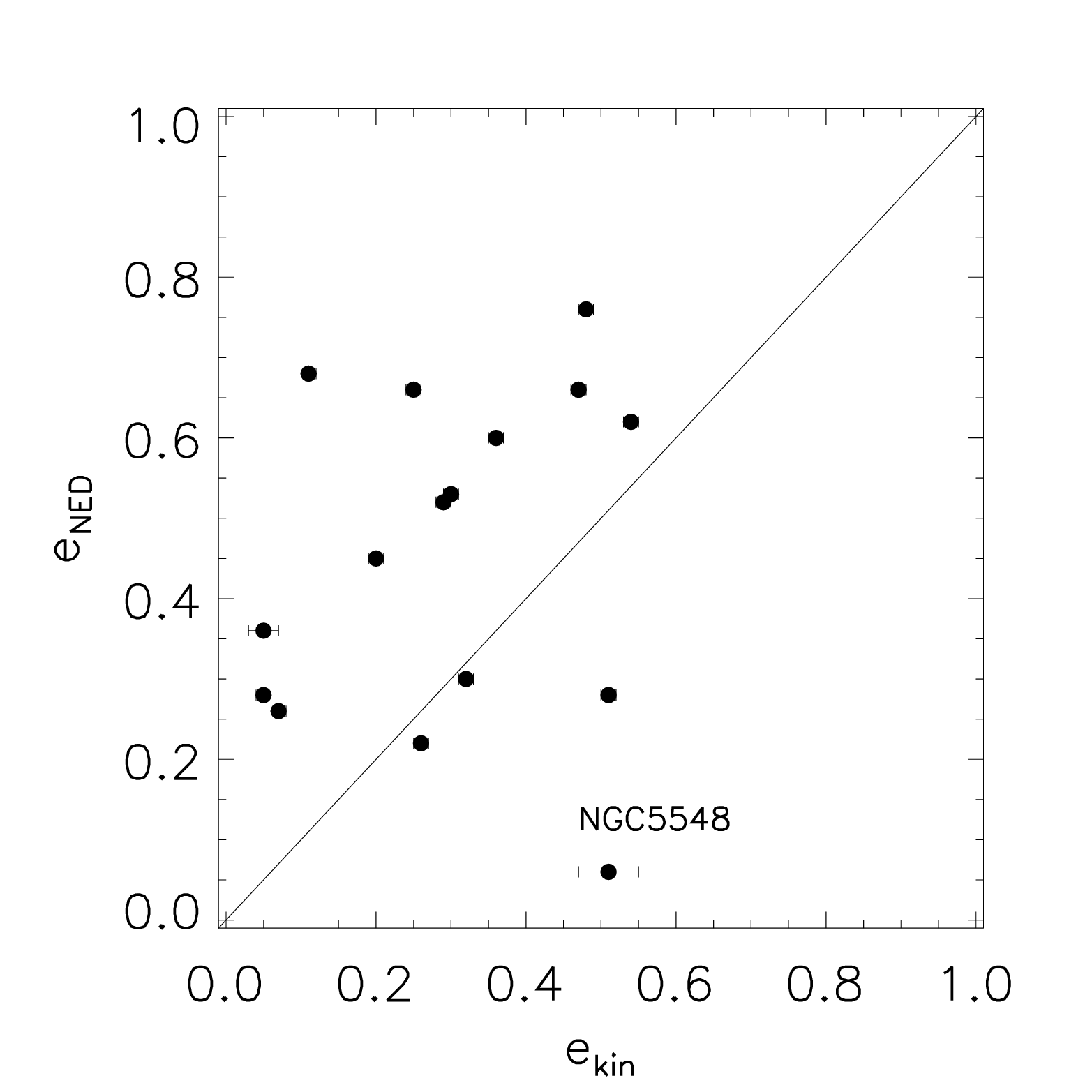}
\caption{Comparison of large scale photometric (y-axis) and small scale kinematic (x-axis) PAs (left), inclination (middle) and ellipticity (right). Continuous lines show one-to-one relations.}
\label{comp_large}
\end{figure*}

In the left panel of Figure~\ref{comp_large} we present a plot of the large scale photometric PA of the major axis {\it versus} the kinematic PA derived from our NIFS velocity fields, constructed using the values of $\Psi_0$ shown in Table~\ref{tab_mod}. This plot shows that the small scale kinematic PA is in approximate agreement with the large scale photometric one.  The mean PA offset is $<\Psi_0-\Psi_{0NED}>=3.9^\circ\pm5.7^\circ$.  Only for three galaxies (NGC\,5548, Mrk\,766 and NGC\,3516) there are significant discrepancies between small scale kinematic and large scale photometric major axes. 
NGC\,5548 and Mrk\,766 are almost face on galaxies and thus it is hard to obtain a precise determination of $\Psi_0$, both from photometry and kinematics, justifying the discrepancy. For NGC\,3516 the $\Psi_{0NED}$ is distinct from that observed in optical images \citep[$\Psi_0=56^\circ$,][]{schmitt00} and from the stellar kinematics based on optical IFS \citep[$\Psi_0=53^\circ$,][]{arribas97}. On the other hand, the $\Psi_{0NED}$ for NGC\,3516 is very similar to the orientation of the bar of the galaxy \citep[$\Psi_0=-10^\circ$,][]{veilleux93} and thus the value of $\Psi_{0NED}$ may be biased due to the bar, that is stronger in near-IR bands. 

 Previous studies have found similar results. For example, the morphological study by \citet{malkan98} of the inner kiloparsecs of nearby active galaxies showed that the resulting classification of the small scale structure was very similar to the one given in the Third Reference
Catalog \citep[RC3,][]{corwin94}, showing that not only the kinematic PA at small scale but also the photometric PA at small scale is similar to that at large scale. 

\citet{barbosa06} used the Gemini Multi-Object Spectrograph (GMOS) IFU to map the stellar kinematics of the inner 200--900 pc of six nearby active galaxies by fitting the stellar absorption lines of the Calcium triplet around 8500\AA\ and also found that the kinematic position angle at small scale is in good agreement with the large scale photometric measurements. \citet{dumas07} used optical IFS to map the stellar and gas kinematics of the central kiloparsec of a matched sample of nearby Seyfert and inactive galaxies at angular resolutions ranging from 0\farcs9 to 2\farcs5, using the SAURON IFU on the William Herschel Telescope. They found that the orientations of the kinematic line of nodes are very similar with those derived from large scale photometry for both active and inactive galaxies. \citet{fb06} present the stellar kinematics of a sample of 24 spiral galaxies using the SAURON IFU. Their sample includes only 5 active nuclei and they found misaligned photometric and kinematic axes for 9 objects in their sample (only one being an active galaxy), interpreted as being due to non-axisymmetric structures (as bars) and more easily detected at low galaxy inclinations. 



The middle and right panels of Figure~\ref{comp_large} show the comparison between the large photometric (y-axis) and small kinematic (x-axis) scale disk inclination and ellipticity, respectively.  In contrast to the orientation of the major axis of the galaxy, these parameters do not follow the same distribution at small and large scale. For most cases, both the inclination and the ellipticity of the disk at large scale are larger than that derived for the inner 3$^{\prime\prime}\times$3$^{\prime\prime}$. This result can be interpreted as being due to the fact that the large scale measurements are dominated by the disk component as they are estimated from the apparent major and minor axis measurements obtained from large scale K$_S$ images, while at small scale, the near-IR emission is dominated by emission from evolved red stars at the bulge of the galaxy \citep[e.g.][]{maraston05,rogerio15}, which play an important role in the observed morphology and kinematics observed with NIFS.

\subsection{Gravitational potentials and comparison with previous studies}

The velocity residual maps for all galaxies of our sample show small values, indicating that the stellar velocity fields are well reproduced by the rotating disk model, with kinematic axes that follows the same orientation of large scale measurements. In addition, the deprojected rotation velocity amplitude is always larger than the mean velocity dispersion,
indicating the the stellar kinematics of the galaxies of our sample are dominated by regular rotation. We can compare our results with previous studies of the stellar kinematics of active and inactive galaxies.

 \citet{dumas07} present the stellar kinematics of a  sample of 39 active galaxies and a matched control sample, selected to have similar blue magnitudes, Hubble type and inclinations. They found that the stellar kinematics of both active and inactive galaxies show regular rotation patterns typical of disc-like systems. A similar result was found by \citet{barbosa06} using higher angular resolution ($<$1\farcs0) IFS with Gemini Telescopes of a sample of six nearby Seyfert galaxies. In addition, they found partial rings of low-$\sigma$ values at 200--400 pc from the nucleus for three galaxies, interpreted as tracers of recently formed stars that partially keep the cold kinematics of the original gas from which they have formed.  \citet{fb06} present kinematic maps for a sample of nearby spiral galaxies obtained with the SAURON IFU, which show regular stellar rotation for most galaxies. However, kinematic decoupled components are frequently seen in the inner region, as sudden changes in the velocity field, which are often associated with a drop in the $\sigma$ and anti-correlated $h_3$ values with respect to the $V_{LOS}$. In addition, they found kinematic signatures of non-axisymmetric structures for 37\% of the galaxies of their sample (only one harboring an AGN). For 20\,\% of their sample (5 galaxies, none of them harboring an AGN) they found kinematic signatures of bars as predicted in N-body simulations of barred potentials \citep[e.g.][]{kuijken95,bureau05}. 

Hubble Space Telescope (HST) H-band images up to 10$^{\prime\prime}$ radius and ground-based near-infrared and optical images of a matched Seyfert versus non-Seyfert galaxy sample of 112 nearby galaxies show a statistically significant excess of bars among the Seyfert galaxies at practically all length-scales \citep{laine02}. In addition, they also found that Seyfert galaxies always show a preponderance of ``thick" bars compared to the bars in non-Seyfert galaxies. On the other hand, recent results show that AGN hosts at $0.2 < z < 1.0$ show no statistically significant enhancement in bar fraction compared to inactive galaxies \citep[e.g][]{cheung15}. Large scale bars are seen for  28.5\% of face-on spiral hosts of AGN, as obtained from the study of more the 6.000 AGN hosts of from the  Sloan Digital Sky Survey \citep[SDSS;][]{alonso13}.

Our kinematic maps suggest the presence of nuclear bars in only 2 galaxies: Mrk\,1066 and NGC\,5899, as revealed by the presence of an {\it S} shape zero velocity line  observed in the $V_{LOS}$ maps \citep[e.g.][]{combes95,emsellem06}. This corresponds to only 12.5\% of our sample, although the statistics is low so far and this result is preliminary. The rest of the galaxies are dominated by rotation. The difference in the proportion of barred galaxies or non-axisymmetric structures in our study relative to that of \citet{fb06} may be due to the small number of objects in both studies, and to the difference in the field-of-view (FoV) of the two studies. Our FoV ($3^{\prime\prime}\times3^{\prime\prime}$) is smaller than those of previous studies, making it more difficult to identify kinematic signatures of bars as predicted in N-body simulations \citep[e.g.][]{bureau05}, as double-hump rotation curves, broad $\sigma$ profiles with a plateau at moderate radii and $h_3 - V_{LOS}$ correlation over the projected bar length. On the other hand, the photometric detection of bars mentioned above are mainly obtained using large scale images. 
Thus, our results suggest that the motion of the stars is dominated by the gravitational potential of the bulge, as the FoV of our observations is smaller than the bulge length for all galaxies.

\begin{figure}
\includegraphics[scale=0.45]{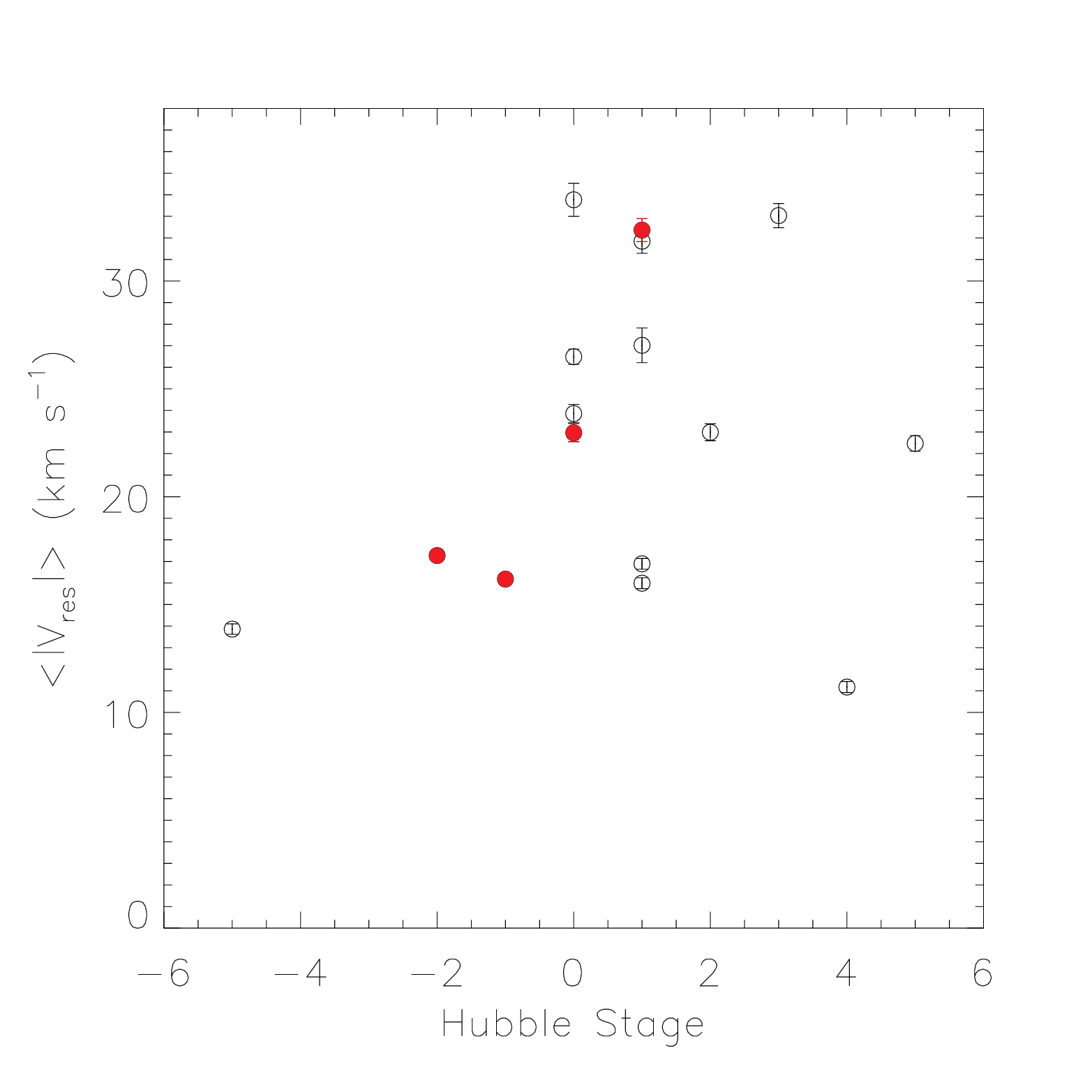}
\includegraphics[scale=0.45]{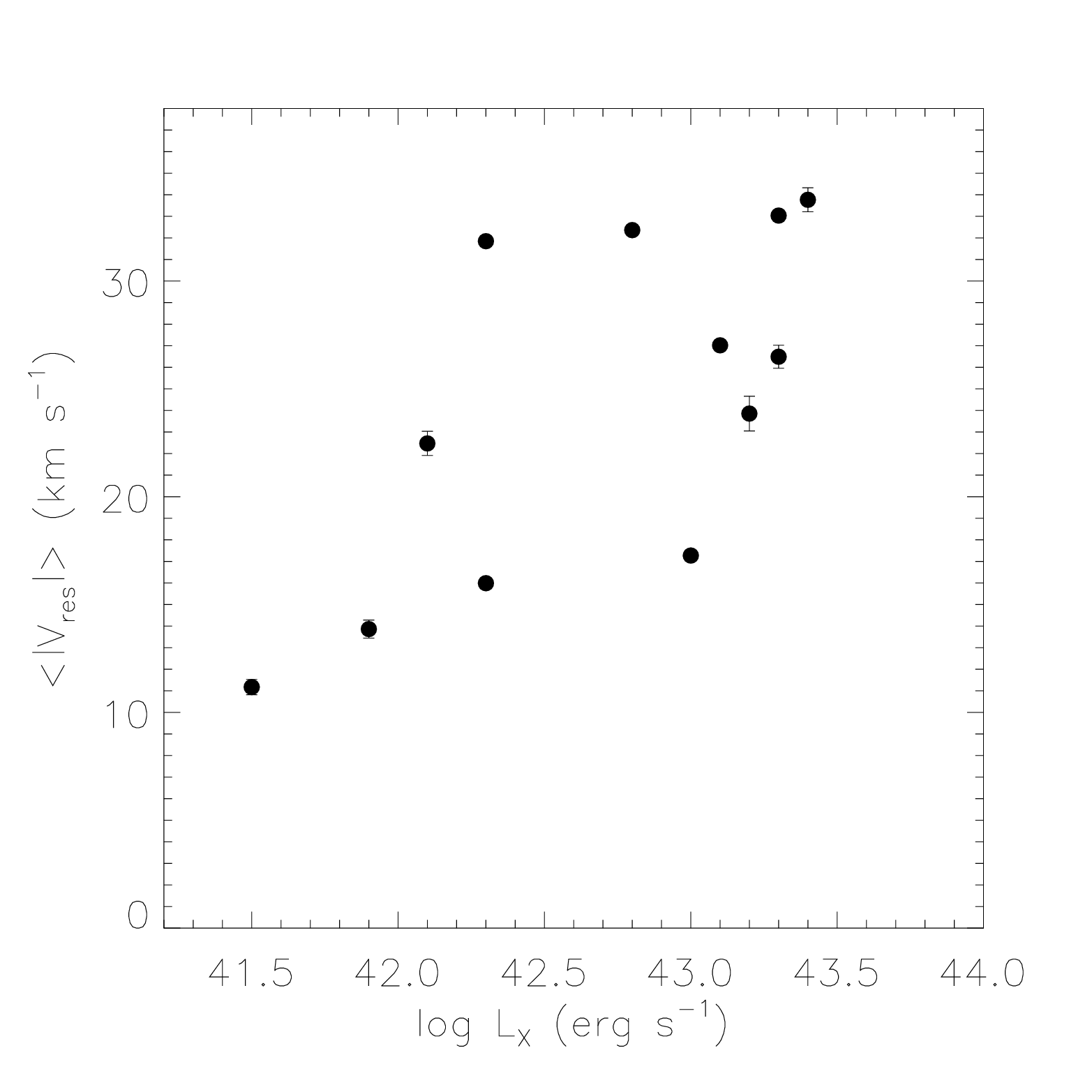}
\caption{Plots of the standard deviation of the residual maps  ($<|V_{res}|>$) vs. Hubble index (top) and X-ray luminosity ($L_X$, bottom). In the top panel, filled circles correspond to barred galaxies.}
\label{correlations}
\end{figure}

In order to further investigate how the galactic potentials and deviations from ordered rotation are related to the host galaxy and AGN, we plotted 
the mean value of the modulus of the residual velocities ($<|V_{res}|>$), where $V_{res}= V_{LOS}-V_{mod}$, against the Hubble index and hard X-ray (14-195 keV) luminosity ($L_X$) from the Swift-BAT 60-month catalogue \citep{ajello12}, which measures direct emission from the AGN.  These plots are shown in Figure~\ref{correlations}.
For four galaxies of our sample (MRK1066, MRK1157, MRK607 and NGC5929), there are no X-ray luminosities available in the  BAT catalogue and thus the $<|V_{res}|>$ vs. plot contains only 12 points. 
The $<|V_{res}|>$  was estimated as the mean value of 10.000 bootstrap realizations in which for each interaction the $|V_{res}|$ is calculated for a sample selected randomly among the values observed in the residual map. The standard deviation in the simulated $|V_{res}|$ represents the intrinsic scatter of each residual map and is used as the uncertainty for $<|V_{res}|>$. 

The top panel  of Fig.~\ref{correlations} shows that there is no correlation between $<|V_{res}|>$ and Hubble index, with a Pearson correlation coefficient of only $R=0.12$. This result shows that large scale structures do not affect significantly the stellar kinematics of the inner kiloparsec of the galaxies of our sample.

On the other hand, the bottom panel of Fig.~\ref{correlations} suggests that $<|V_{res}|>$ is correlated with $L_X$. We computed a Pearson correlation coefficient of $R=0.74$, with less than 1\,\% of probability that this distribution of points can be generated by a random distribution. Although the number of points is small, this  trend may mean that more luminous AGN have a larger impact in the surrounding stellar dynamics. As a speculation, we propose that strong AGN may quench circum-nuclear star formation in the galaxy disk and thus the stellar motions have a more important component of bulge star kinematics. On the other hand, for lower luminosity AGNs, the active nuclei may not be powerful enough the quench the star formation and thus the stellar dynamics has a stronger contribution from stars more recently formed in the plane of the galaxy.

\subsection{Implications to AGN feeding and star formation}

The velocity dispersion maps show structures of lower $\sigma$ than the surroundings ($\sim$50--80\,\kms) for 10 galaxies (62\,\%) of our sample.
Such velocity dispersion drops are commonly reported in the literature \citep[e.g.][]{bottema89,bottema93,fisher97,gl99,emsellem01,marquez03,fb06,barbosa06} and have been interpreted as being tracers of relatively recent star formation as compared to the bulge stellar population \citep[e.g.][]{emsellem01,emsellem06,marquez03,barbosa06,n4051}. Indeed, stellar population synthesis using near-IR IFS with NIFS reveal that the low-$\sigma$ rings seen in Mrk1066 and Mrk1157 are associated to an intermediate-age stellar population \citep[$<$700\,Myr;][]{mrk1066_pop,mrk1157_pop}. 

The low-$\sigma$ structures may be related to accretion of gas to the inner kiloparsec of galaxies as a result of streaming motions towards the nucleus along nuclear bars or dust spirals, observed for several active galaxies by our group \citep[e.g.][]{fathi06,n4051,sb09,m79,n2110,sm14,sm16,sm17,lena15} and other groups \citep[e.g.][]{vandeVen09,sanchez09,smajic15}. Several works have been aimed to investigate the presence of nuclear bars and dust spirals in active and inactive galaxies using high resolution HST images \citep[][]{laine02,pogge02,sl07,martini13}. These studies reveal an excess of bars in Seyfert galaxies, when compared to a matched sample of inactive galaxies \citep{laine02} and dust structures seem to be present in all early type AGN hosts, while only 26\,\% of inactive early type show significant dust in the nuclear region \citep{sl07}. For late type galaxies, large amount of dust is observed for both active and inactive galaxies \citep{sl07}. 

The observed gas inflows mentioned above can lead to the accumulation  of large reservoirs of gas that can feed both star formation and the AGN. In such scenario, it would be expected that  low-$\sigma$ structures should be more frequent in active than in inactive galaxies. However, several studies report the presence of low-$\sigma$ structures in inactive galaxies. For example, \citet{fb06} found that at least 46\,\% of their sample of spiral galaxies show $\sigma$-drops, most of them being inactive. A similar result is reported by \citet{ganda06}, who found central $\sigma$ drops for many objects of their sample of 18 spiral galaxies. On the other hand, these drops are not commonly observed in elliptical galaxies \citep{emsellem04}. Thus, the presence of low-$\sigma$ may be related to recent star formation in the inner kiloparsec of the galaxies of our sample, but possibly unrelated to the nuclear activity.

\subsection{The stellar kinematics and AGN Feeding and Feedback processes}

This paper is the first of a series in which we will investigate the AGN feeding and feedback processes using J and K band NIFS observations of a sample of nearby active galaxies selected using as main criteria the hard X-ray luminosity. The results presented here will be used to compare gas and stellar kinematics in order to isolate non-circular motions, by constructing residual maps between the observed velocity fields for the ionized (traced by [Fe\,{\sc ii}] and H recombination lines) and molecular (traced by H$_2$ emission lines) gas and the rotating disk models presented here. The analysis of the residual maps, together with velocity channel maps along the emission-line profiles, will allow us to identify possible gas inflows and outflows. Similar methodology has already been successfully used by our AGNIFS group \citep[e.g.][]{n4051,m1066_kin,m79,m1157_kin,n2110}. The gas inflow and outflow rates can be compared with AGN properties (e.g. bolometric luminosity and accretion rate) 
 to draw a picture of the feeding and feedback processes in AGNs.

\section{Conclusions}\label{conc}

We used near-IR integral field spectroscopy to map the stellar kinematics of the inner 3$^{\prime\prime}\times$3$^{\prime\prime}$ of a sample of 16 nearby Seyfert galaxies. We present maps for the radial velocity, velocity dispersion and higher order Gauss Hermite moments, obtained by fitting the CO stellar absorptions in the K-band. The observed velocity fields were symmetrized and modeled by a thin rotating disk in order to derive kinematical parameters. The main results of this work are:

\begin{itemize}

\item The observed velocity fields for all galaxies show regular rotation. In addition, for two galaxies (Mrk\,1066 and NGC\,5899) the velocity field shows an  {\it S} shape zero velocity line which is  interpreted as a signature of nuclear bars.

\item  The residuals of the modeling of the stellar velocity field are correlated with the hard X-ray luminosity, suggesting that the nuclear source plays a role on the observed stellar dynamics of the inner kiloparsec of the galaxies, with stronger AGNs showing less ordered stellar orbits than weak AGNs.
 
\item The velocity dispersion maps show low-$\sigma$  ($\sim50-80$\,\kms) rings for 4 galaxies (Mrk\,1066, Mrk\,1157, NGC\,5929 and NGC\,788) or ``patches" of low-$\sigma$ structures (for Mrk\,607, NGC\,2110, NGC\,3516, NGC\,4051, NGC\,4235 and NGC\,5899) at typical distances of 200~pc, interpreted as being originated in young/intermediate age stellar populations. Centrally peaked $\sigma$ maps are observed for three galaxies (NGC\,1052, NGC\,3227 and NGC\,4388).

\item  The $h_{3}$ moment is anti-correlated with the velocity field for 8 galaxies (NGC\,1052, NGC\,2110, NGC\,3227, NGC\,3516, NGC\,4051, NGC\,5506, Mrk\,607 and Mrk\,1066) -- positive $h_{3}$ values seen at locations where the velocity field shows blueshifts and $h_{3}<0$ for locations where the velocity field shows redshifts. The presence of these wings are attributed to the contribution of stars from the galaxy bulge that present lower rotation velocities.

\item The $h_{4}$ maps show small values at most locations for all galaxies. For the galaxies with low-$\sigma$ rings, higher $h_{4}$ values are observed co-spatially with the ring, being interpreted as an additional signature of young/intermediate age stars at these locations.

\item The observed velocity fields are well reproduced by a rotating disk model, with deprojected velocity amplitudes in the range $\sim$60--340\,\kms.

\item The orientations of the line of nodes derived from the small scale velocity fields are similar to the photometric major axis orientations of the large scale disks, while the disk ellipticity and inclination are smaller at small scale, as compared to those at large scale.

\end{itemize} 

The stellar kinematics and rotating disk models derived in this work will be compared to the gas kinematics and flux distributions in future studies with the aim of isolating and quantifying non-circular motions in the gas of the galaxies of our sample in order to map and quantify feeding and feedback processes in our AGN sample.

\section*{Acknowledgments}
We thank an anonymous referee for useful suggestions which helped to improve the paper. 
Based on observations obtained at the Gemini Observatory, 
which is operated by the Association of Universities for Research in Astronomy, Inc., under a cooperative agreement with the 
NSF on behalf of the Gemini partnership: the National Science Foundation (United States), the Science and Technology 
Facilities Council (United Kingdom), the National Research Council (Canada), CONICYT (Chile), the Australian Research 
Council (Australia), Minist\'erio da Ci\^encia e Tecnologia (Brazil) and south-eastCYT (Argentina).  

This research has made use of the NASA/IPAC Extragalactic Database (NED) which is operated by the Jet Propulsion Laboratory, California Institute of Technology, under contract with the National Aeronautics and Space Administration. We acknowledge the usage of the HyperLeda database (http://leda.univ-lyon1.fr).
{\it R.A.R.} and {\it R. R.} thank to CNPq and FAPERGS for financial support.

\appendix

\section{Stellar kinematics based on already published data}\label{appendix_a}

\begin{figure*}
\includegraphics[scale=0.7]{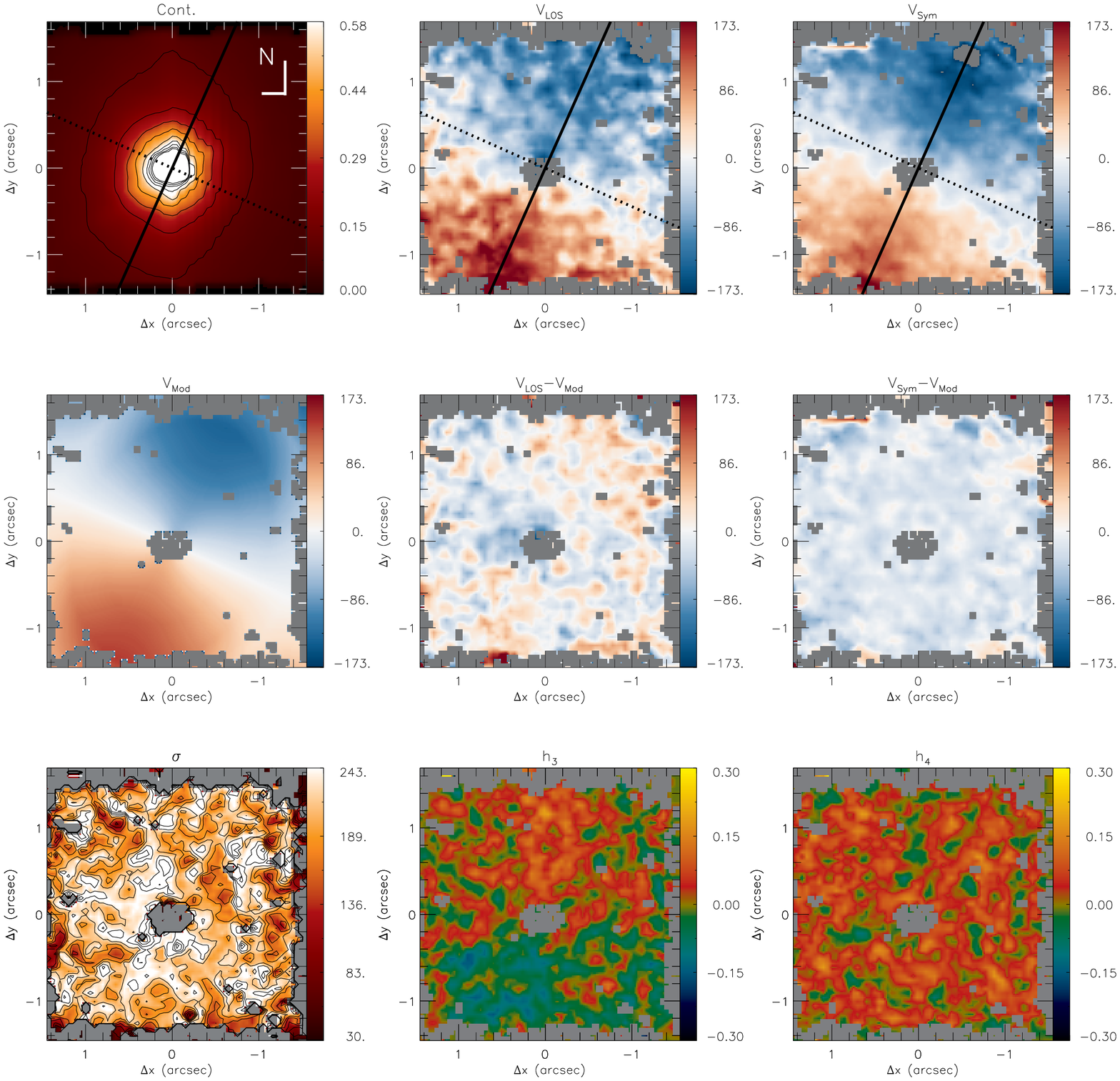}
\caption{Same as Fig.~\ref{n788} for NGC\,2110. The original stellar kinematics measurements are presented in \citet{n2110}}
\label{n2110_fig}
\end{figure*}

\begin{figure*}
\includegraphics[scale=0.7]{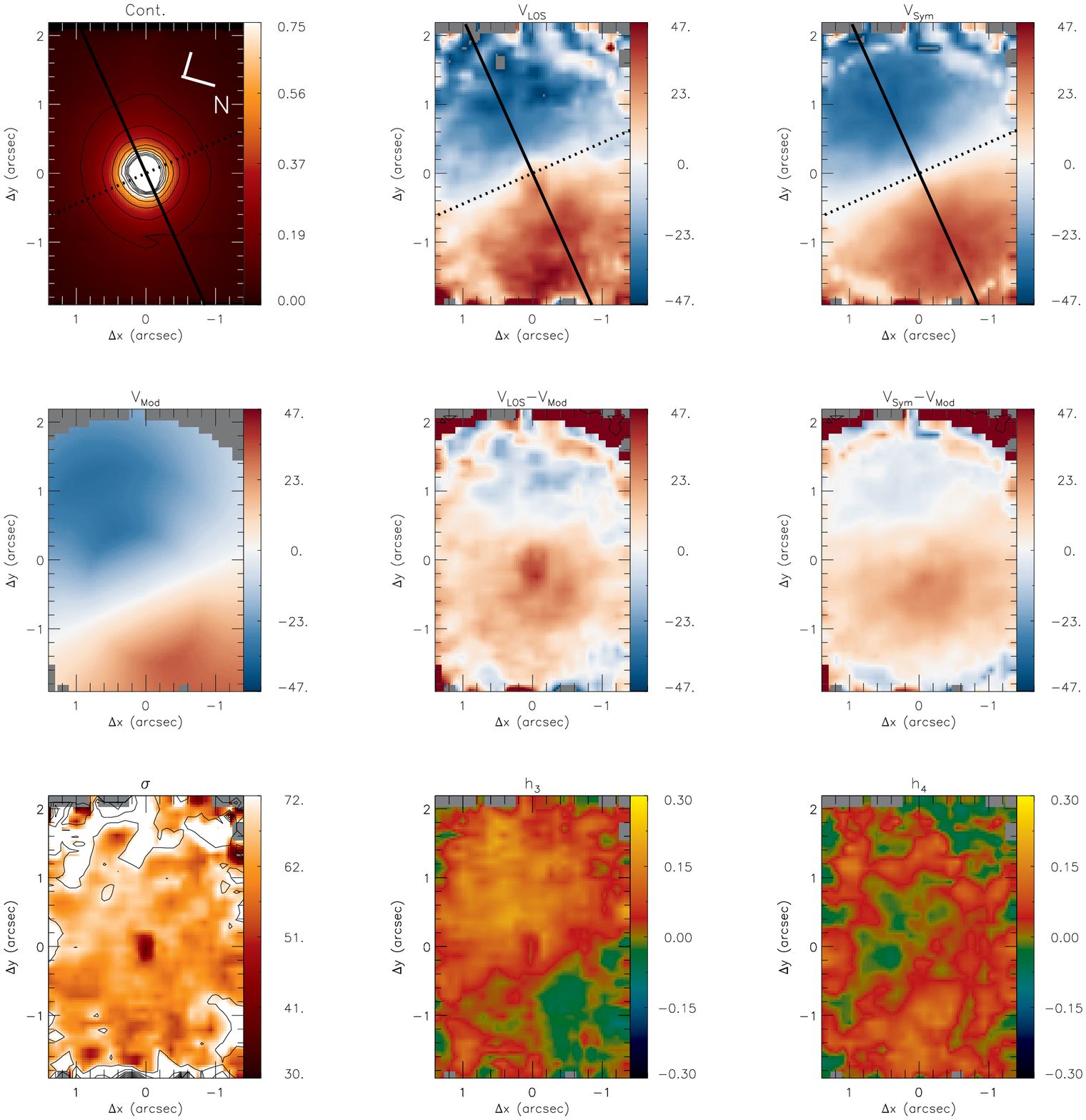}
\caption{Same as Fig.~\ref{n788} for NGC\,4051. The original stellar kinematics measurements are presented in \citet{n4051}}
\label{n4051_fig}
\end{figure*}

\begin{figure*}
\includegraphics[scale=0.7]{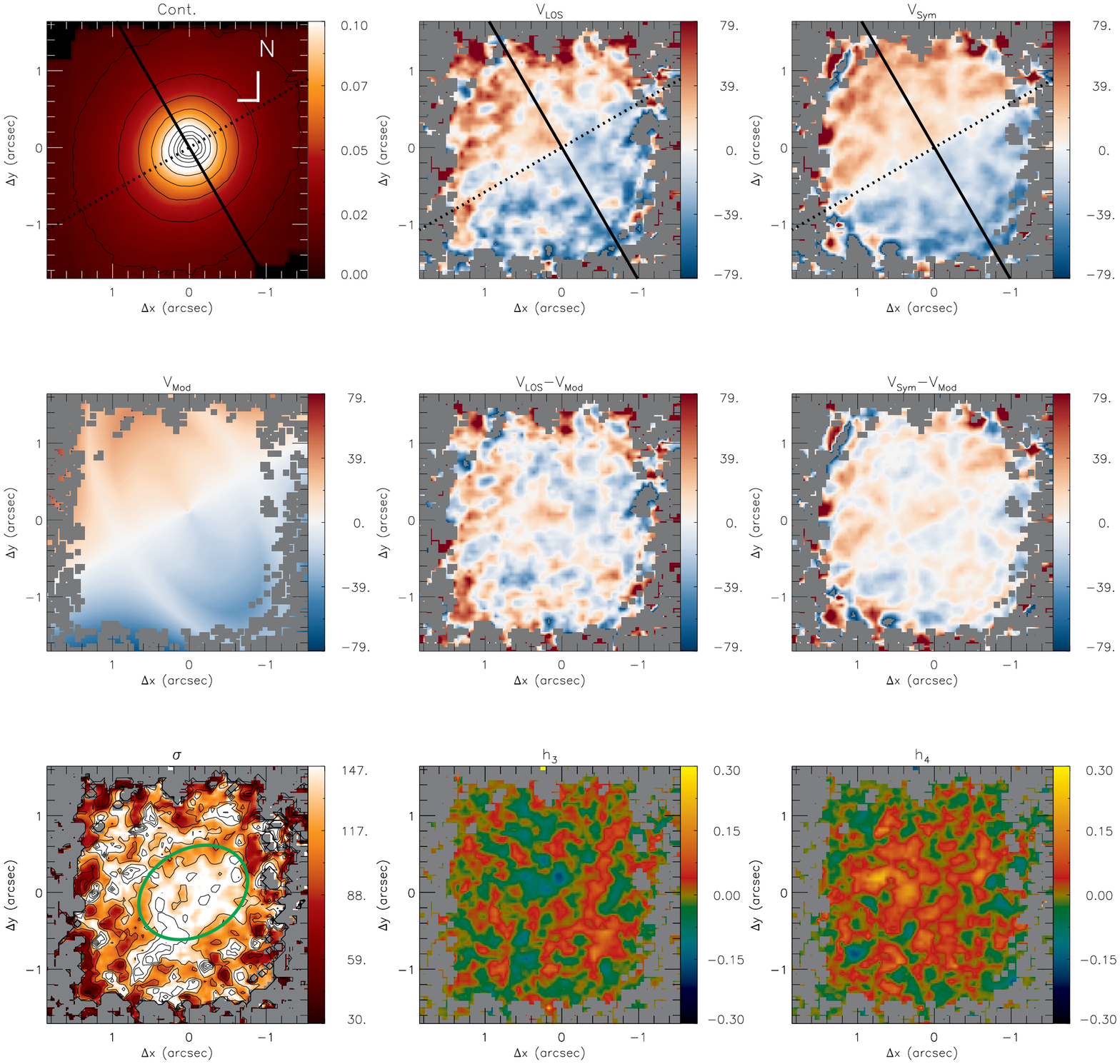}
\caption{Same as Fig.~\ref{n788} for NGC\,5929. The original stellar kinematics measurements are presented in \citet{n5929}}
\label{n5929_fig}
\end{figure*}

\begin{figure*}
\includegraphics[scale=0.7]{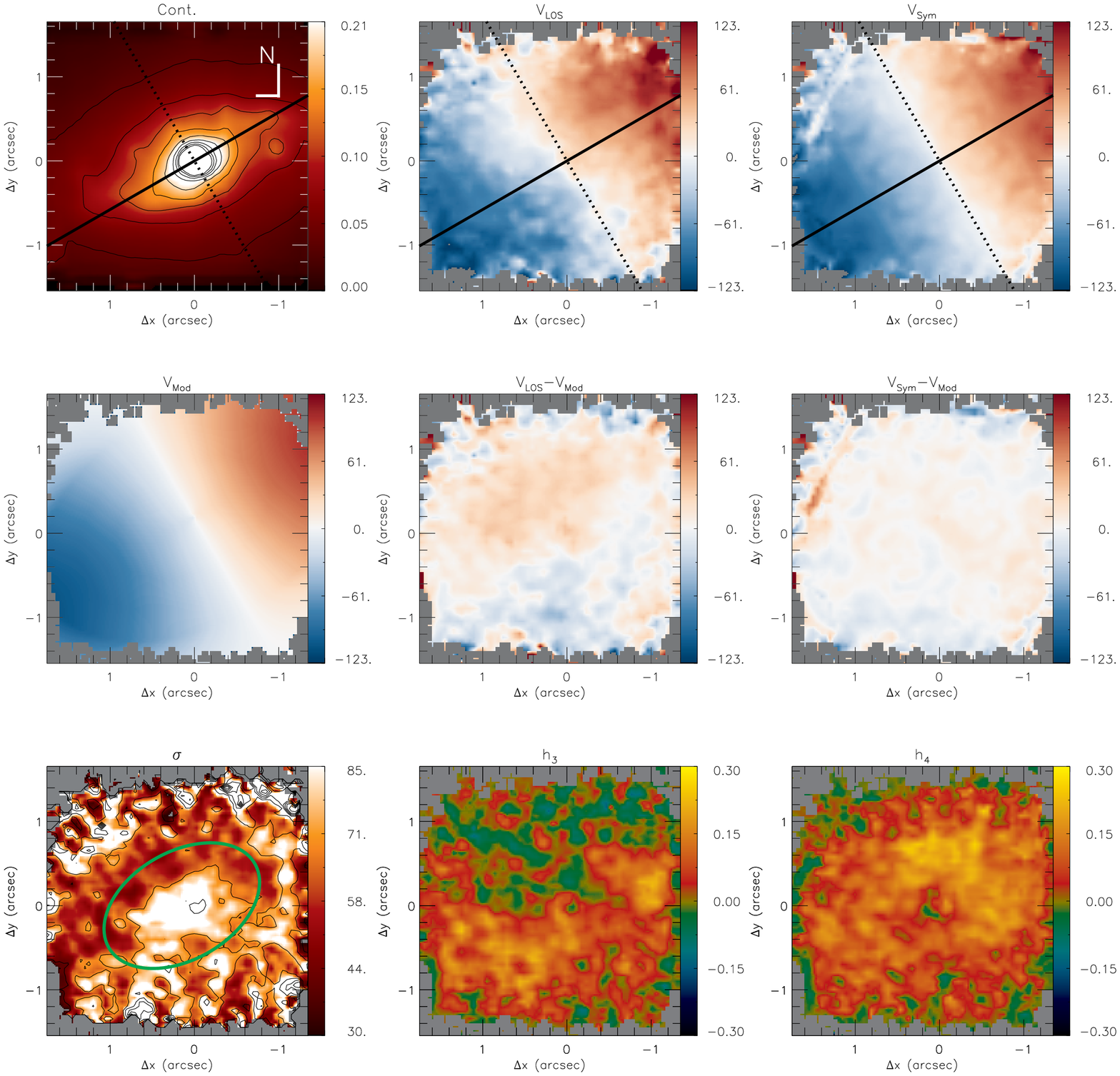}
\caption{Same as Fig.~\ref{n788} for Mrk\,1066. The original stellar kinematics measurements are presented in \citet{m1066_kin}}
\label{mrk1066_fig}
\end{figure*}

\begin{figure*}
\includegraphics[scale=0.7]{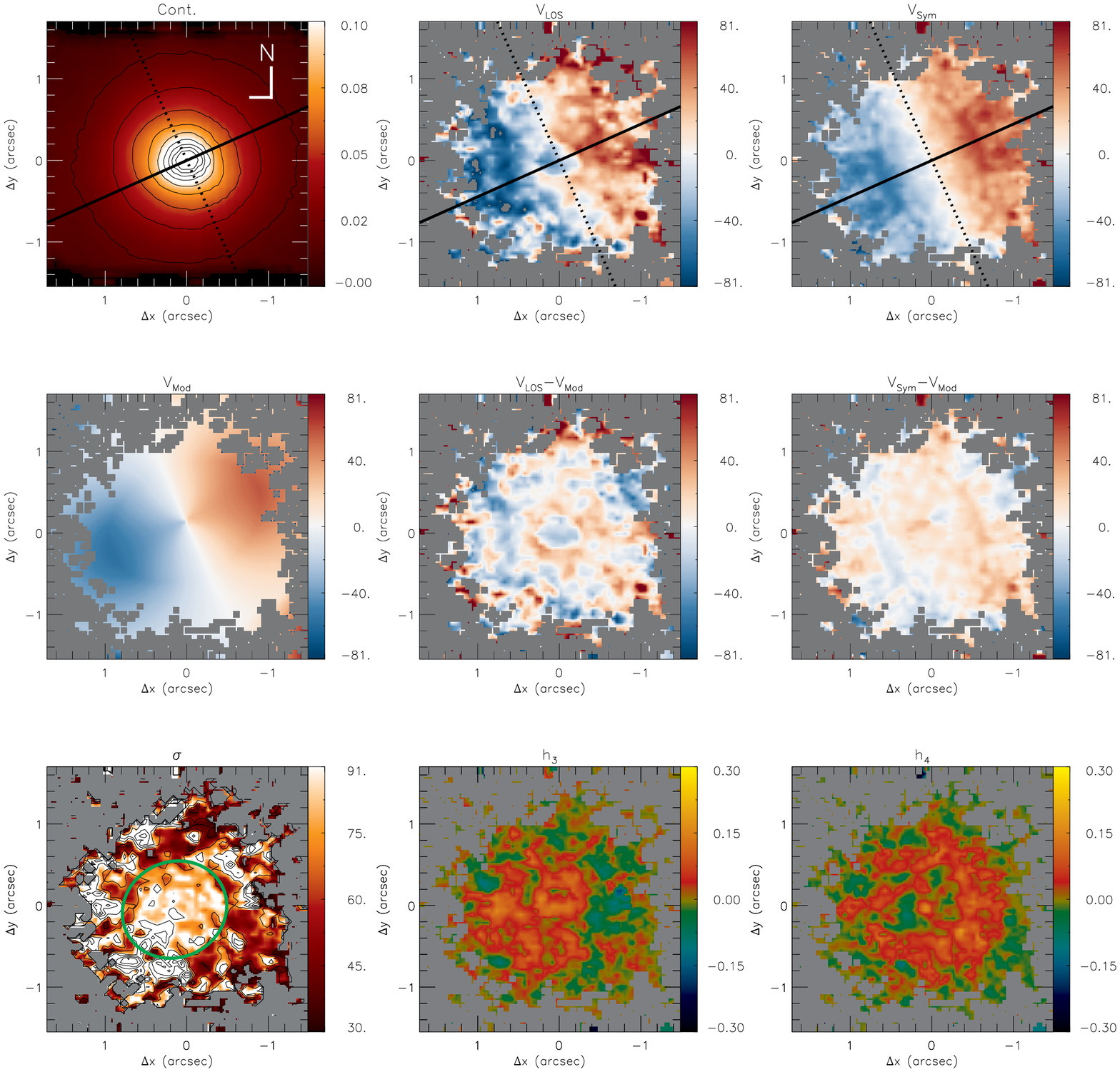}
\caption{Same as Fig.~\ref{n788} for Mrk\,1157. The original stellar kinematics measurements are presented in \citet{m1157_kin}}
\label{mrk1157_fig}
\end{figure*}

Figures \ref{n2110_fig} -- \ref{mrk1157_fig} show maps for the stellar kinematics of galaxies with previous measurements already published by our group.

\section{One-dimensional cuts for the stellar kinematics}\label{appendix}
\begin{figure*}
\centering
\small
\vspace{0.3cm}
\begin{tabular}{r l}
\includegraphics[scale=0.35]{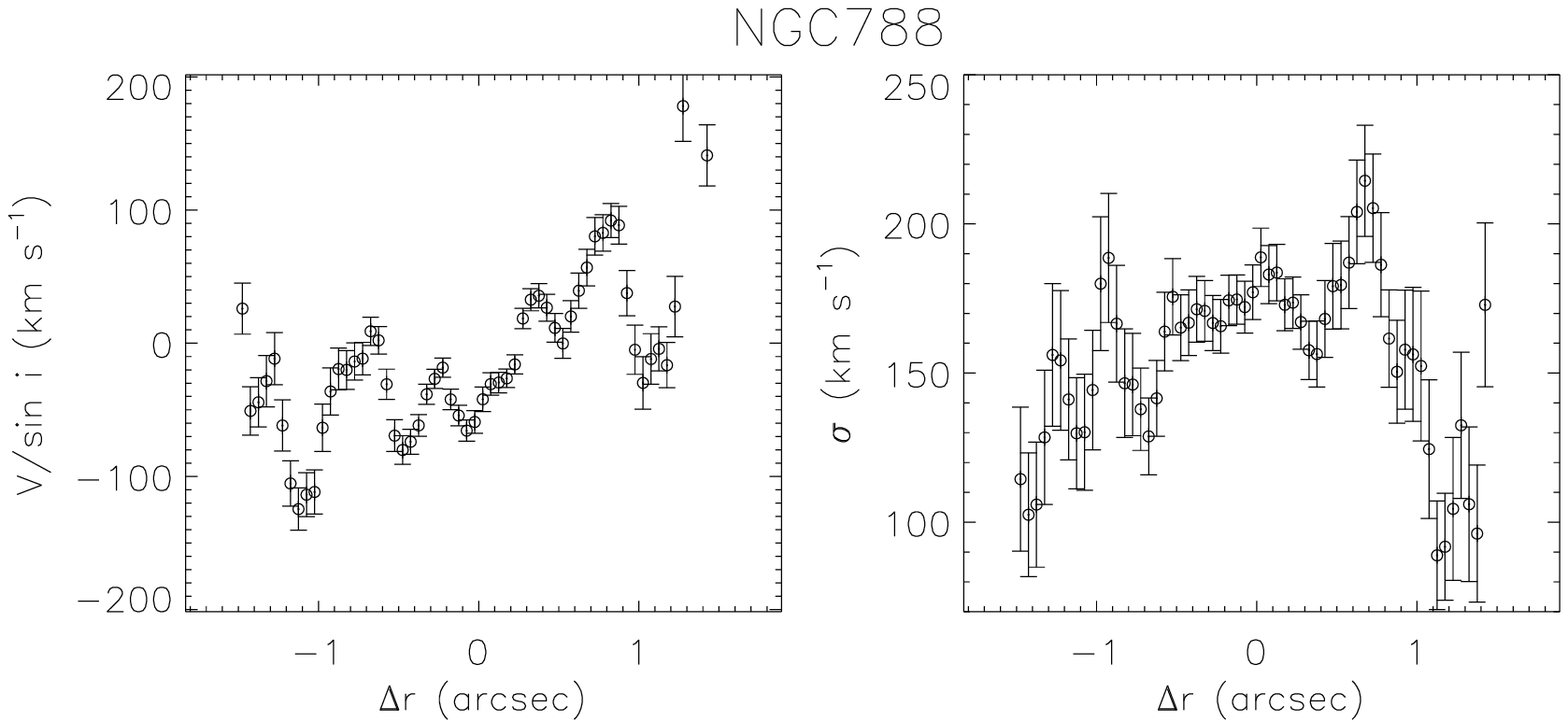} &
\includegraphics[scale=0.35]{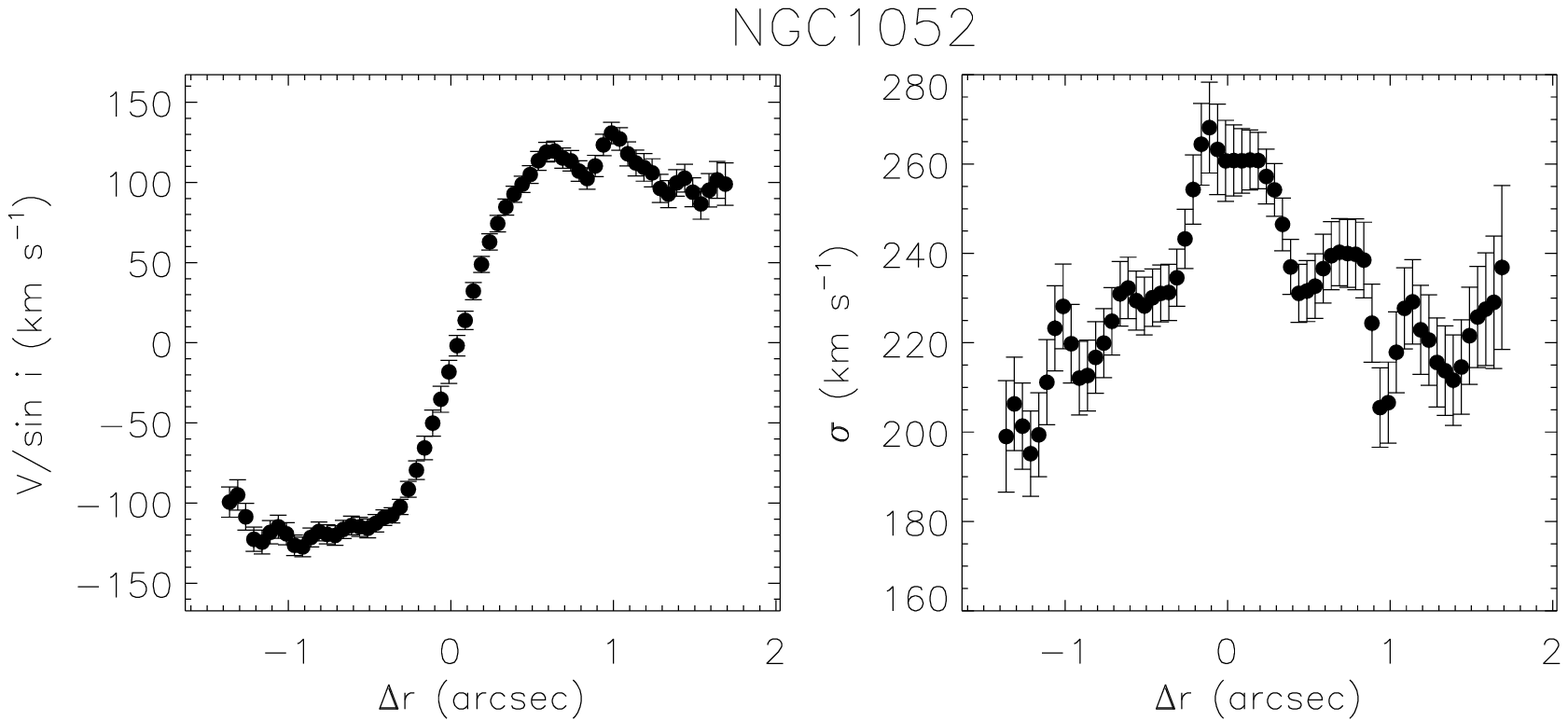} \\

\includegraphics[scale=0.35]{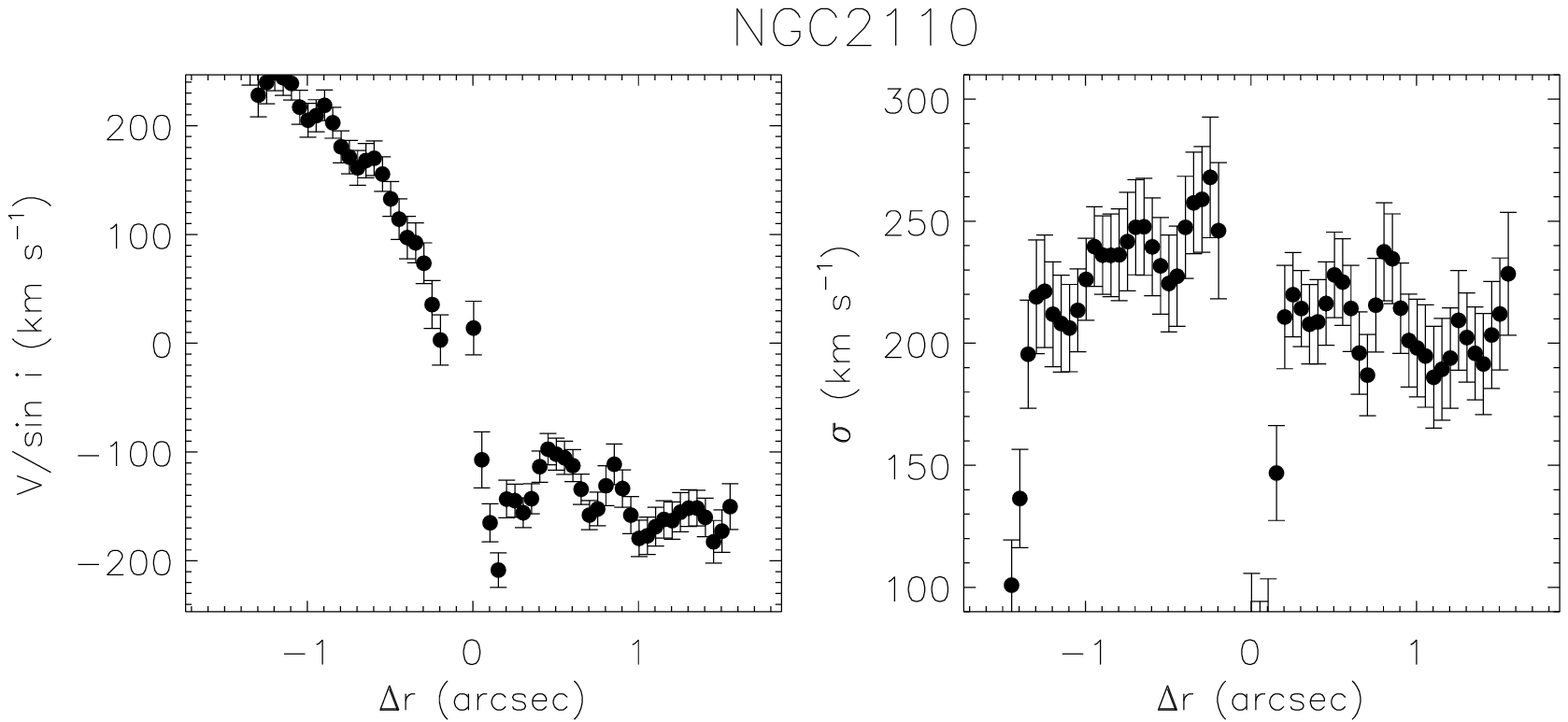} &
\includegraphics[scale=0.35]{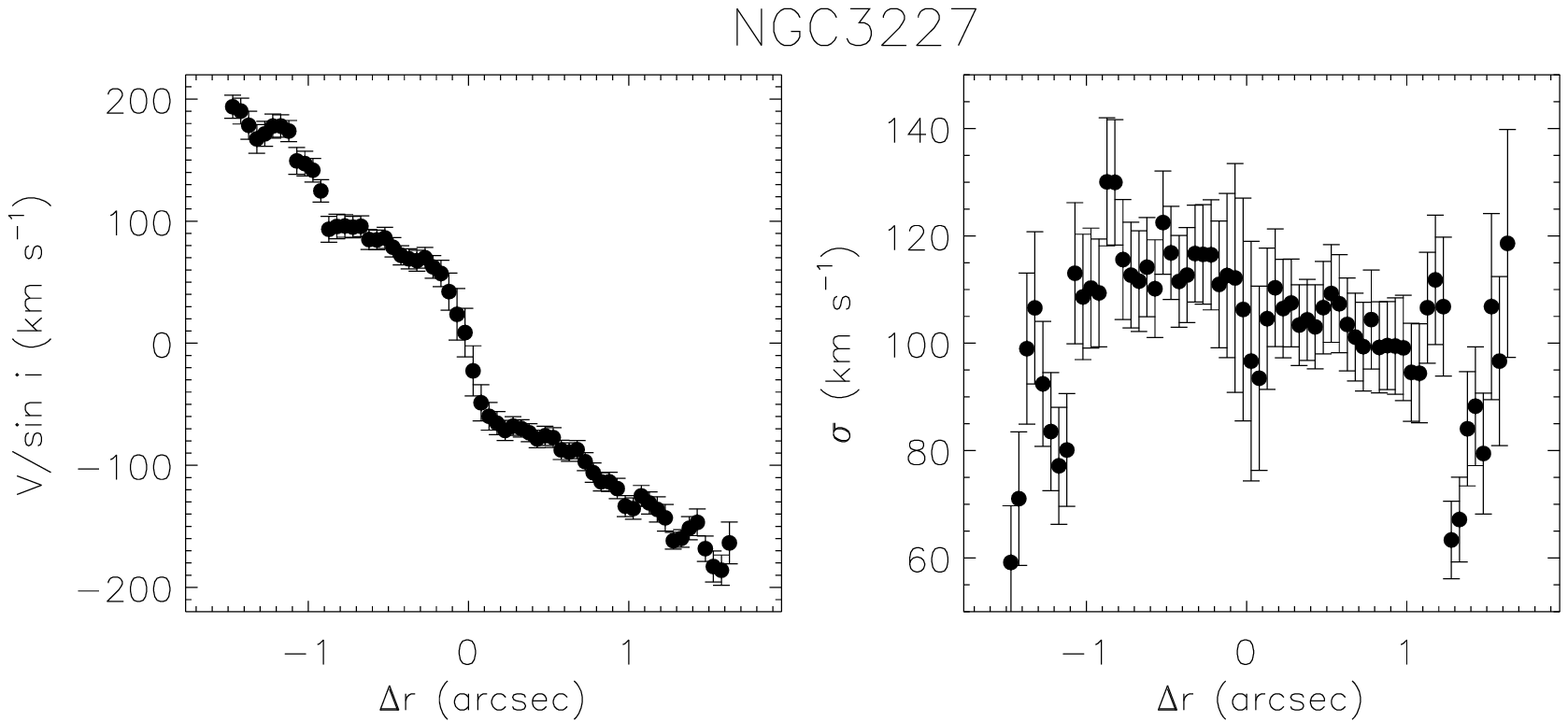} \\

\includegraphics[scale=0.35]{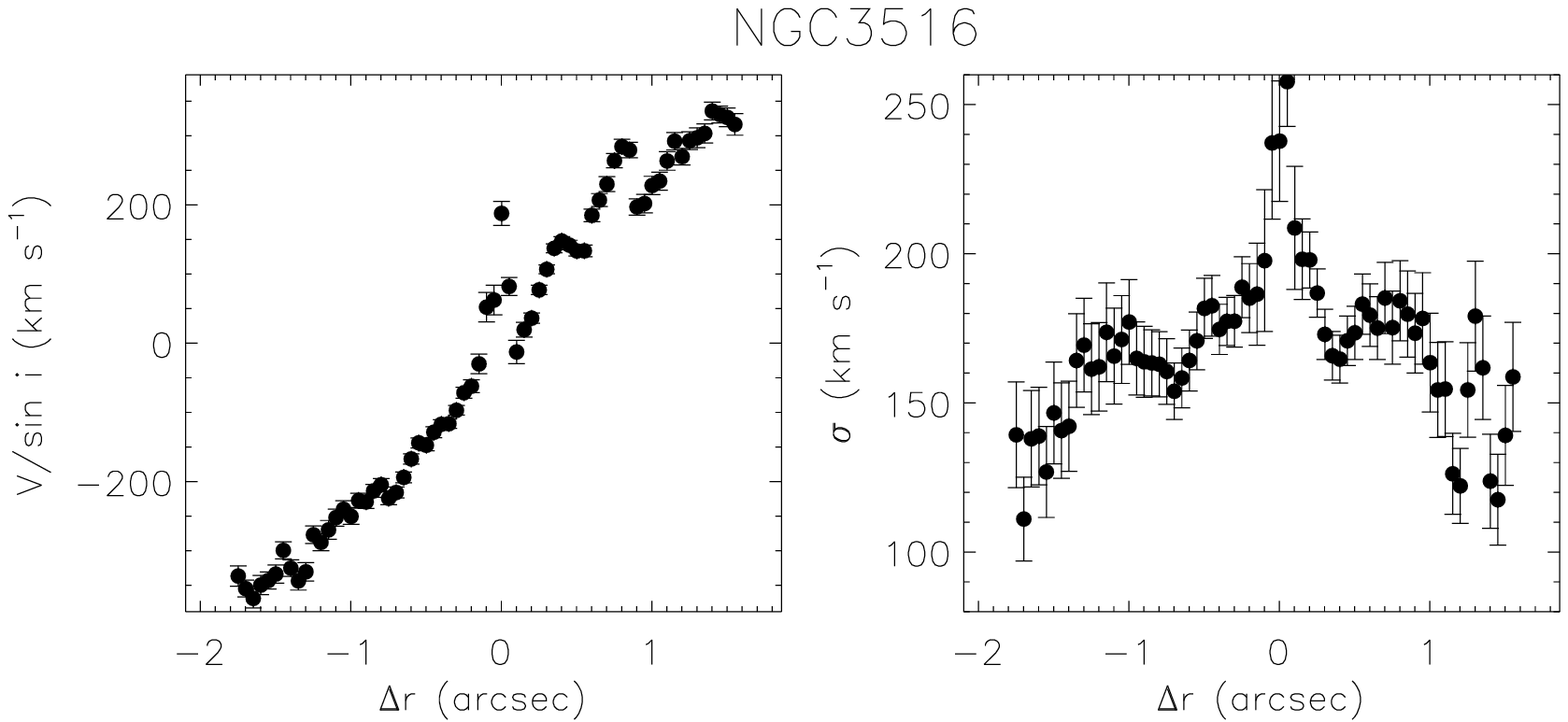} &
\includegraphics[scale=0.35]{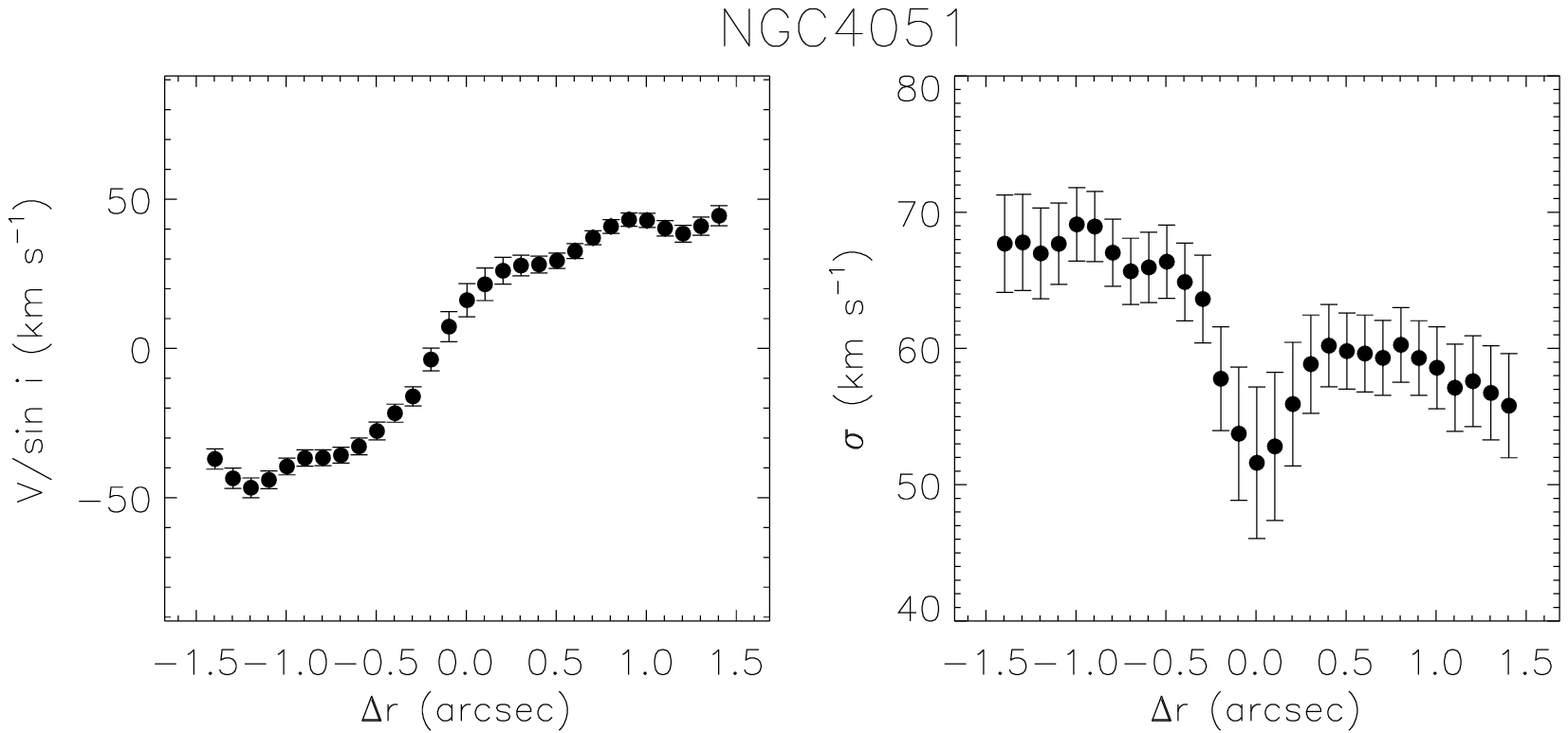} \\

\includegraphics[scale=0.35]{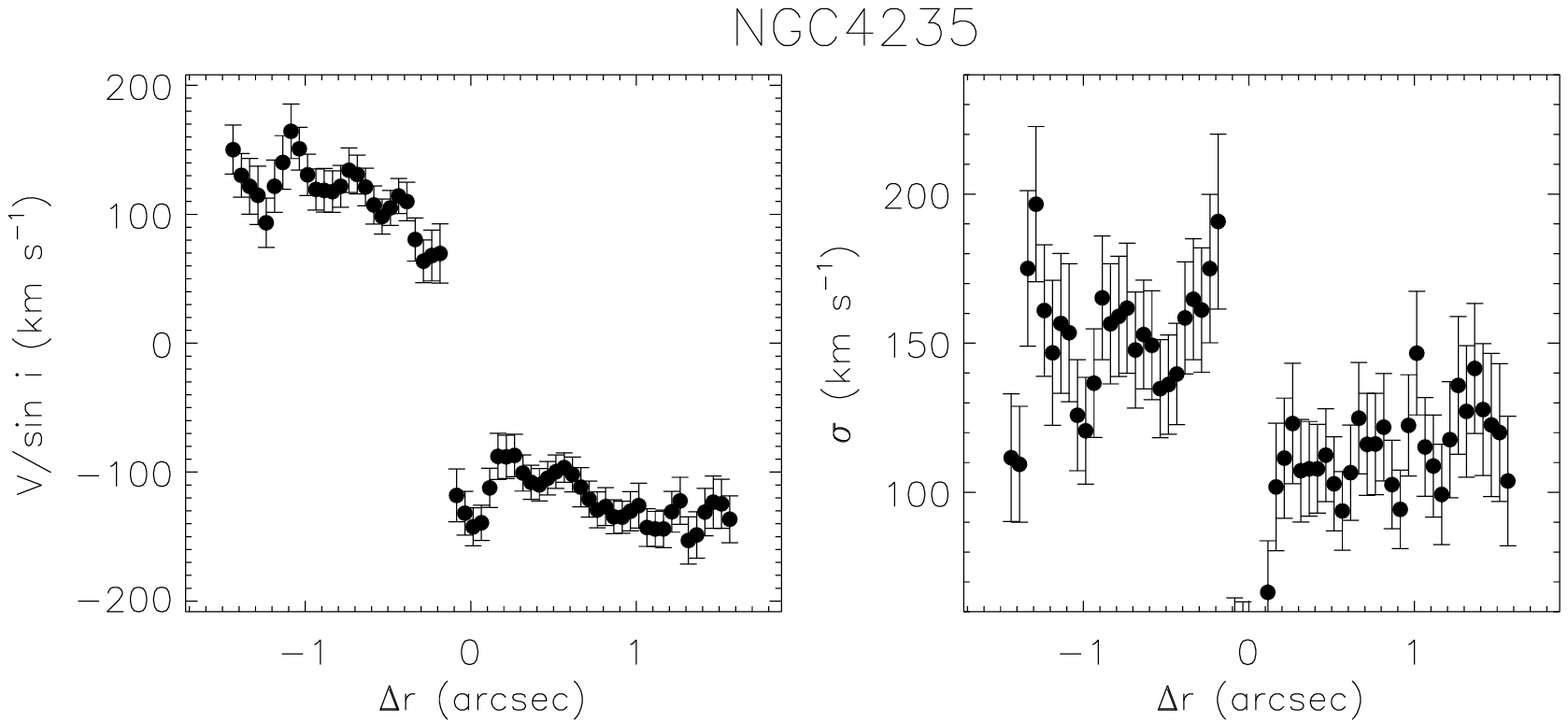} &
\includegraphics[scale=0.35]{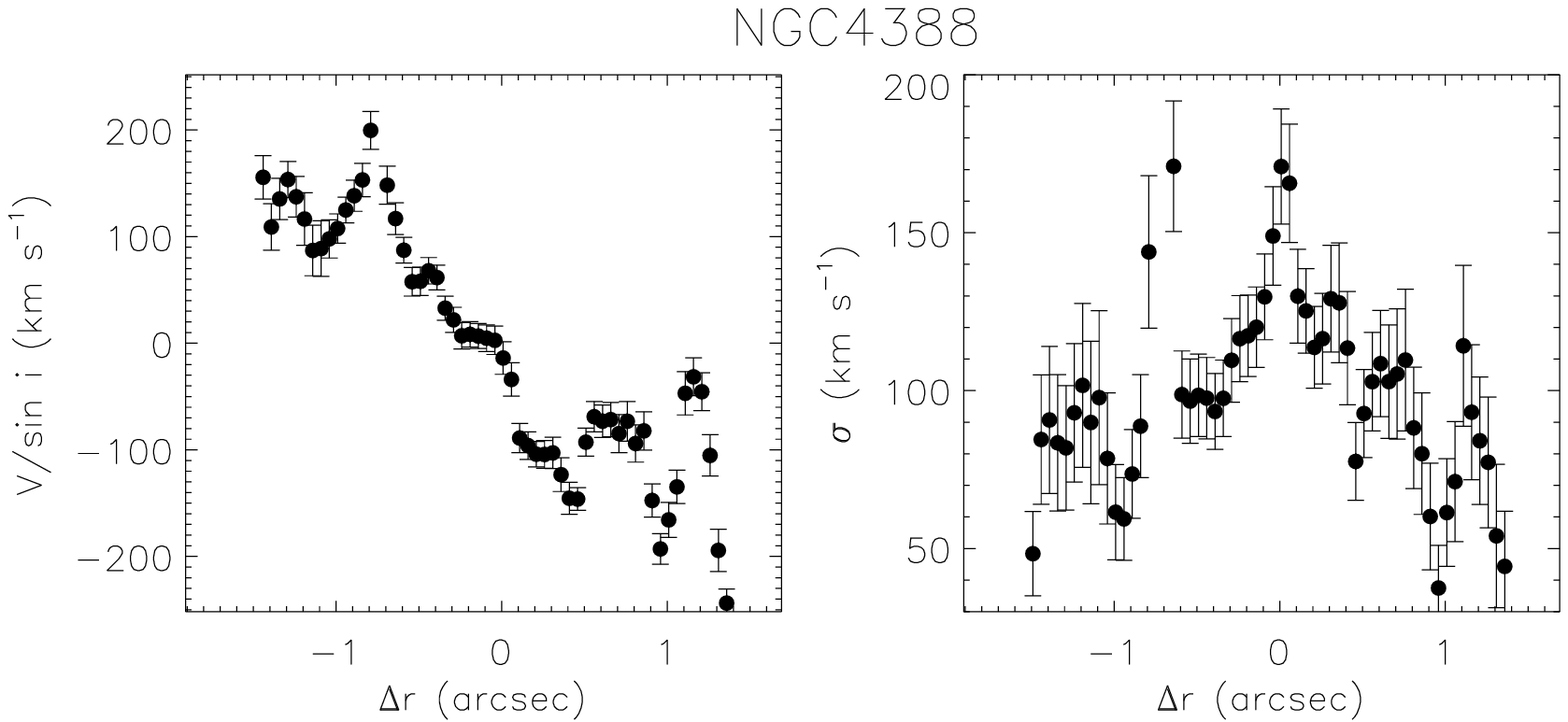} \\

\includegraphics[scale=0.35]{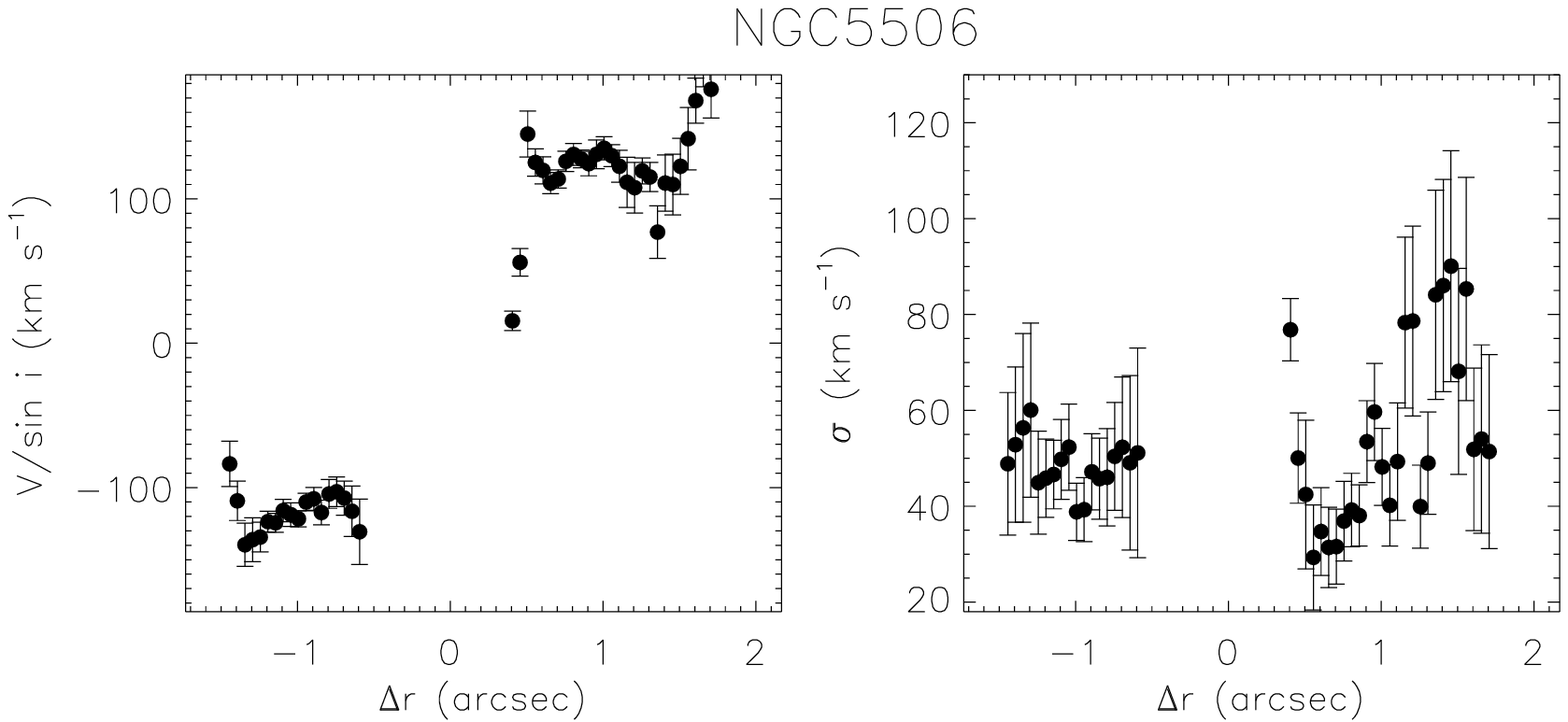} &
\includegraphics[scale=0.35]{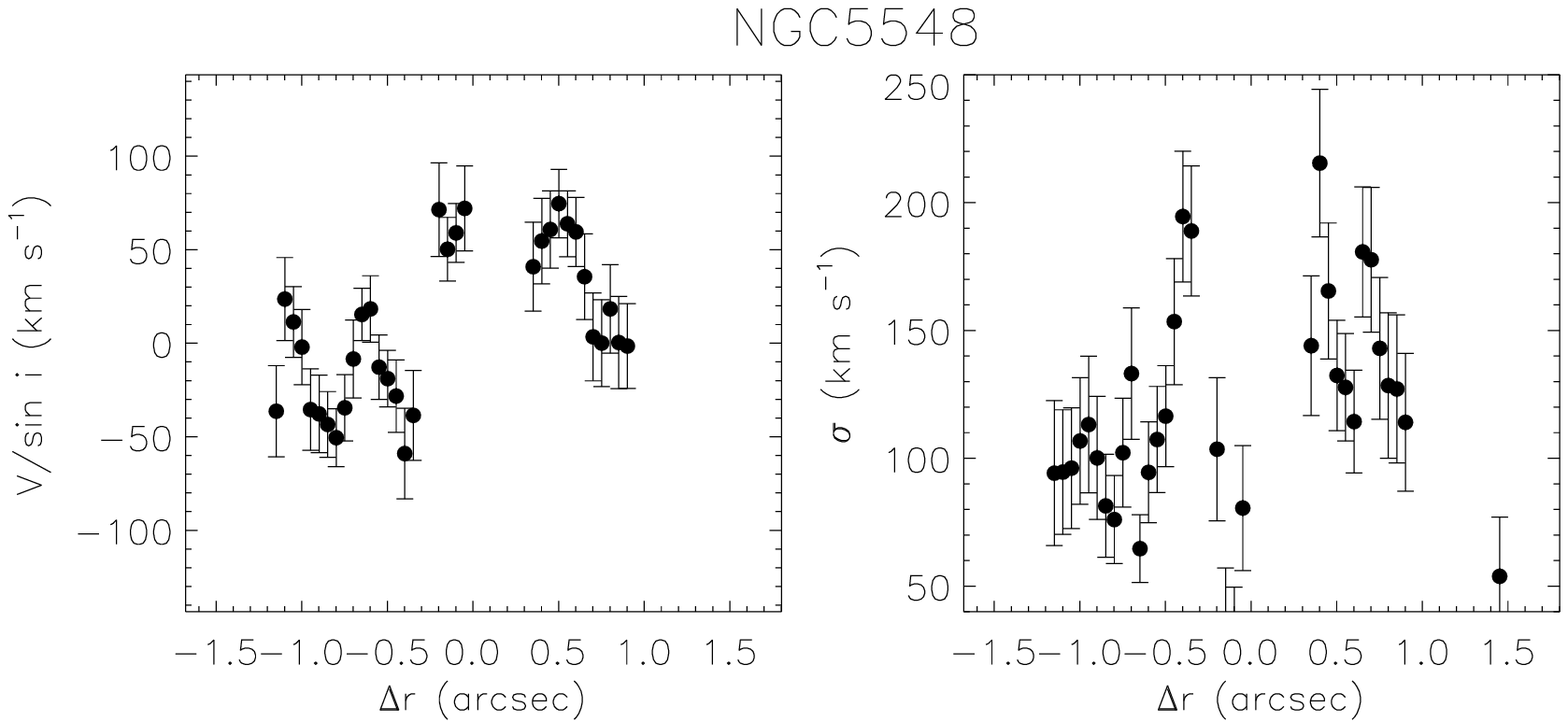} \\

\includegraphics[scale=0.35]{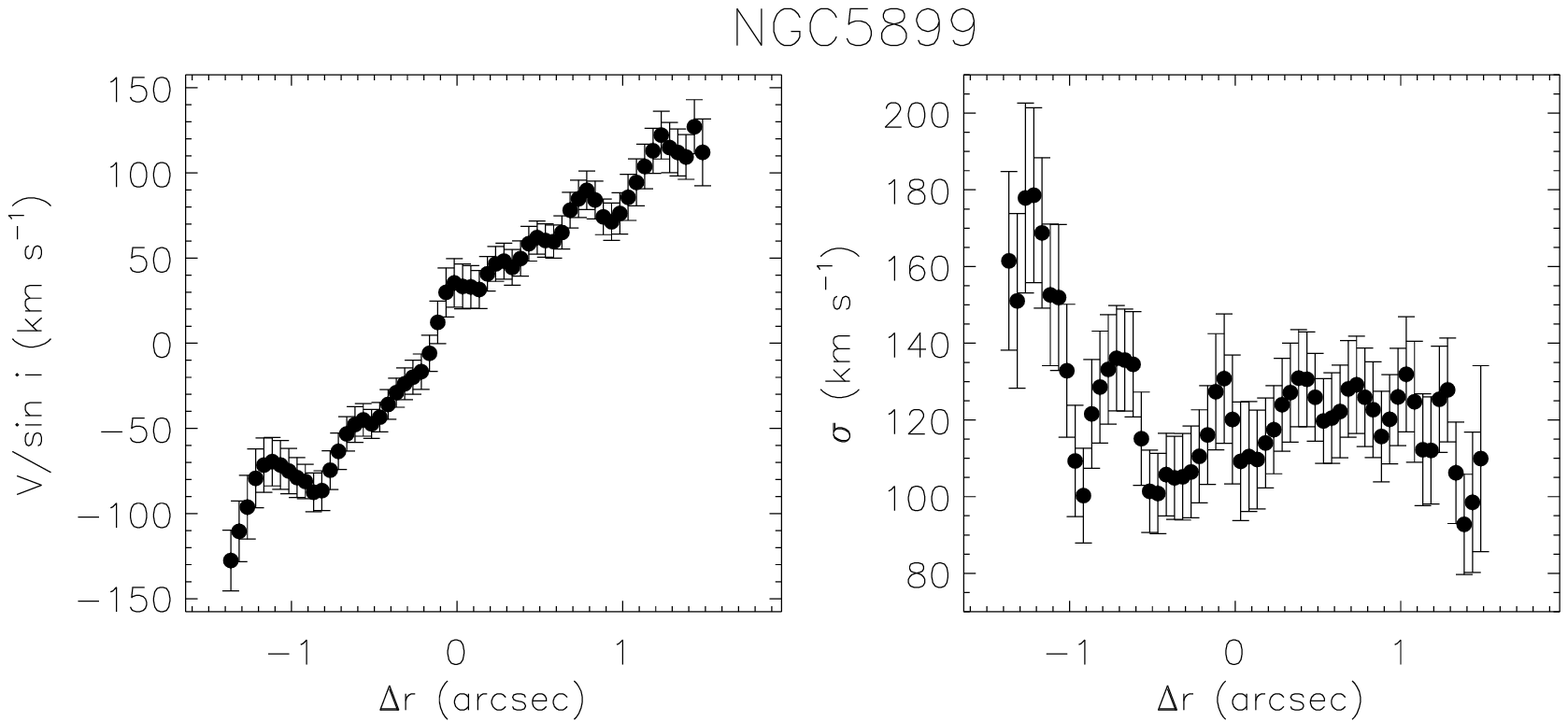} &
\includegraphics[scale=0.35]{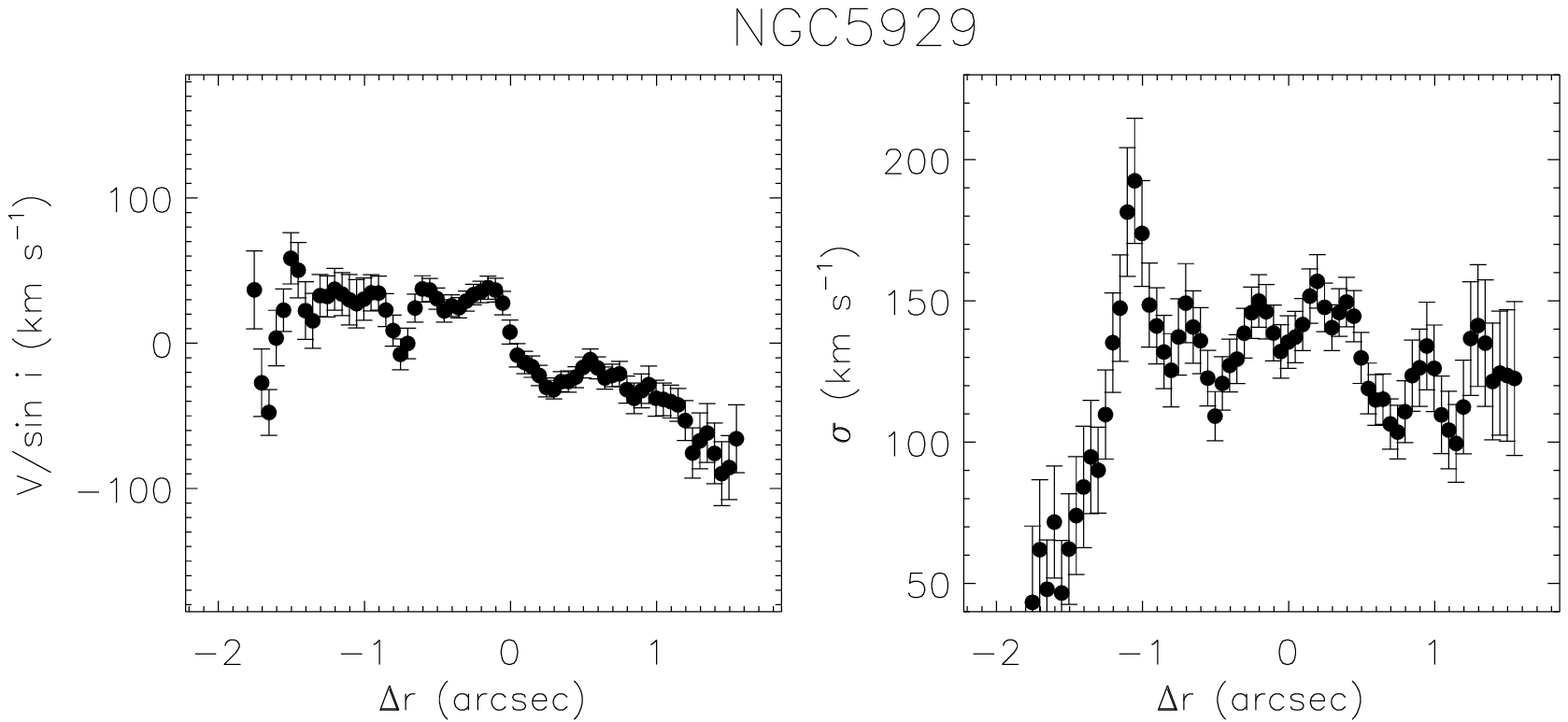} \\

\includegraphics[scale=0.35]{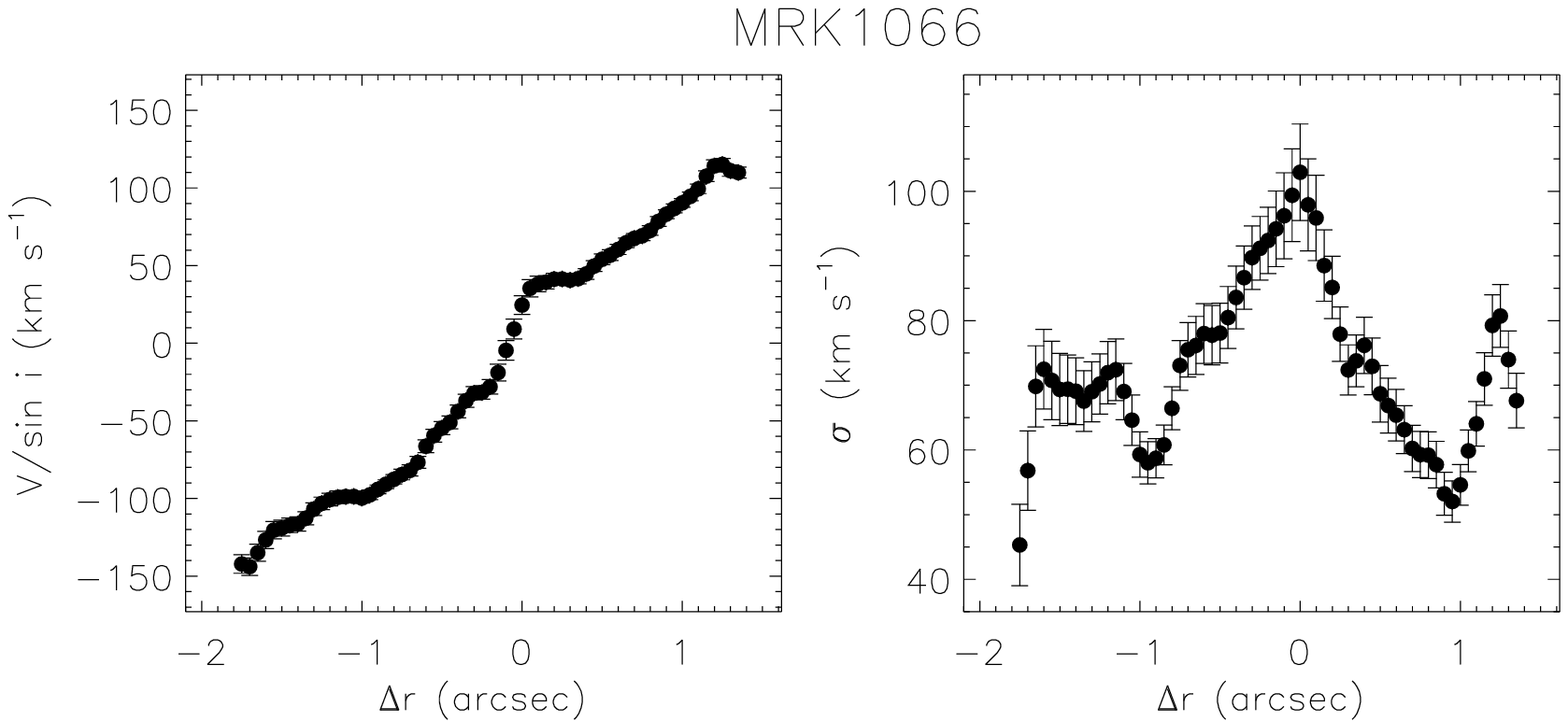} &
\includegraphics[scale=0.35]{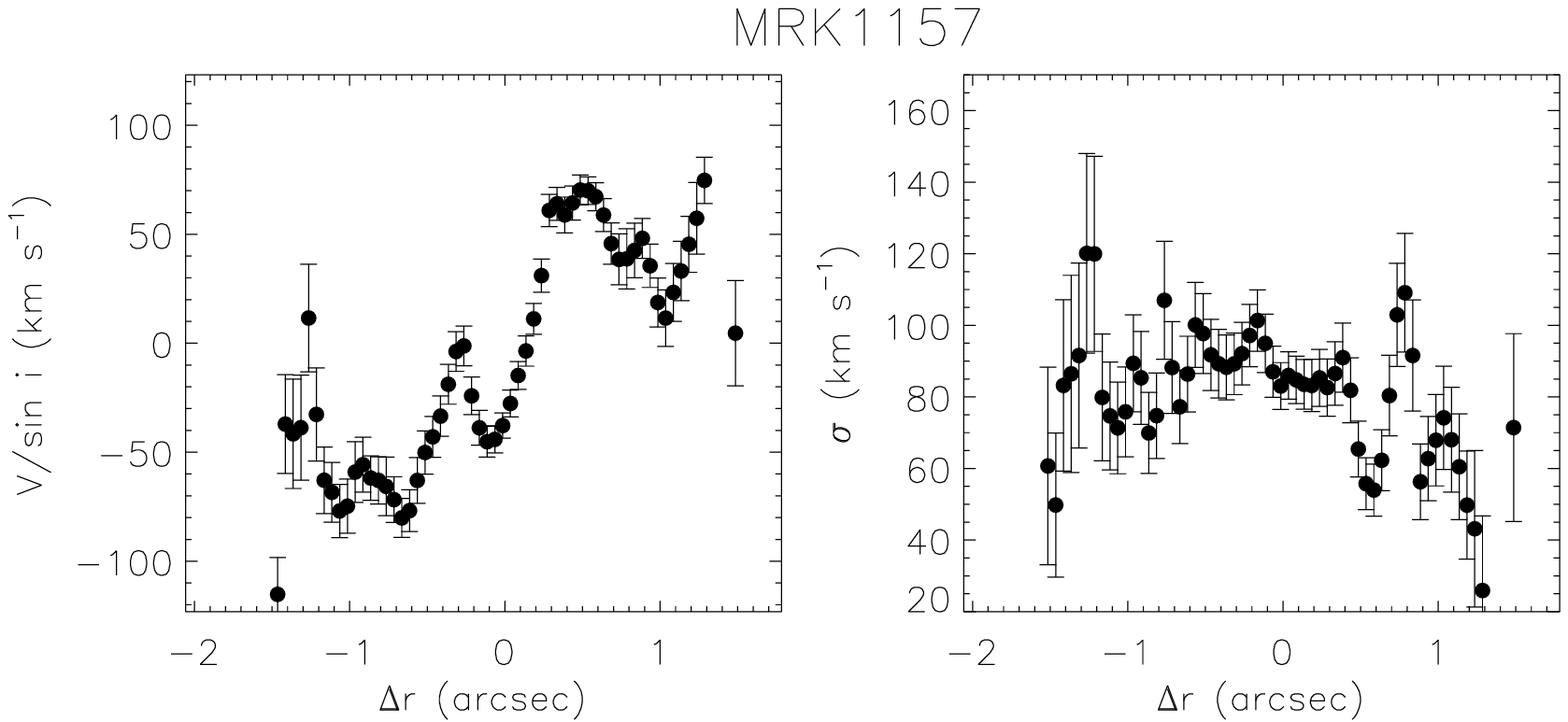} \\

\includegraphics[scale=0.35]{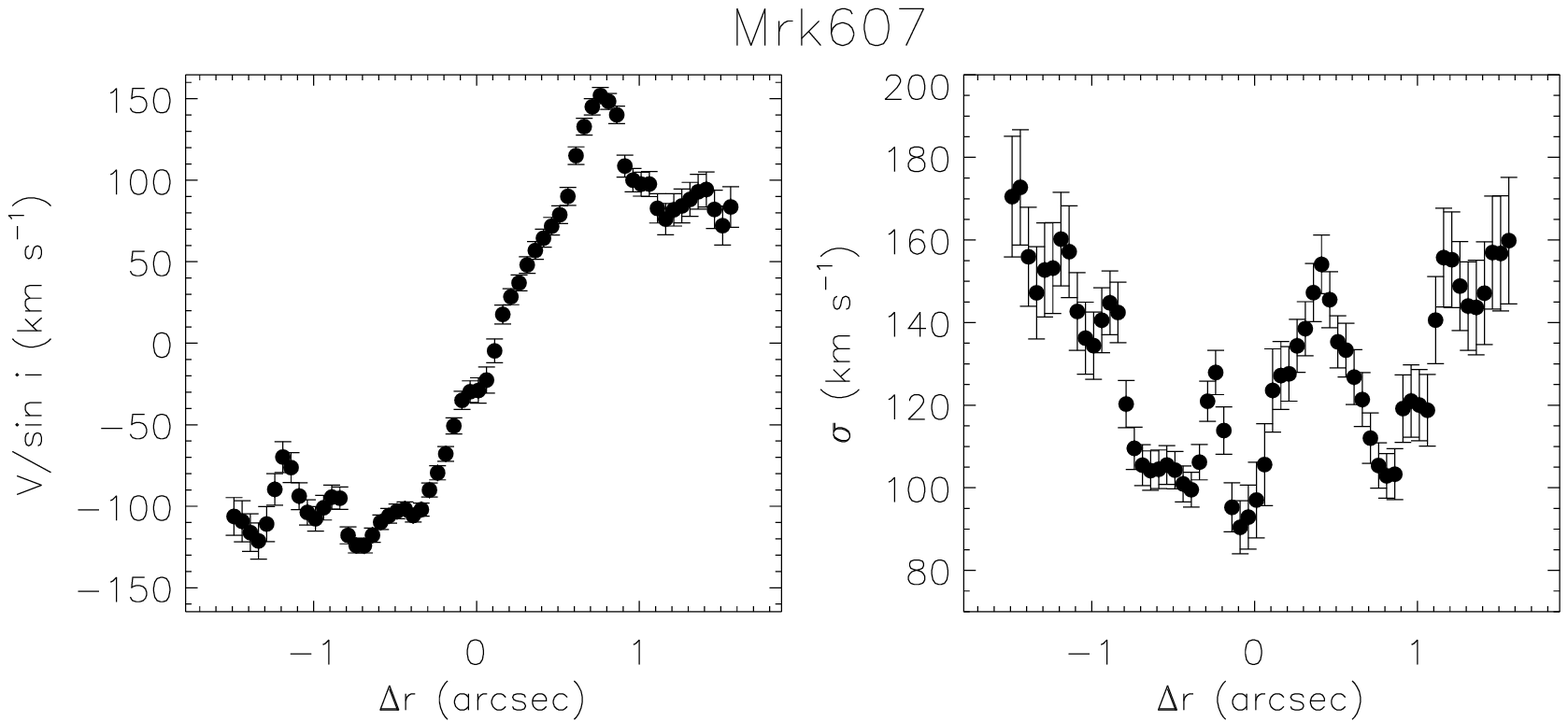} &
\includegraphics[scale=0.35]{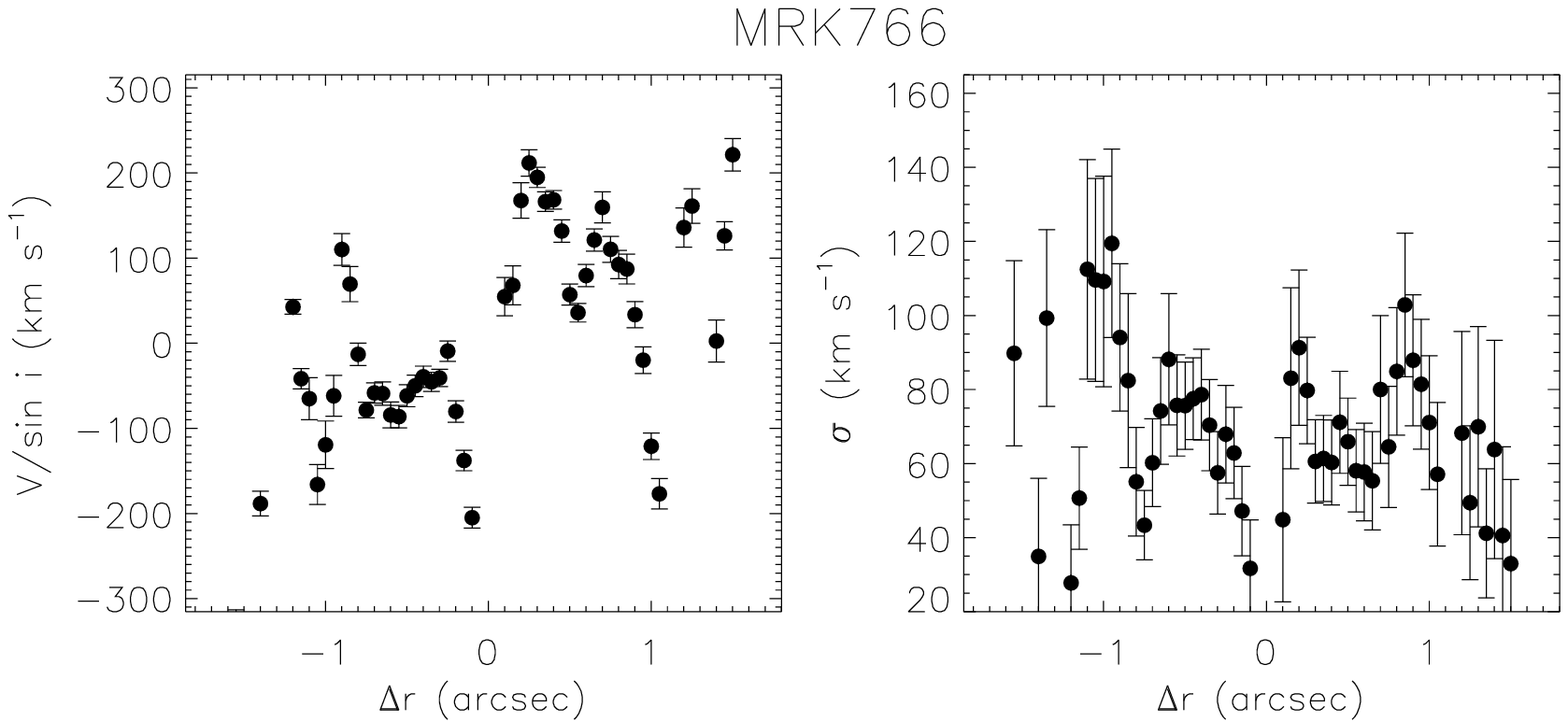} \\
\end{tabular}
\caption{One-dimensional cuts along the major axis of the galaxies obtained by averaging the observed velocities within a pseudo-slit with 0\farcs25 width. The orientation of the major axis  and the inclination of the disk used in these plots are from Table.\ref{tab_mod}.}
\label{vcuts}
\end{figure*}

\begin{figure*}
\centering
\includegraphics[scale=1]{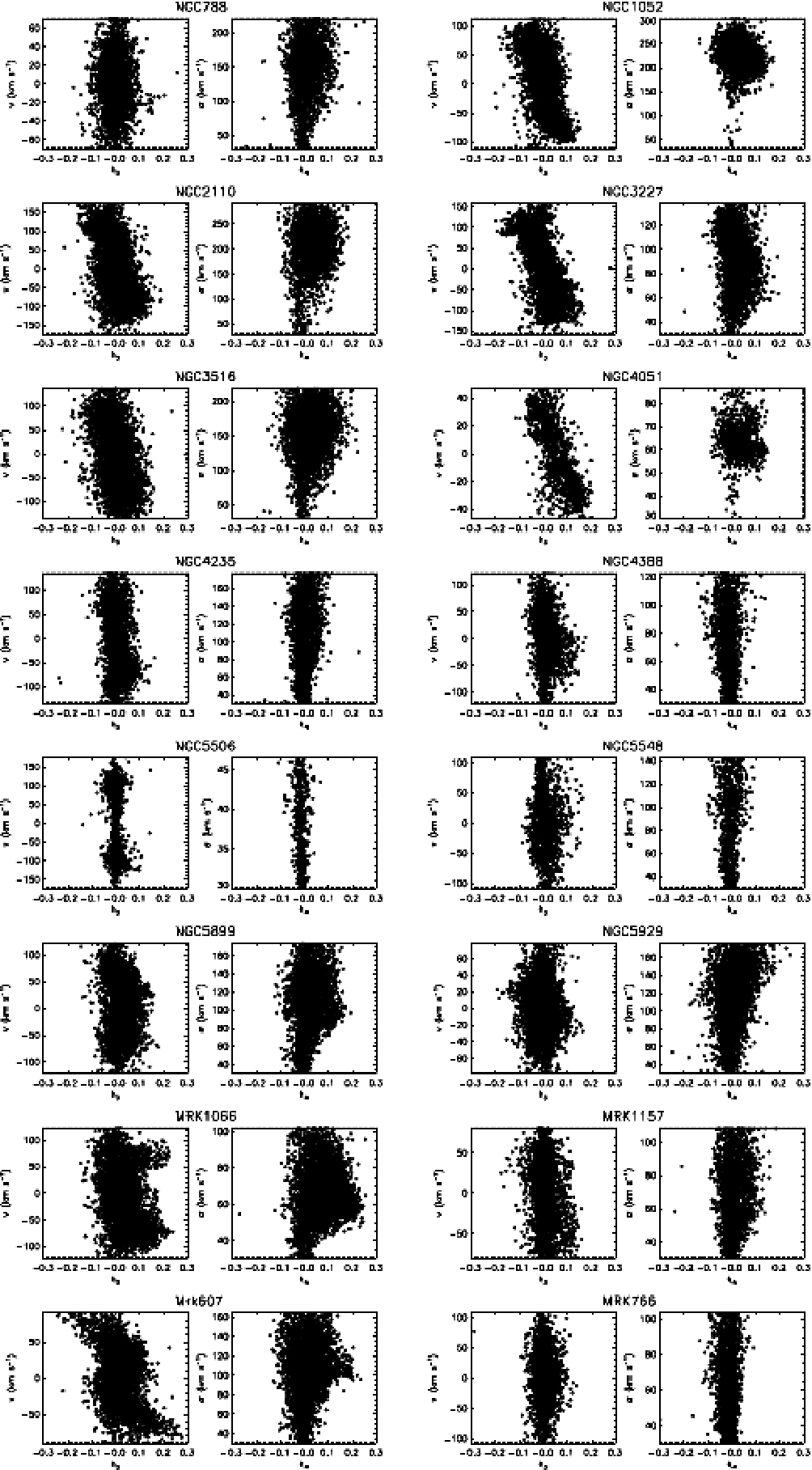} 
\caption{Plots of the LOS velocity ($V_{ LOS}$) vs. $h_3$  and $\sigma$ vs. $h_4$ for the galaxies of our sample. For most galaxies $V_{LOS}$ and $h_3$ are anti-correlated and a trend of higher values of $\sigma$ being observed at locations with negative $h_4$ values and smaller $\sigma$ values associated to positive $h_4$ values.}
\label{h3vel}
\end{figure*}

Figure \ref{vcuts} shows one-dimensional cuts along the major axis of the galaxies for the LOS velocity (left) and $\sigma$ (rigth). Plots of the LOS velocity ($V_{ LOS}$) vs. $h_3$  and $\sigma$ vs. $h_4$ using all spaxels are shown in Figure~\ref{h3vel}.

\label{lastpage}

\end{document}